\documentclass[review,3p,twocolumn]{elsarticle}
\usepackage{amsmath,comment}
\usepackage[colorlinks]{hyperref}

\newcommand{\rhom}{\overline\rho}
\newcommand{\At}{$A$ }

\journal{Journal of \LaTeX\ Templates}

\begin{document}

\begin{frontmatter}

\title{Variable-density buoyancy-driven turbulence with asymmetric initial density distribution}
\tnotetext[mytitlenote]{Asymmetric HVDT}


\author[DA,DL]{Denis Aslangil}
\ead{denis.aslangil@gmail.com}

\author[DL]{Daniel Livescu}
\ead{livescu@lanl.gov}

\author[DA]{Arindam Banerjee}
\ead{arb612@lehigh.edu}

\address[DA]{Department of Mechanical Engineering \& Mechanics, Lehigh University,
Bethlehem, PA 18015, USA}
\address[DL]{Los Alamos National Laboratory, Los Alamos, NM 87545, USA}

\begin{abstract}
The effects of different initial density distributions on the evolution of buoyancy-driven homogeneous variable-density turbulence (HVDT) at low (0.05) and high (0.75) Atwood numbers are studied by using high-resolution direct numerical simulations. HVDT aims to mimic the acceleration-driven Rayleigh-Taylor and shock-driven Richtmyer-Meshkov instabilities and reveals new physics that arise from variable-density effects on the turbulent mixing. Here, the initial amounts of pure light and pure heavy flows are altered primarily to mimic the variable-density turbulence at the different locations of the Rayleigh-Taylor and Richtmyer-Meshkov instabilities' mixing layers where the amounts of the mixing fluids are not equal. It is found that for the low Atwood number cases, the asymmetric initial density distribution has limited effects on both global and local flow evolution for HVDT. However, at high Atwood number, both global flow evolution and the local flow structures are strongly affected by the initial composition ratio. The flow composed of more light fluid reaches higher turbulent levels and the local statistics reach their fully-developed behavior earlier in the time evolution. During the late time decay, where most of the flow is well-mixed, all parameters become independent of the initial composition ratio for both low and high Atwood number cases.
\end{abstract}


\end{frontmatter}

\section{Introduction}\label{Sec:intro}

In this study, we investigate, using direct numerical simulations (DNS), the mixing of two miscible fluids with different densities (or molar masses) in the idealized flow termed as homogeneous variable-density turbulence (HVDT) \cite{batchelor1992,Sandoval1997,livescu2007,livescu2008,aslangil2019}. The occurrence of the variable-density (VD) mixing in atmospheric and oceanic flows \citep{MOLCHANOV2004559,Adkins1769,Wunsch_2004_doi:10.1146/annurev.fluid.36.050802.122121}, supernova formations \citep{Gull_1975_doi:10.1093/mnras/171.2.263,Colgate_1966_ApJ...143..626C}, combustion applications in ramjet engines \citep{GIVI1989,clemens_mungal_1995,Sellers_Chandra_1997_doi:10.1108/02644409710157596} and high energy density processes like inertial confinement fusion \citep{Lindl_1995_doi:10.1063/1.871025,lindl1998inertial,Nakai_1996_0034-4885-59-9-002,Nakai_2004_0034-4885-67-3-R04} makes HVDT a fundamental flow to investigate VD dynamics.
        In HVDT, the triply periodic domain contains the heterogeneous mixture of pure light and pure heavy fluids as random patches \cite{batchelor1992,Sandoval1997,livescu2007,aslangil2019}. When the acceleration field is applied to the domain, these pure fluids start to move in opposite directions, similar to the case of acceleration-driven Rayleigh-Taylor Instability. These buoyancy-driven motions generate turbulent kinetic energy ($E_{TKE}$) which significantly enhances the molecular mixing of these two fluids. Eventually, the flow becomes turbulent and, subsequently, the turbulent dissipation starts to overcome $E_{TKE}$ generation while the flow mixes and buoyancy-forces weaken. HVDT evolution is thus highly non-equilibrium and comprises birth, growth, and gradual decay of the VD turbulence \cite{batchelor1992}. This dynamic behavior allows us to connect this idealized flow to various engineering applications and natural phenomena similar to the core region of the mixing layer of acceleration driven Rayleigh Taylor (RT/RTI) and shock driven Richtmyer-Meshkov (RM/RMI) instabilities. Moreover, HVDT has the capability to capture most of the important dynamics that are observed in VD jets, VD mixing layers, Rayleigh-B\'enard Instability (RBI) and RTI with acceleration reversals \cite{aslangil2019,livescu2020,Dimonte_2007,Ramaprabhu_ADA,Denis_PRE}.

We define the initial composition ratio ($\overline{\chi}_0$) of the flow as:

\begin{equation}\label{Eq:Initial_ratio}
    \overline{\chi}_0=\frac{<\chi_{h}>_0}{<\chi_{l}>_0}
\end{equation}
where $<\chi_{h}>_0$ and $<\chi_{l}>_0$ are the initial mole fractions of the pure heavy and pure light fluids, respectively. Most of the published works in scientific literature \citep{batchelor1992,SandovalPhd,livescu2007,aslangil2019} have only investigated cases in which the initial amounts of pure light and pure heavy fluids are equal ($<\chi_{h}>_0$=$<\chi_{l}>_0\approx0.5$) and the initial composition ratio is unity ($\overline{\chi_0}\approx0.5/0.5=1$). Most natural and engineering applications seldom have balanced amounts of pure light and heavy fluids. For example, in combustion applications, the initial amounts of pure fluids can vary significantly from fuel rich to lean conditions. Different initial amounts of heavy and light fluids also pose a new challenge for mix models, and have not been tested before \cite{ZHOU2017_1,ZHOU2017_2}. 

We thus choose to alter the initial composition ratio to identify possible effects of differential initial density distributions in HVDT mixing with the low ($1.1:1$) and high ($7:1$) density ratios. 
    Atwood number (\At) is another important non-dimensional number, which represents the ratio of the densities of heavy and light fluids and is defined as:
\begin{equation}\label{Eq:At}
    A = \frac{\rho_2-\rho_1}{\rho_2+\rho_1}\qquad \Rightarrow \qquad \frac{\rho_2}{\rho_1}=\frac{1+A}{1-A}
\end{equation}
where, $\rho_2$ and $\rho_1$ are the densities of the heavy and light fluids, respectively. Here, we study the low \At number case ($A=0.05$), which is close to the Boussinesq case, and the high \At number case ($A=0.75$), which is the non-Boussinesq case. In VD turbulence, at high \At numbers, the mixing behavior and the turbulence features of different density regions differ from each other \citep{livescu2008,livescu2009mav,banerjee_kraft_andrews_2010,livescu2013nst,aslangil2019}; for example, pure light fluid mixes faster than pure heavy fluid \cite{livescu2008}.
Recently, Aslangil et al. \cite{aslangil2019} showed that, at high density ratios, conditional expectations of $E_{TKE}$, its dissipation, and the enstrophy also become asymmetric when with respect to the density field. They also showed that during $E_{TKE}$ growth, the lighter fluid regions move faster compared to the heavier fluid regions. This asymmetric behavior has been attributed to the smaller inertia of the lighter fluid regions. 
        Here, we aim to test the generality of such results by changing the initial amounts of heavy and light fluids and to answer the following questions: 
        
        (1) Is it possible to achieve a more symmetric late time state, or to enhance or suppress the turbulence growth starting with asymmetric initial conditions? 
        
        (2) Are the local statistics of the flow and the flow topology affected by the initial composition of the flow? 
        
\begin{figure*}
 \noindent(\emph{a}) \hspace{4cm}  (\emph{b}) \hspace{4cm}  (\emph{c})\\
    \centering
    \includegraphics[width=4.6cm]{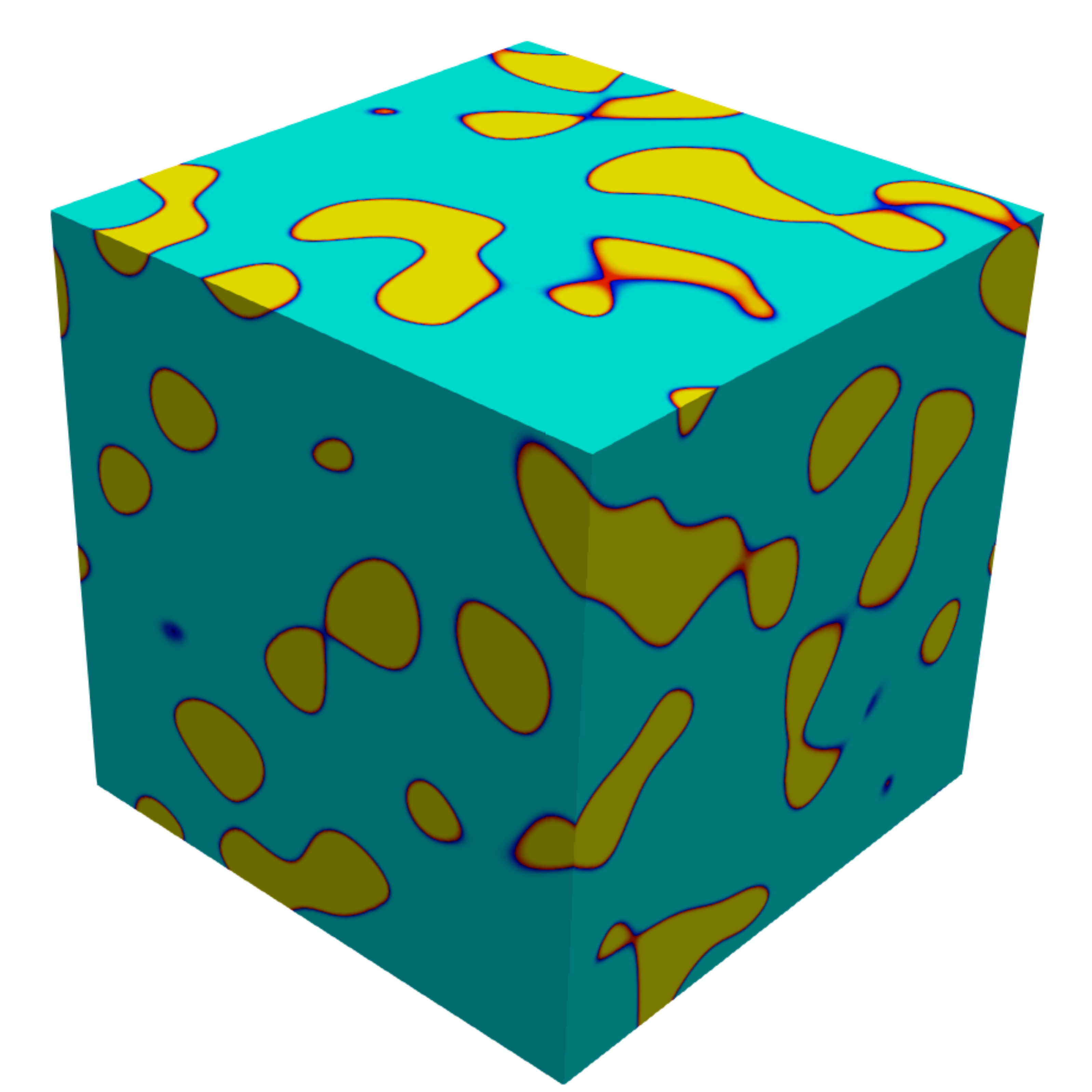}\includegraphics[width=4.6cm]{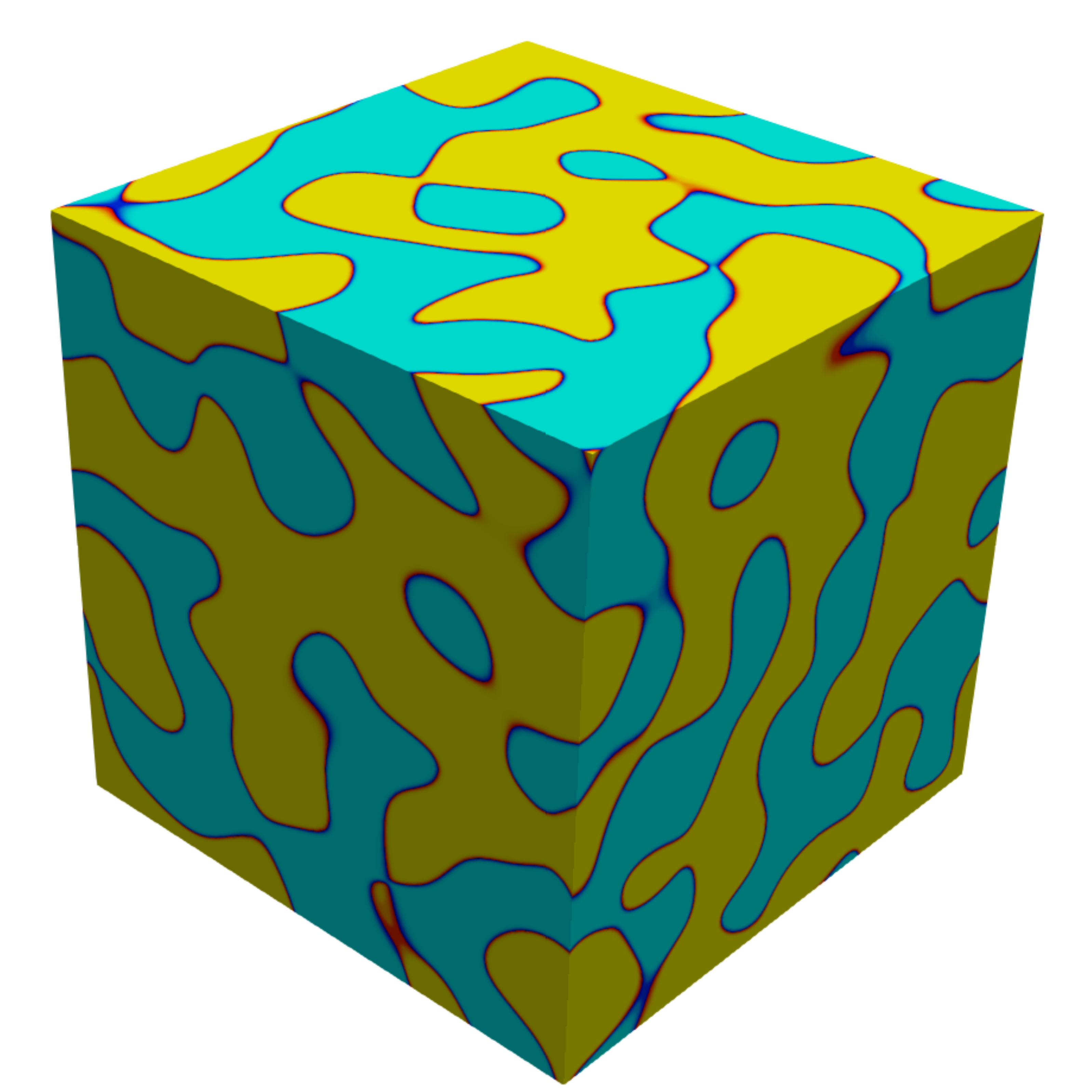}\includegraphics[width=4.6cm]{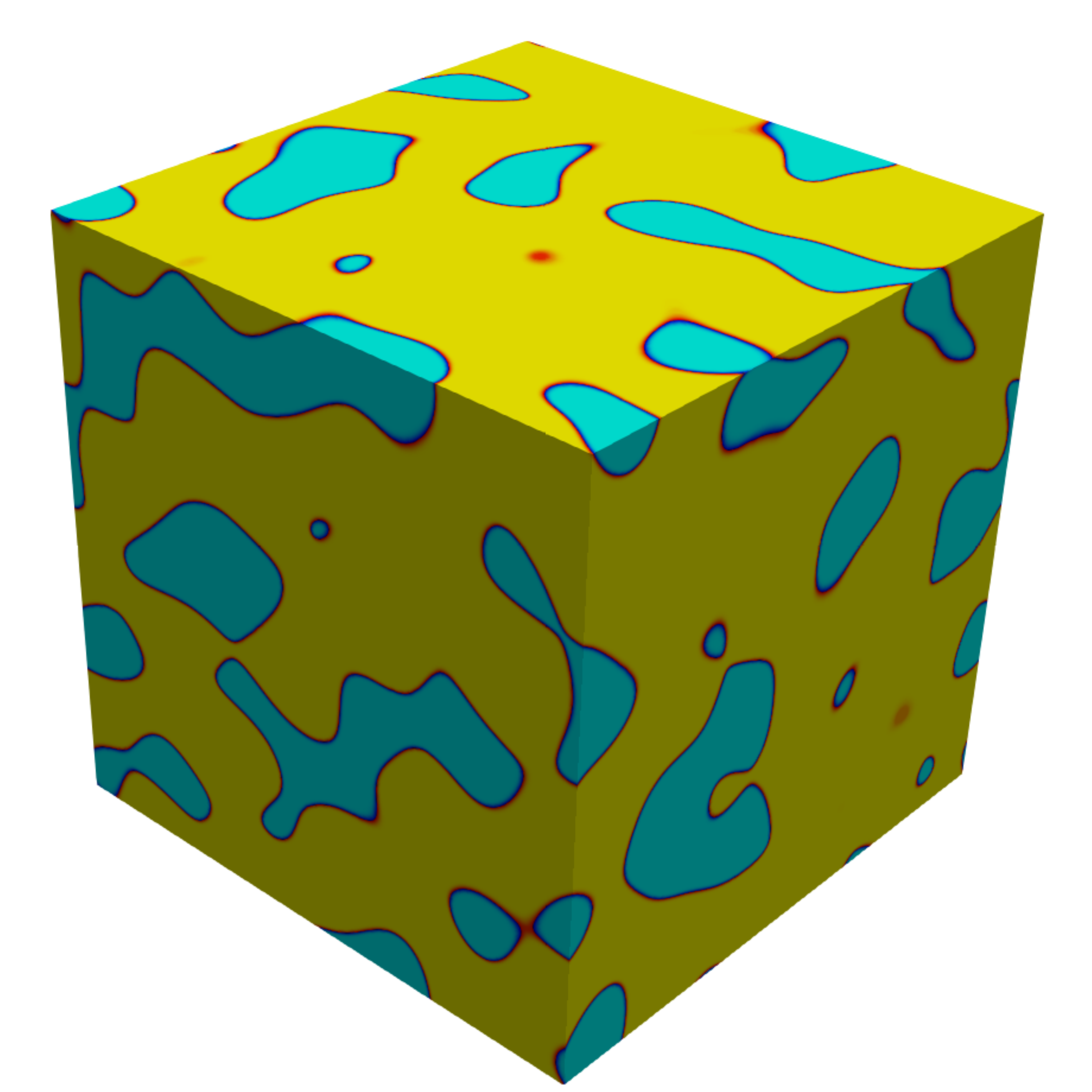}
    \caption{Initial configuration of the density field for (a) heavy fluid dominated flow ($\overline{\chi}_0=3$), (b) symmetric distribution ($\overline{\chi}_0=1$), and (c) light fluid dominated flow ($\overline{\chi}_0=1/3$), where yellow color represents the pure light fluid ($\chi_h=0, \chi_l=1$) and blue color represent the pure heavy fluid ($\chi_h=1, \chi_l=0$).}
    \label{fig:init_dens}
\end{figure*}

        To answer these questions we have performed direct numerical simulations (DNS) with (a) symmetric initial density distribution , for comparison purposes, where the amounts of lighter and heavier fluids are similar ($\overline{\chi}_0\approx1$) which we call initially symmetric distributed flow (SF); and with (b) asymmetric initial density distributions (non-SF) (b-i) heavy-fluid dominated flow (HF) where around three-quarters of the domain is initially composed of heavier fluid ($\overline{\chi}_0\approx3$); as well as (b-ii) light-fluid dominated flow (LF) where around three-quarters of the domain is initially composed of lighter fluid ($\overline{\chi}_0\approx1/3$), as seen in figure \ref{fig:init_dens}. (In this paper, the L-F and H-F cases are referred to as non-SF as well.) Moreover, the influence of low (0.05) and high (0.75) \At number effects are investigated.
\section{Governing equations and computational approach} \label{Sec:Eqs}

The superscript $^*$, in this paper, denotes instantaneous values, capital Roman letters or angle brackets denote mean values, and lower-case Roman letters or primes denote Reynolds fluctuations. For example, the velocity decomposition in index notation is written as: $u^*_i=U_i+u_i$; while the density decomposition is written as: $\rho^*=\rhom +\rho$. Moreover, to investigate VD effects, the Favre (density weighted) averaged values are also presented, denoted by the tilde $~\tilde{}~$ for the Favre-averages and double primes $~^{''}$ for the Favre fluctuations; the velocity (Favre) decomposition is written as: $u^*_i=\tilde{U}_i+u^{''}_i$, with $\tilde{U}_i=\langle \rho^* u^*_i\rangle /\rhom $.

The incompressible variable-density limit of the fully compressible Navier-Stokes equations with two miscible fluids is used to investigate the mixing of two fluids with different micro-densities. The equations have full diffusion and heat flux operators, under the limitation of infinite speed of sound \citep{livescu2013nst,livescu2020}, with the stress tensor being assumed to be Newtonian such that $\tau^*_{ij}=(\rho^*/Re_0)(u^*_{i,j}+u^*_{j,i}-(2/3)u^*_{k,k}\delta_{ij}$). Non-dimensional forms of these equations can be written as \citep{cook_dimotakis_2001,livescu2007,livescu2020}:

\begin{equation} \label{Eq:continuity}
\rho^*_{,t}+(\rho^*u^*_j)_{,j}=0,
\end{equation}

\begin{equation} \label{Eq:moment}
(\rho^*u^*_i)_{,t}+(\rho^*u^*_iu^*_j)_{,j}=-p^*_{,i}+\tau^*_{ij,j}+\frac{1}{Fr^2}\rho^*g_i,
\end{equation}
where, $u_i^*$ is the velocity in direction $i$, $p^*$ is the pressure, and $g_i$ is the gravity (acceleration) in direction $i$. Assuming the speed of sound to be infinite, in incompressible VD turbulence, specific volume changes during mixing and the divergence of velocity is non zero: 
\begin{equation} \label{Eq:divergence}
u^*_{j,j}=-\frac{1}{Re_0Sc}ln\rho^*_{,jj}.
\end{equation}
This relation can be derived by using either the mass fraction transport equations or the incompressible energy transport equation, together with the requirement that the micro-densities, $\rho_2$ and $\rho_1$, of the fluids are constant, which leads to \cite{Joseph-EJMB90}:

\begin{equation}
\label{Eq:massfraction}
    \frac{1}{\rho^*}=\frac{Y_1^*}{\rho_1}+\frac{Y_2^*}{\rho_2},
\end{equation}
where $Y_1$ and $Y_2$ are the mass fractions of the two fluids.
This relation also represents the infinite speed of sound limit of the equation of state for fluid mixtures obeying the ideal gas equation of state \citep{livescu2013nst}. Since $Y_1^*+Y_2^*=1$, relation (\ref{Eq:massfraction}) becomes a diagnostic equation for the mass fractions. A more general derivation is given in Ref. \cite{livescu2020}. Moreover, due to the homogeneity, the mean pressure gradient, $P_{,i}$, needs to be specified and it is chosen to give the maximally unstable flow similarly to the previous studies \cite{livescu2007,livescu2008,aslangil2019}:
\begin{equation}
    \label{Eq:mean_Pgrad}
    P_{,i}=\frac{1}{V}\Big(\frac{g_i}{Fr^2}-\langle vp_{,i} \rangle + \langle u_iu_{j,j}\rangle + \langle v\tau_{ij,j}\rangle \Big),
\end{equation}
where $V$ is the mean specific volume ($v^*=1/\rho^*=V+v$). This also leads to $U_i=0$; hence, in this study $u^*_i=u_i$.
 
In equations \ref{Eq:moment} and \ref{Eq:divergence}, $Re_0$ is the computational Reynolds number, $Sc$ is the Schmidt number and $Fr$ the Froude number, and are defined as:

\begin{equation} \label{Eq:Reynolds}
\begin{split}
    Re_0&=\rho_0L_0U_0/\mu_0, \\
    Sc~&=\mu_0/\rho_0D_0, \\
    Fr^2&=U^2_0/gL_0.
\end{split}
\end{equation}
where, $\rho_0$ is the mean density [$\rho_0=\int_V \rho^* dV$], $g$ is the magnitude of acceleration field, $\mu_0$ is the reference dynamic viscosity, $D_0$ is the diffusion coefficient, $L_0$ and $U_0$ are the reference length and velocity scales. The instantaneous dynamic viscosity is defined as $\mu^*=\mu_0 \rho^*/\rho_0=\nu_0 \rho^*$, where $\nu_0$ is the kinematic viscosity and equal to $D_0$; as $Sc=1$ for all the cases investigated in this paper.  In addition, $Re_0$ equal to $4000$ and 556 for the cases with $A=0.05$ and $A=0.75$, respectively.
Equations (\ref{Eq:continuity}) and (\ref{Eq:moment}), together with the divergence condition (\ref{Eq:divergence}), are solved in a triply periodic, $(2\pi)^3$, domain using the CFDNS code \cite{cfdns}, as described in \cite{livescu2007}. The spatial derivatives are evaluated using Fourier transforms and the time advancement is performed with the variable time step third order Adams-Bashforth-Moulton scheme, coupled with a fractional time method. To minimize the aliasing errors, the advection terms are written in the skew-symmetric form. 

Table \ref{Table:cases} lists the various cases that were chosen to investigate the influence of Atwood number and the initial composition ratio on HVDT. In the nomenclature chosen for the case names, $A005$ denotes the low Atwood number $0.05$ and $A075$ denotes the high Atwood number $0.75$. In addition, the initial flow composition is represented by the last 2 letters: LF denotes $\overline{\chi}_0=1/3$, SF, $\overline{\chi}_0=1$ and HF, $\overline{\chi}_0=3$; additionally, in all line plots in this paper, the red, black and blue lines represent  LF, SF and HF cases respectively. 

\begin{table*}
  \begin{center}
\def~{\hphantom{0}}
\setlength{\tabcolsep}{18pt}
  \begin{tabular}{ccccc}
      Cases	  & $A$   	& $\overline{\chi}_0 \approx $ 		&   $Re_0$	& $Re_{\lambda,max}$	\\      [2pt]
      A$005$LF & 0.05  & $ 1/3$       & $4000$	&	$124$ \\
      A$005$SF & 0.05  & $ ~~1$       & $4000$	&   $141$ \\
      A$005$HF & 0.05  & $ ~~3$       & $4000$	&   $120$ \\
      A$075$LF & 0.75  & $ 1/3$       & ~$556$  &   $~80$ \\
      A$075$SF & 0.75  & $ ~~1$	      & ~$556$	&   $~60$ \\
      A$075$HF & 0.75  & $ ~~3$	      & ~$556$	&   $~48$ \\
  \end{tabular}
  \caption{Parameters for the DNS cases.}
  \label{Table:cases}
  \end{center}
\end{table*}

The density field in all simulations is initialized as a Gaussian random field with top-hat energy spectrum between wave numbers $3$ to $5$ similar to the previous studies \cite{livescu2007,aslangil2019}. After transforming into the real space, for the S-F cases, the negative values are assigned the value of $\rho_1 (=1)$, and the positive values are assigned as $(1+A)/(1-A)=\rho_2$. As a result, the pure fluid densities yield the desired Atwood number. Similarly, in the real space, the values larger than $(1.5)$ and $(-1.5)$ are assigned as $\rho_1 (=1)$, and the values smaller than $(1.5)$ and $(-1.5)$ are assigned as $(1+A)/(1-A)=\rho_2$ to compose the H-F and L-F cases. 
        In addition, the initial density field is smoothed using a Gaussian filter with a width of $1.1 \Delta x$, which ensures that the mixing layer between the pure fluid regions is captured on the grid.  The convergence test is conducted with the same initial conditions but with a higher resolution domain mesh $1024^3$ for the large density ratio case (A075SF) where the results from the $512^3$ and $1024^3$ domain meshes are similar. The resultant non-dimensional initial density integral length-scale, which is calculated from the initial density spectra \cite{aslangil2019}, is $1.34-1.36$ for all cases. The initial density variance ($\rho^2$) values are $0.002$ and $6.55$ for non-SF cases with \At $0.05$ and $0.75$, respectively, and have values of $0.00258$ and $8.4$ for the A005SF and A075SF cases. 
In order to keep the initial time steps reasonably small, similar to our previous \cite{aslangil2019} study, a $5^{th}$ order polynomial equation is used to gradually apply non-dimensional acceleration to the flow between $t/t_r=0$ to $0.1$. Moreover, for all simulations, $\eta k_{max} > 2$ at all times during the flow evolution, where $\eta=\left(1/[Re_0^3 (\epsilon/\rho_0)]\right)^{1/4}$ is the Kolmogorov microscale, and $k_{max}= \pi N /\mathcal{L} = N/2 $ is the largest resolved wave number, which indicates that all simulations are well resolved.

In HVDT, the time evolution of the global parameters collapses better for the different cases if they are normalized by the following normalizations \cite{livescu2007,aslangil2019}: time ($t$) is normalized by $t_r=\sqrt{Fr^2/A}$, and so $U_r=\sqrt{A/Fr^2}$. Thus, in time evolution figures, the x-axis, which presents the time, is divided by $t_r$.
\section{Results}

\subsection{Flow evolution and regimes}

In HVDT, triply periodic domain is initially occupied by pure light and heavy fluids, which leads to differential buoyancy-forces within the flow while the acceleration is applied to the domain. These forces push the fluids in the opposite direction, which generates turbulent kinetic energy. Aslangil et al. \cite{aslangil2019} showed that for high \At number cases with SF conditions, it is more difficult to stir the flow and generate turbulence in the heavy fluid regions due to larger inertia within those regions. In this paper, we have altered the initial composition of the flow, leading to smaller or larger regions of flow that are occupied by each pure fluid.
The flow evolution is divided into the four different regimes according to the behavior of the $E_{TKE}$ (see Table \ref{Table:E_TKE_beh}) to streamline the flow analysis and connect the idealized flow with the real applications with different $E_{TKE}$ behavioral scenarios \cite{aslangil2019,Aslangil_book_ch}.

\begin{table*}
  \begin{center}
\def~{\hphantom{0}}
\setlength{\tabcolsep}{18pt}
  \begin{tabular}{ccc}
     ~ & $\dfrac{dE_{TKE}}{dt}$	  &  $\dfrac{d^2E_{TKE}}{dt^2}$ \\
     Explosive growth  &  $>0$   &  $>0$\\
     Saturated growth  &  $>0$   &  $<0$\\
     Fast decay        &  $<0$   &  $<0$\\
     Gradual decay     &  $<0$   &  $>0$
       \end{tabular}
  \caption{$E_{TKE}=<\rho^*u_i^{''}u_i^{''}>$ behaviour during different HVDT regimes.}
  \label{Table:E_TKE_beh}
  \end{center}
\end{table*}

Explosive growth refers to the regime with rapid increase of $E_{TKE}$. During this regime, the motions of the large scales accelerate, whereas the stirring is not adequately evolved to be able to stir these large structures. Hence, due to the lack of wide range of motions, the molecular mixing is relatively slow and localized at the interface between the pure fluids \cite{aslangil2019}.
        Saturated growth starts when the growth of the $E_{TKE}$ slows down. During this regime, a wider range of scales starts to attend to the mixing process, so mixing reaches its fastest rate. This leads to distinguishable differences between the evolution of the density PDFs of the low and high \At number cases.  
Fast decay, on the other hand, refers to the following regime, in which $E_{TKE}$ decays rapidly. During this regime, the flow becomes fully-developed and is mostly well-mixed at the end of this regime. 
        Gradual decay is characterized by a slower decay than the canonical isotropic turbulence decay as in this case, the decay is assisted by continuous buoyancy-production \cite{aslangil2019}. The turbulence generation continues to assist the flow even as it is mostly mixed, as the production term never becomes negligible compared to the dissipation term. Moreover, our previous study \cite{aslangil2019} showed that both density and velocity PDFs tend to reach symmetric shapes for both low and high \At number cases.
     
\begin{figure}
    \centering
(\emph{a}) \\
\includegraphics[height=4.6cm]{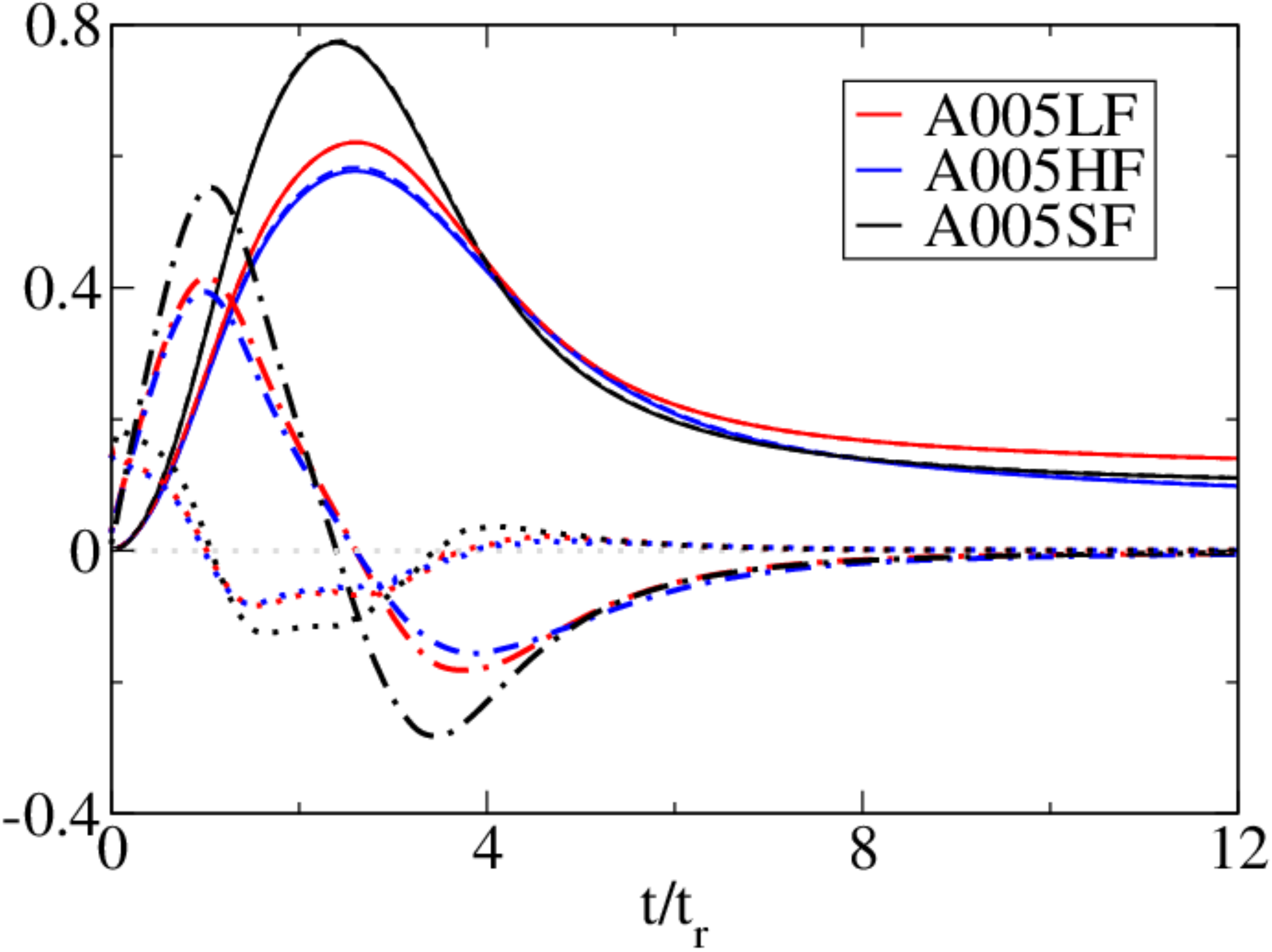}\\ 
(\emph{b}) \\
\includegraphics[height=4.6cm]{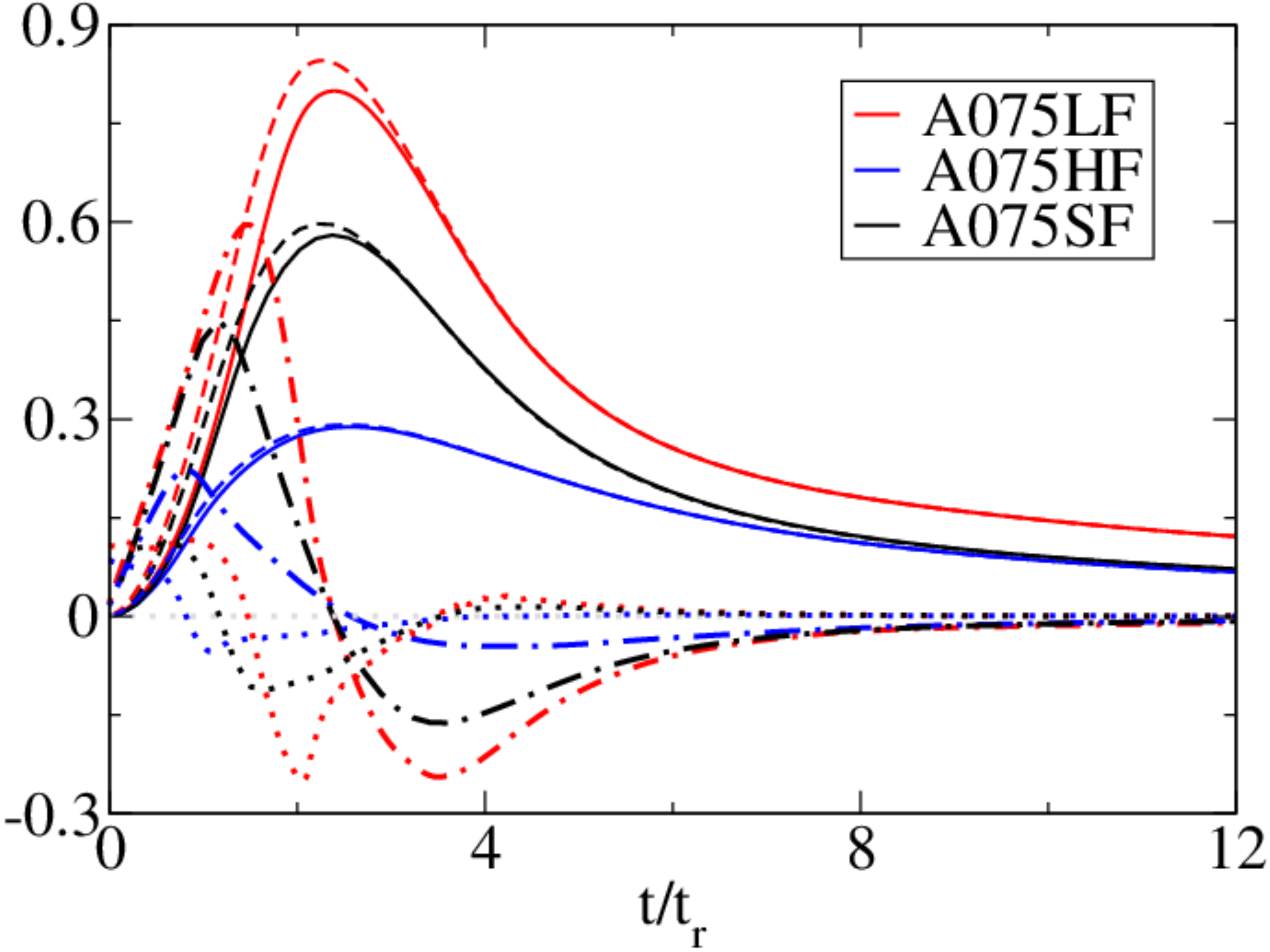}
    \caption{Evolution of the $E_{KE}$ (dashed-lines), $E_{TKE}$ (solid lines), $d(E_{TKE})/dt$ (dashed-dots), and $d^2(E_{TKE})/dt^2/4$ (dots) for the (a) \At=0.05 and (b) \At=0.75 cases.}
    \label{fig:tke}
\end{figure}

To further aid in understanding of HVDT evolution, the kinetic energy ($E_{KE}$) and the Favre averaged turbulent kinetic energy ($E_{TKE}$) are used and defined as:
\begin{equation}
    E_{KE}=\frac{1}{2}\langle\rho^*u_i^{*}u_i^{*}\rangle;~~~~E_{TKE}=\frac{1}{2}\langle\rho^*u_i^{''}u_i^{''}\rangle.
\end{equation} 
Figure \ref{fig:tke} presents the normalized $E_{KE}$, $E_{TKE}$ and its time derivatives (note that $d^2E_{TKE}/dt^2$ is divided by 4 for better illustration) where the energies are normalized by $0.5\rhom U_rU_r$. 
As it is seen in Figure \ref{fig:tke}, at low \At number, $E_{TKE}$ evolution reaches higher values for the SF case compared to non-SF cases. This is attributed to the larger initial mean variance of the density field ($\sigma_0=\rho^2$) for the SF case. Thus, $\sigma_0$ reaches its maximum value for the initially symmetric distributed flows where $\overline{\chi_0}=1$ and is $\approx 1.28$ times larger than $\sigma_0$ for both LF and HF cases, which reaches maximum values where $\overline{\chi_0}=1/3$ and $\overline{\chi_0}=3$, respectively.
Meanwhile, at high \At number, $E_{TKE}$ reaches much higher values for the LF compared to the SF, whereas HF reaches much lower $E_{TKE}$ values compared to SF cases.
        This is consistent with the notion that flows with smaller inertia can be stirred more easily than flows with larger inertia. As a result, for higher \At numbers, the lighter flow regions move much faster compared to the heavier flow regions. These faster motions within the lighter fluid regions are able to generate high enough $E_{TKE}$ to compensate the initial handicap of the flow due to having smaller $\sigma_0$. Thus, for the higher \At number case, even if $\sigma_0$ is smaller than for the SF case, LF reaches the highest $E_{TKE}$ levels during the flow evolution.
 In addition, compared to the other two cases, explosive growth is observed for a longer duration for the A075LF case. Meanwhile, the explosive growth duration is the shortest for the A075HF case. For A075SF case, explosive growth duration is intermediate, and is similar to the cases with $A=0.05$. However, $E_{TKE}$ maxima occur almost at the same normalized time instant for all cases. Moreover, during gradual decay, the time derivative of the $E_{TKE}$ converges for all \At number cases, indicating similar long time decay behavior for all cases with different initial composition ratios.

\begin{figure*}
    \centering
(\emph{a}) \hspace{3.5cm}  (\emph{b}) \\
    \includegraphics[height=4.4cm]{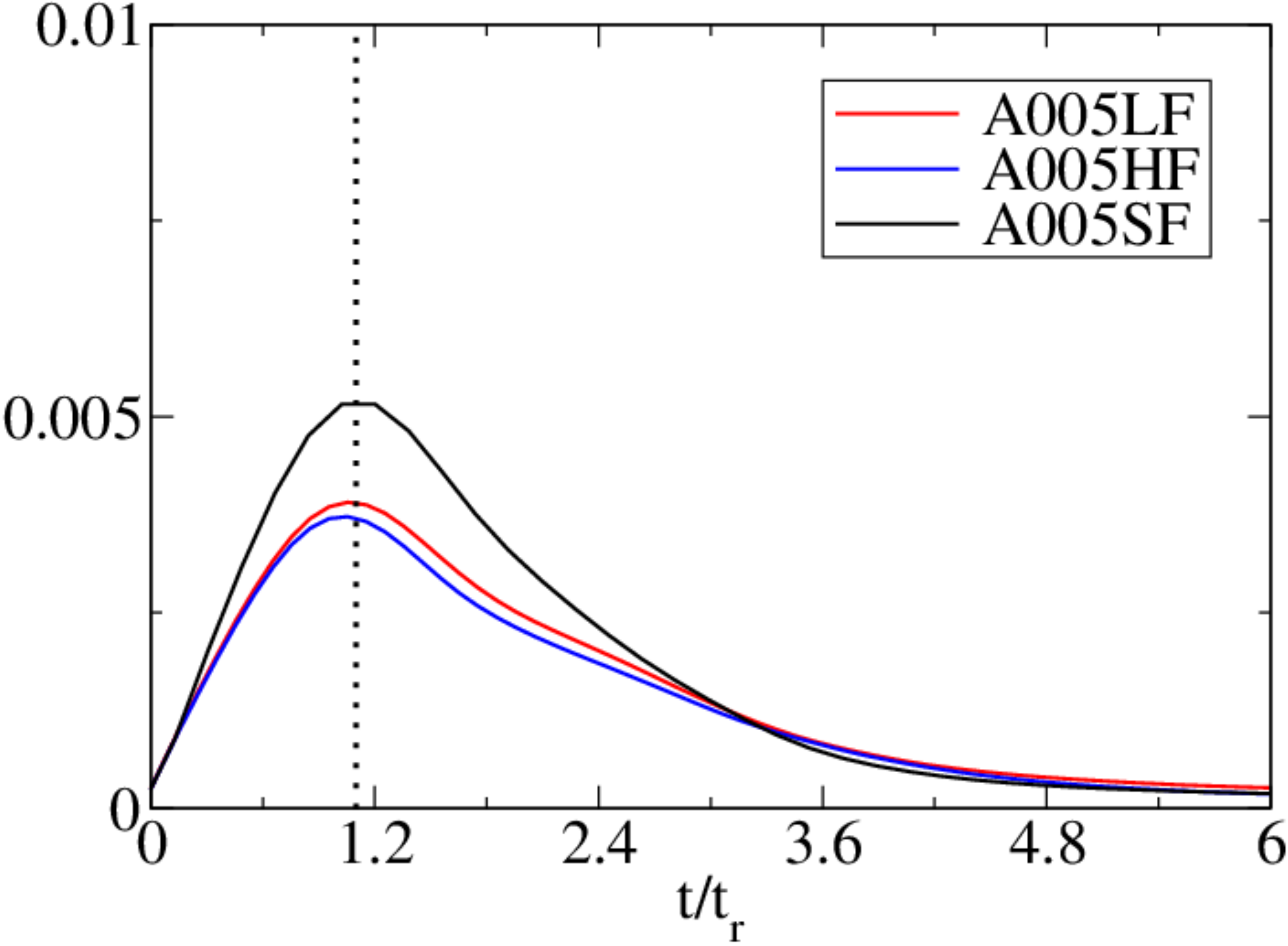}\includegraphics[height=4.4cm]{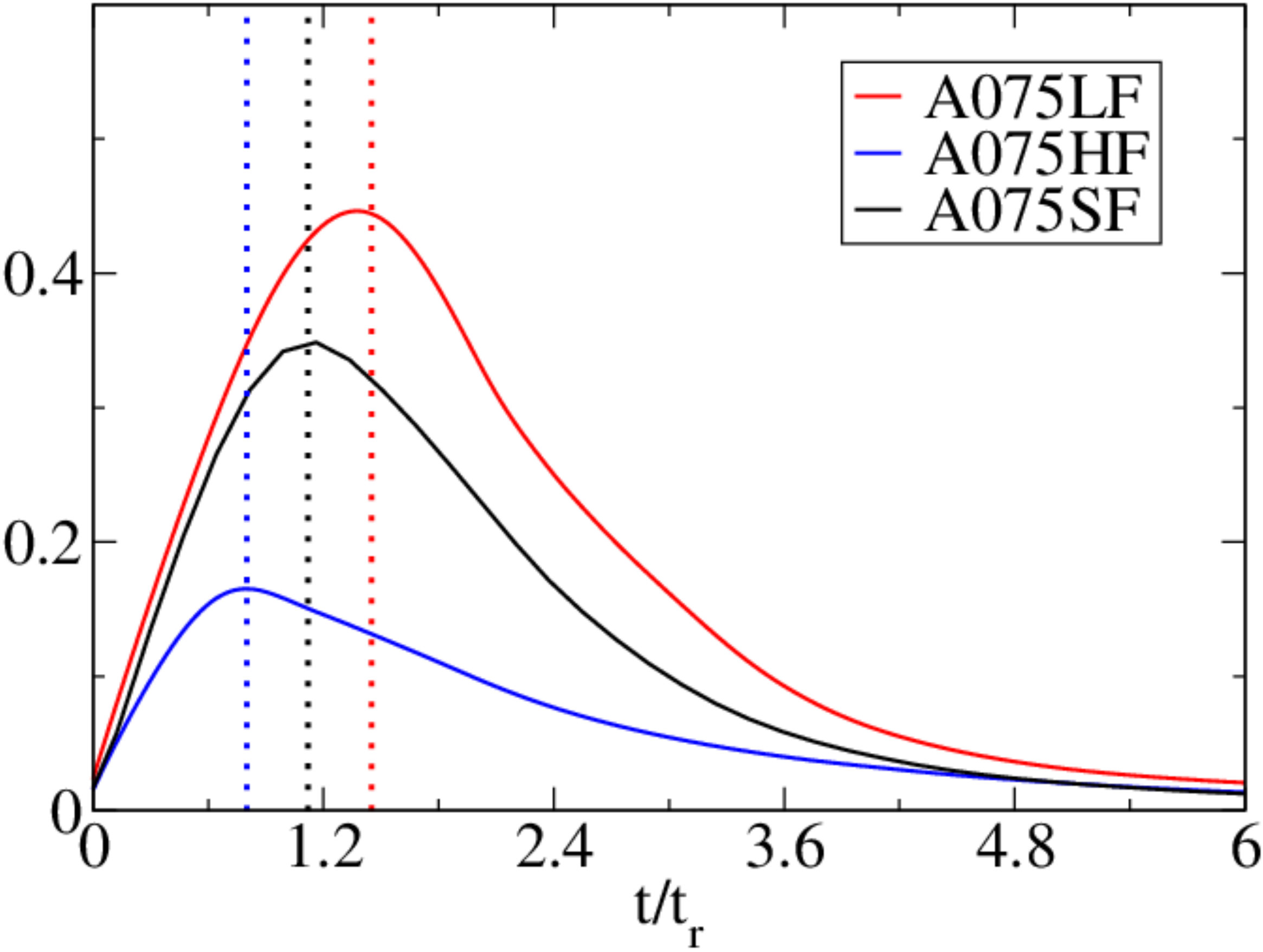}\\
(\emph{c}) \hspace{3.5cm}  (\emph{d}) \\
    \includegraphics[height=4.4cm]{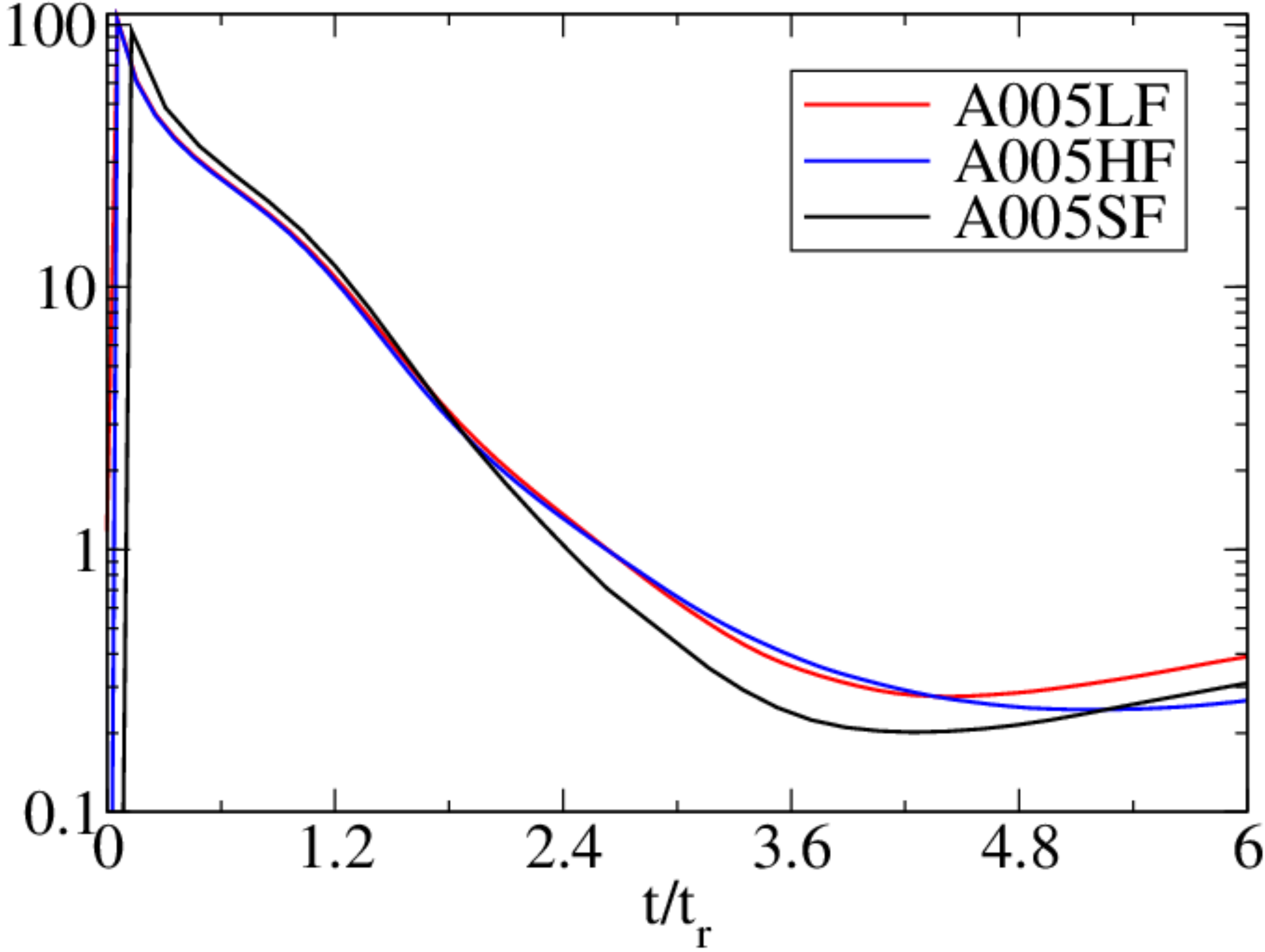}\includegraphics[height=4.4cm]{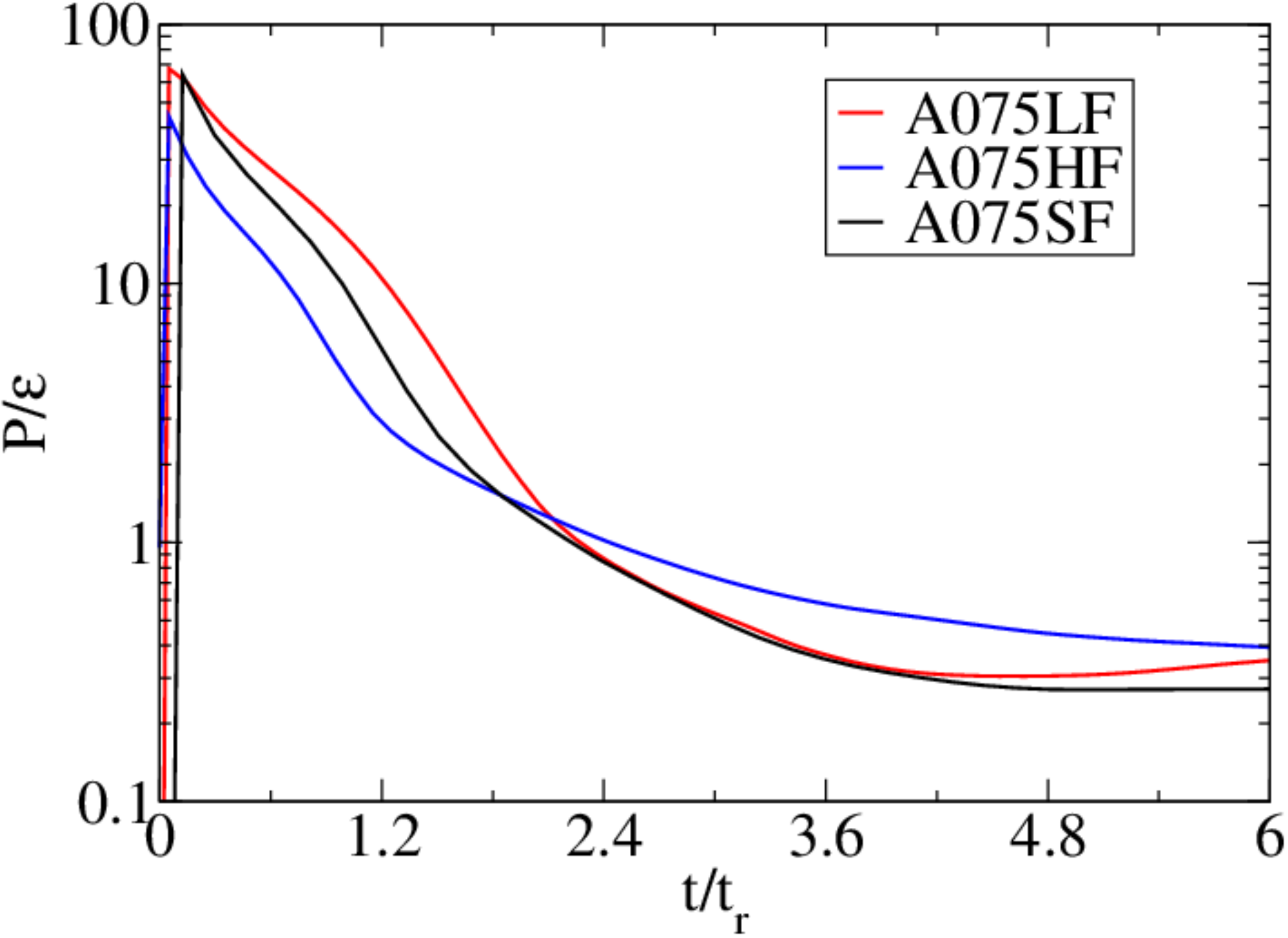}\\
    \caption{Evolution of $a_1$ for the (a) \At=0.05 and (b) \At=0.75 cases and evolution of $a_1\rhom/\epsilon$ for the (c) \At=0.05 and (d) \At=0.75 cases. The doted vertical lines presents the end of the explosive growth regime.}
    \label{fig:ai}
\end{figure*}

 Transport equations for the $E_{TKE}$ and $E_{KE}$ can be written as \cite{livescu2007}:
\begin{equation}
     E_{TKE,t}=a_iP_{,i} +\langle pu^{''}_{j,j}\rangle -\epsilon^{''},
     \label{Eq:TKE_evolv}
\end{equation}
\begin{equation}
     E_{KE,t}=\frac{g_i}{Fr^2}\rhom a_i +\langle pu^*_{j,j}\rangle -\epsilon^{*},
    \label{Eq:KE_evolv}
\end{equation}
where $a_i=\langle \rho u_i \rangle/\rhom$ is the mass flux. $\epsilon^{''}$ and $\epsilon^{*}$ are the dissipation of $E_{TKE}$ and $E_{KE}$, respectivly. Due to homogeneity, $\epsilon^{''}=\epsilon^{*}=\langle u^*_{i,j}\tau^*_{ij}\rangle$ \cite{aslangil2019}. Similarly, $\langle pu^*_{j,j}\rangle=\langle pu^{''}_{j,j}\rangle=\langle pu_{j,j}\rangle$. In addition, pressure-dilatation is not equal to zero in variable density turbulence and is associated with the gain or loss of energy through molecular mixing. The vertical mass flux [$a_1$] has been demonstrated to be the sole mechanism to convert potential energy to the kinetic energy in buoyancy-driven HVDT \cite{livescu2007}. Figure \ref{fig:ai} presents the vertical mass flux [$a_1$] and the production over dissipation ratio [$a_1\rhom/\epsilon=P/\epsilon$] for low and high $A$ number cases. As it is seen, the mass flux term, which is proportional to the $E_{KE}$ production, reaches slightly higher values for the $A005SF$ case than the non-SF cases at low Atwood number, a feature consistent to the behavior of $E_{TKE}$. 
        In addition, mass flux has a maximum occurring at the end of the explosive growth regime for all cases reported in this paper. The differential duration of explosive growth regime for different cases with $A=0.75$ leads to differences in the time instants where mass flux maximum occurs. For example, since the A075LF case has the longest explosive growth, the mass flux maximum occurs later for this case compared to the other cases of $A=0.75$. Due to this relation between the mass flux maximum and the duration of the explosive growth, the mass flux maximum occurs earlier for the A075HF case, as the explosive growth regime is shorter.
Similar to mass flux, the production over dissipation ratio ($\rhom a_i/\epsilon$) also remains at higher levels and for longer time for the A075LF than the A075HF and A075SF cases (see fig. \ref{fig:ai} b and d). This behavior is similar for low \At number (see fig. \ref{fig:ai}a and \ref{fig:ai}c) cases and is also consistent with the higher $E_{KE}$ and $E_{TKE}$ values for the A075LF case. It is also notable that the ratio never reaches zero and stays at constant values even during gradual decay (the time evolution is only plotted until $t/t_r=6$ for clarity) indicating that production is always an active process during the decay stage which noticeably slows downs the HVDT decay \cite{aslangil2019}.

\begin{figure}
    \centering
    (\emph{a})\\
    \includegraphics[height=4cm]{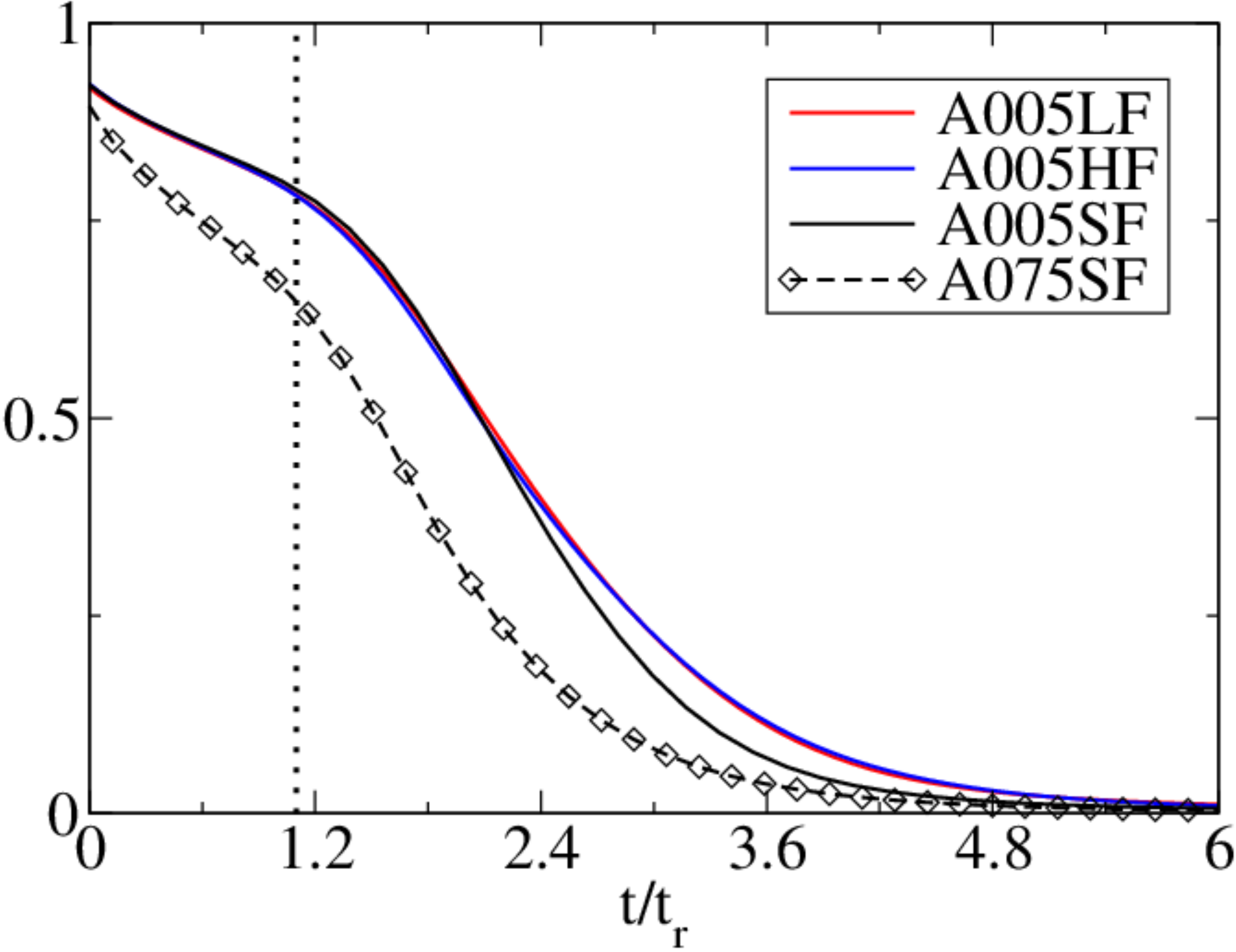}\\
    (\emph{b})\\
    \includegraphics[height=4cm]{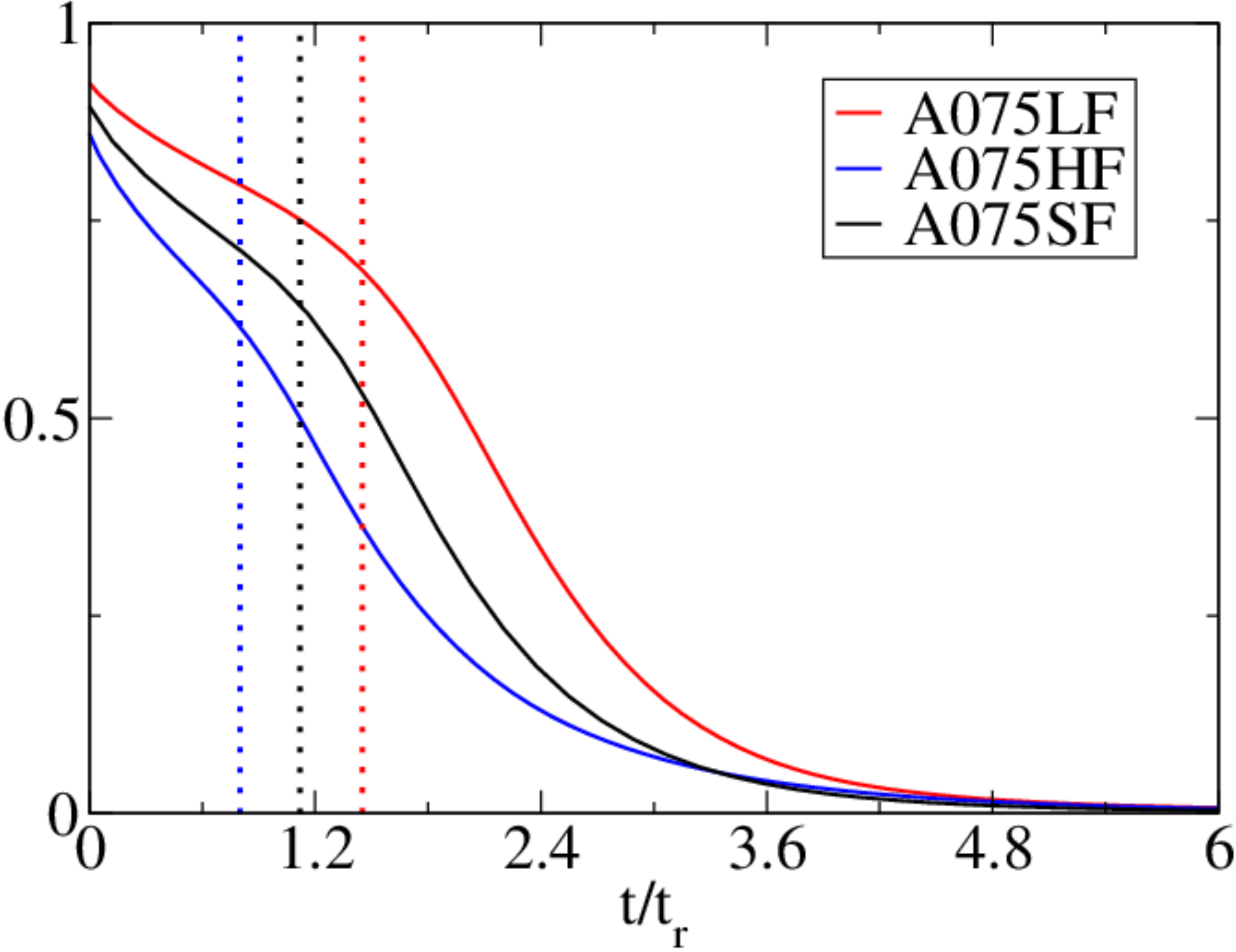}
    \caption{Evolution of the normalized $b$ (a) for the \At=0.05 and A075SF cases, and (b) evolution of the normalized $b$ for the \At=0.75 cases. The doted vertical lines presents the end of the explosive growth.}\label{fig:b}
    \end{figure}

In addition to density variance, the density-specific volume correlation parameter $b$ is an important alternative way to measure the mixture state of the HVDT and appears in turbulence mix models \cite{livescu2008,Banerjee_BHR_2010,BHR_2,Schwarzkopf_Livescu_Gore_etal_2011_b,tomkins_balakumar_orlicz_prestridge_ristorcelli_2013_RMI_b,schwarzkopf2016tls,Pal18}:

\begin{equation} \label{Eq:b}
    b=-\Big\langle\rho\Big(\frac{1}{\rho^*}\Big)^{'}\Big\rangle.
\end{equation}
Figure \ref{fig:b} presents the evolution of $b$ normalized by $(A\rhom)^2\rho_2$. As it is seen, $b$ behaves differently during explosive and saturated growths. During explosive growth, the decay of $b$ is slower compared to its decay during saturated growth, for all cases. During transition from explosive growth to saturated growth, as discussed above, a wider range of scales starts to attend mixing, which leads to a faster decay in the behaviour of $b$. The evolution of the normalized $b$ is similar for all cases with \At $=0.05$ as the duration of explosive growth was also similar for those cases. However, the differential duration of the explosive growth for the cases with \At =$0.75$ is reflected to the behavior of $b$ as well. For example, for the $A075LF$ case, transitional behavior of the $b$ occurs later than for the other two cases. Thus, capturing the transition between the explosive to saturated growth becomes even more important as it does not only represent the transitional behaviour of the momentum within the flow, but also reflects the transitional behavior of the  molecular mixing within HVDT. In addition, during gradual decay, $b$ becomes small ($b<0.1$) indicating that the flow is mostly mixed in this regime for all cases. Minimal differences were observed in the behavior of the density variance ($\sigma$) for the different cases and is not shown for brevity.

    \begin{figure}
    \centering
(\emph{a})\\
    \includegraphics[height=4.2cm]{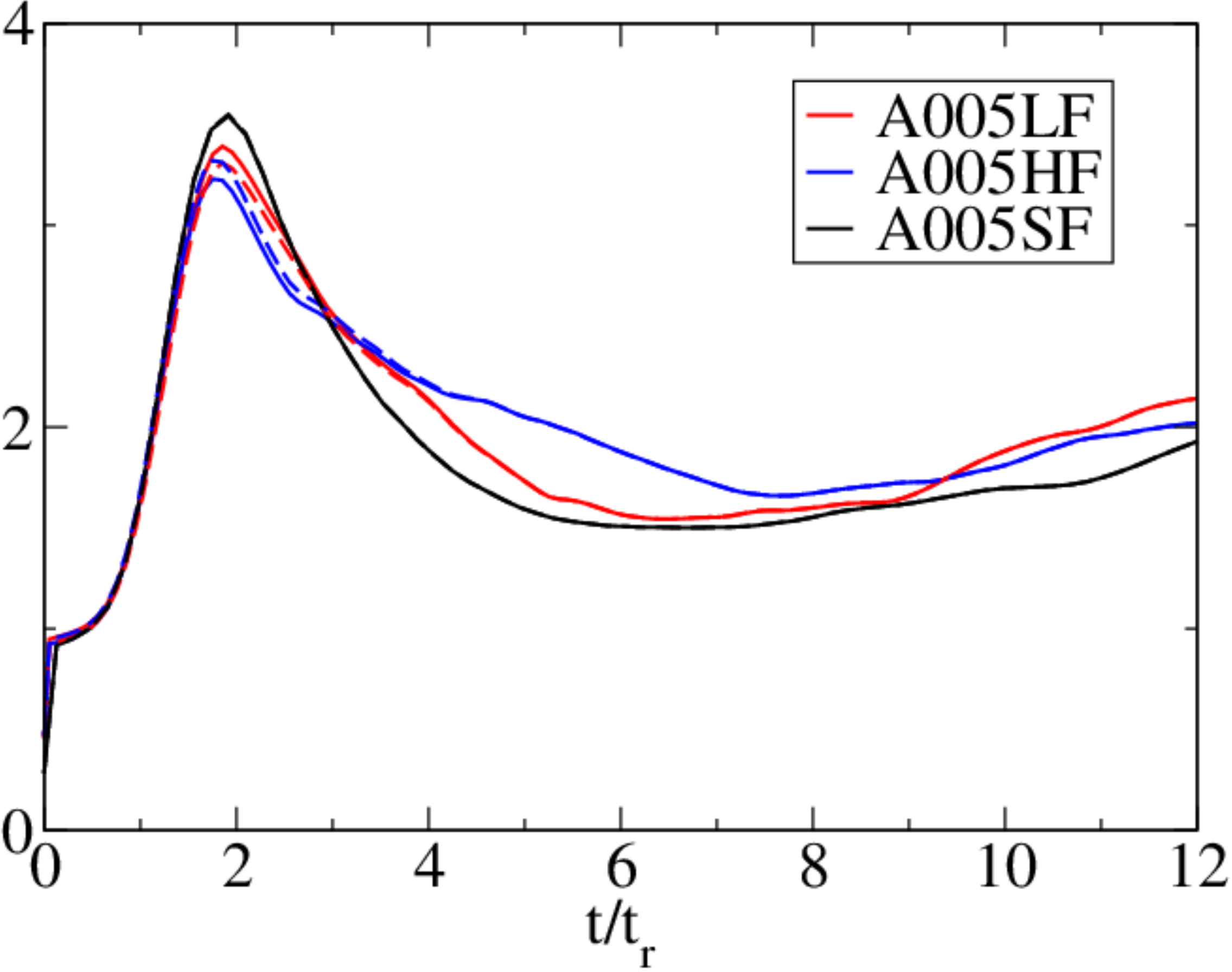}\\
    (\emph{b})\\
    \includegraphics[height=4.2cm]{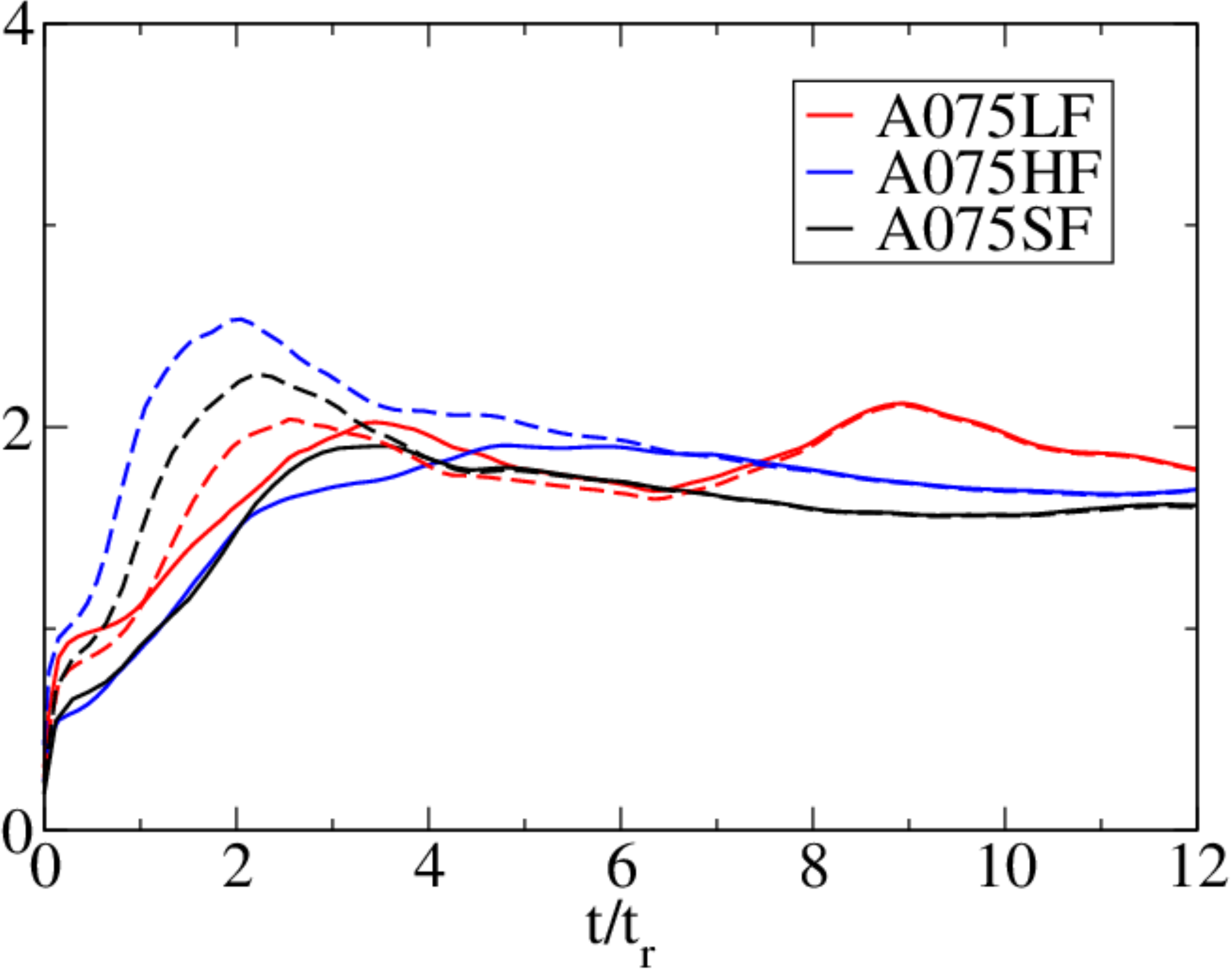}
    \caption{Evolution of $\Upsilon$ (solid lines) and $\Upsilon_b$ (dashed lines) for the (a) \At=0.05 and (b) \At=0.75 cases.}
    \label{fig:upsilon}
\end{figure}

The ratio of the time scales for $E_{TKE}$ and scalar energy ($E_{\rho}=1/2<\rho^2>$) as well as $b$ can be written as \cite{livescu_jaberi_madnia_2000,DonDaniel_ReactionForcing,aslangil2019}: 

\begin{equation}\label{Eq:ratio}
   \begin{split}
    \Upsilon&=\frac{E_{TKE}}{\epsilon}\Bigg/\frac{E_{\rho}}{\xi}=\frac{E_{TKE}\xi}{E_\rho\epsilon};\\
   \Upsilon_b&=\frac{E_{TKE}}{\epsilon}\Bigg/\frac{b}{db/dt}=\frac{E_{TKE}db/dt}{b\epsilon},
   \end{split}
\end{equation}
where $\xi$ is the scalar dissipation ($\xi=D_0<\rho_{,j}\rho_{,j}>$). In low order turbulent mixing models, the scalar dissipation is not explicitly calculated and is predicted by using the turbulent kinetic energy dissipation ($\epsilon$) considering that $\Upsilon$ is a constant \citep{livescu_jaberi_madnia_2000,Kolla_scalar,DonDaniel_ReactionForcing}. However, in HVDT evolution, Aslangil et al. \cite{aslangil2019} showed that it is a dynamic quantity and both $E_{TKE}$ and scalar dissipation have to be captured separately until the gradual decay \cite{aslangil2019}. Here, figure \ref{fig:upsilon} presents $\Upsilon$ and $\Upsilon_b$ values for the different cases with different initial density distributions. At low \At number, these ratios are similar and do not show significant dependency on the initial composition of the flow. However, the behaviors of $\Upsilon_b$ and $\Upsilon$ deviate slightly at high \At number; $\Upsilon_b$ is slightly larger for the HF case during saturated growth, whereas $\Upsilon$ is almost insensitive to the initial composition even at high $A$.

\subsection{Initial composition and high-Atwood number effects on density PDF}

\begin{figure}
    \centering
    (\emph{a})\\
    \includegraphics[height=4.6cm]{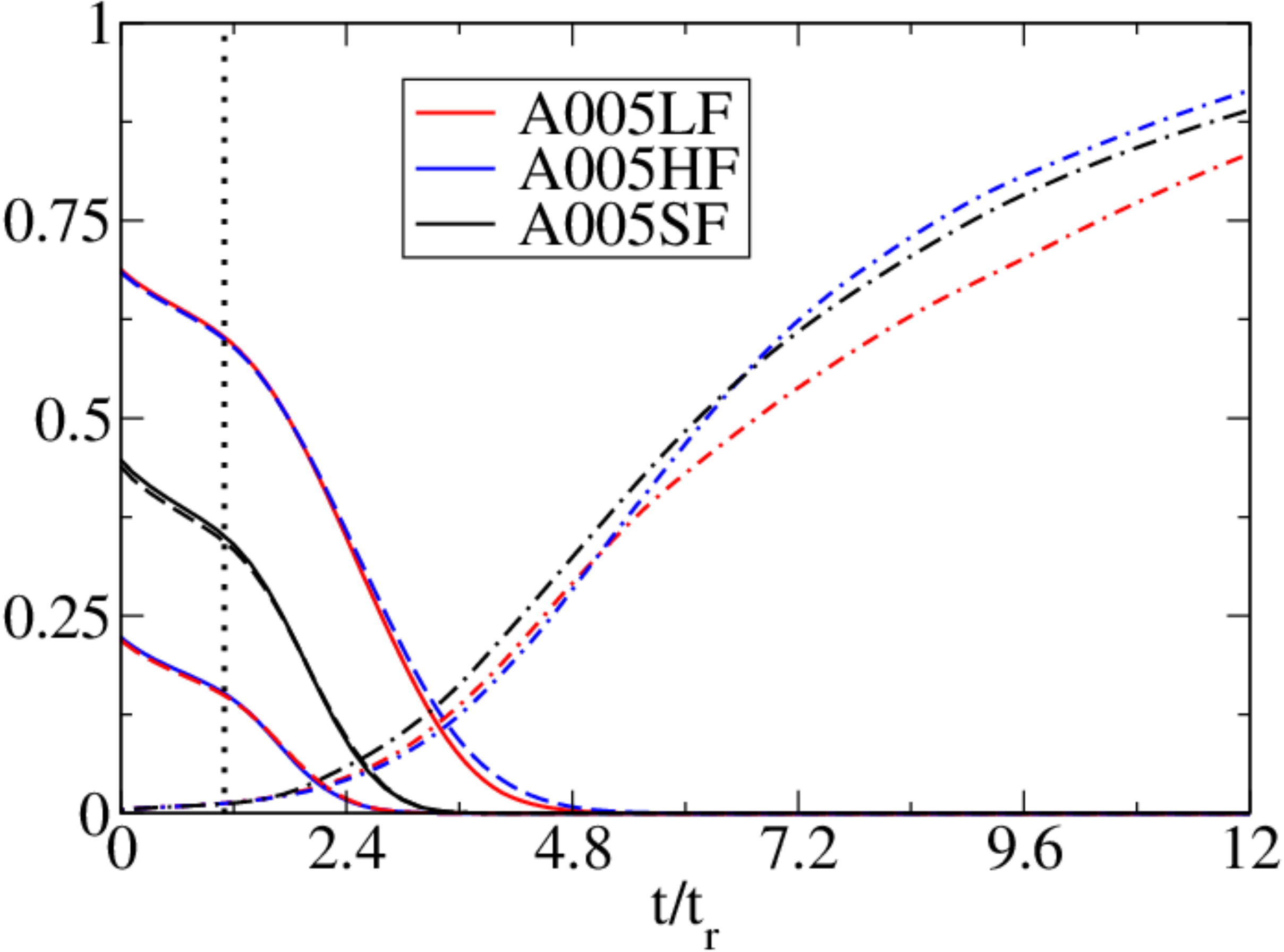}\\
    (\emph{b})\\
    \includegraphics[height=4.6cm]{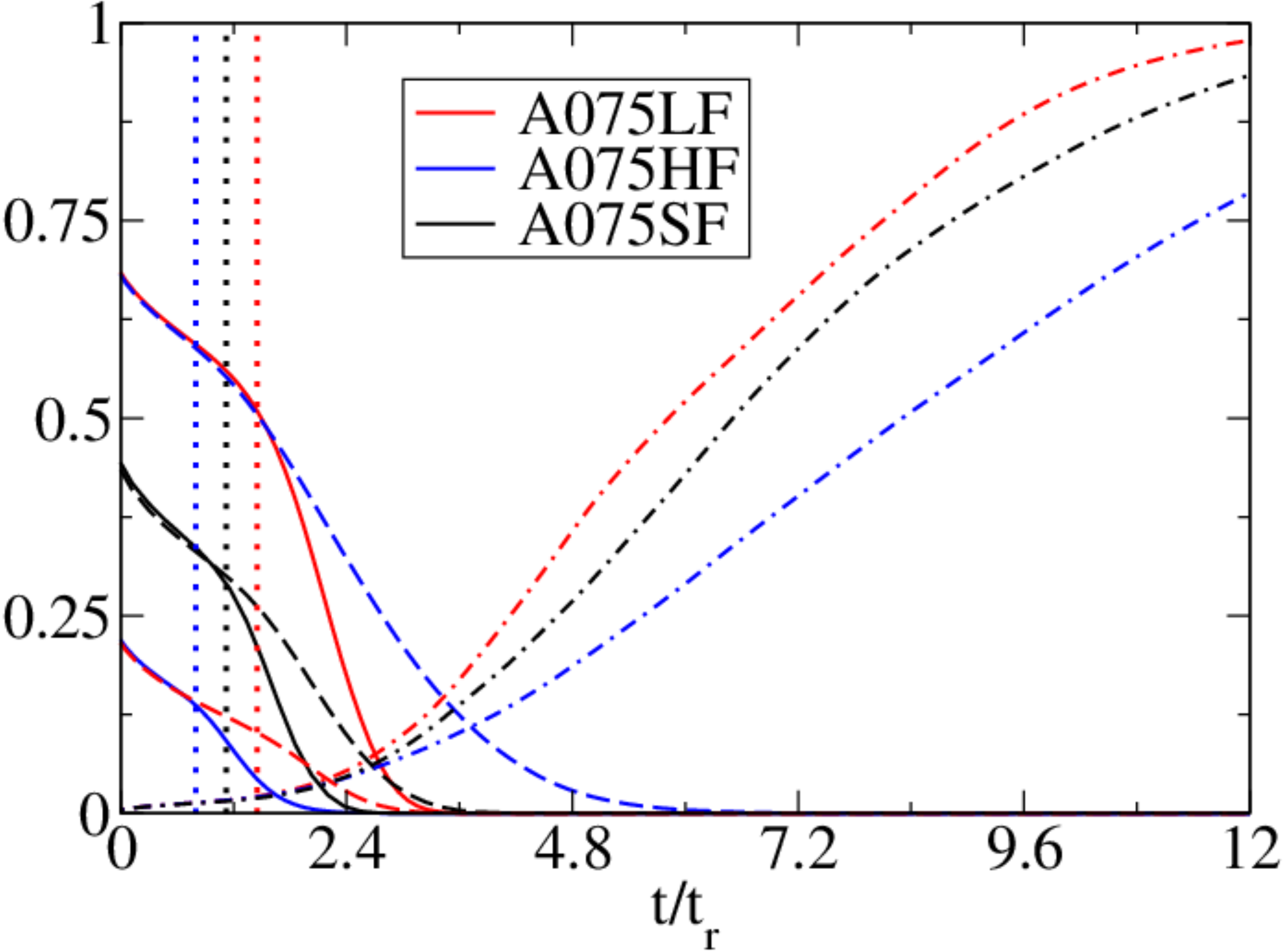}
    \caption{Evolution of the amounts of the pure light (solid line), pure heavy (dashed line), and fully-mixed (dashed-doted line) fluids for the cases with (a) \At=0.05 and (b) \At=0.75. The doted vertical lines presents the end of the explosive growth.}
    \label{fig:PureFluids}
\end{figure}

In the physics of VD turbulence with large density ratios, it is now established that light fluid regions mix faster than the heavy fluid regions, which leads to significant asymmetric evolution of the density PDF and changes the local flow structure \cite{livescu2008,aslangil2019}.  In this subsection, we explore the generality of an asymmetric behavior for the flows with different initial compositions. To better illustrate the mixing behavior of the different flow regions, first, we present the time evolution of the volume fractions of pure light fluid, where $\rho^*\leq \rho_1 +0.05(\rho_2-\rho_1)$, pure heavy fluid, where $\rho^*\geq \rho_1 +0.95(\rho_2-\rho_1)$ and the fully-mixed flow $\vert\rho \vert \leq \rhom \pm 0.025(\rho_2-\rho_1)$. Due to differential initial compositions of the flow, the initial volume fractions of the pure light fluid are $\approx 0.75$, $\approx0.5$ and $\approx0.25$, and the initial volume fractions of the pure heavy fluid are $\approx 0.25$, $\approx0.5$ and $\approx0.75$ for the LF, SF and HF cases, respectively. As shown in Figure \ref{fig:PureFluids}, the mixing rates of the pure fluid regions are similar for the A005SF case, as seen by the matching amounts of the pure fluids.
        Meanwhile, the amount of the pure light fluid within $A005LF$ case matches  the amount of the pure heavy fluid within $A005HF$ case, whereas their initial amounts are $\approx 0.75$, and so they vanish at the same time instant ($t/t_r\approx 4.8$).  This is also valid when we compare the amounts of the pure heavy fluid of $A005LF$ case with the pure light fluid of $A005HF$ case, where their initial amounts are $\approx 0.25$. They also vanish at the same time instant ($t/t_r \approx 2.4$). This indicates that the mixing rates of the pure fluids are similar for all cases and are independent of the initial composition of the flow for the low \At number cases. 

At high \At number, however, for all cases, the pure light fluid mixes faster than the pure heavy fluid. In addition, the transition from the slowest mixing rate to the fastest mixing rate for each pure fluid occurs at the end of the explosive growth. This is similar to the behavior of $b$, as these transitions occur at different time instants consistently with explosive growth end times. There is almost three times more pure light fluid than pure heavy fluid by volume within the A075LF case; however, as the light fluid mixes faster, both pure fluids vanish at almost the same time instant (at around $t/t_r=2.5$), but the pure heavy fluid of the A0755HF case vanishes significantly later (at around $t/t_r=4.8$) than the pure light fluid (which vanishes at around $t/t_r=1.2$). 

\begin{figure*}
\hspace{2.4cm}(\emph{A=0.05}) \hspace{4.8cm}  (\emph{A=0.75}) \\
    \centering{
    \rotatebox{90}{\hspace{1.5cm}$t/t_r=1.2$}\includegraphics[width=6cm]{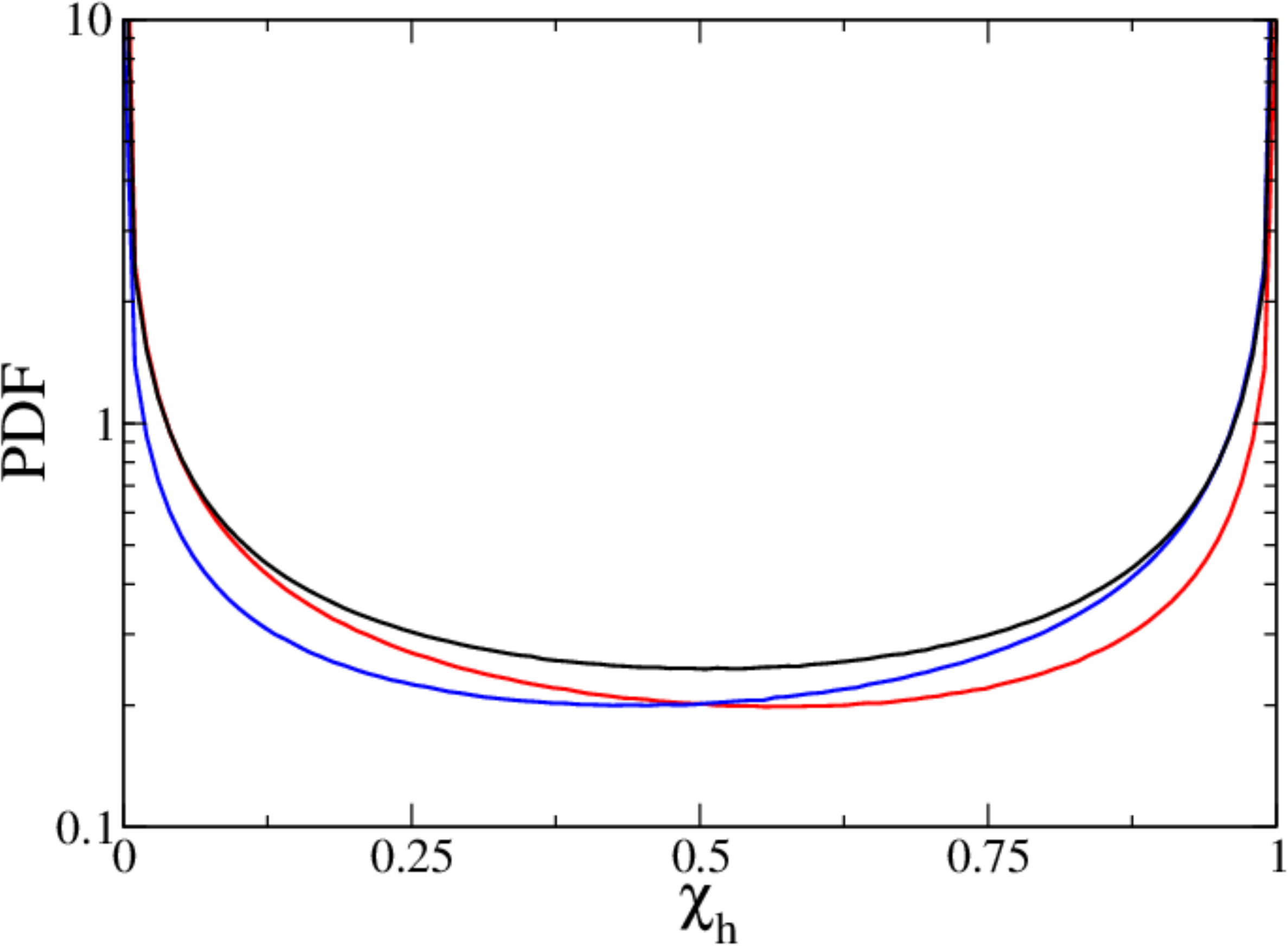}\includegraphics[width=6cm]{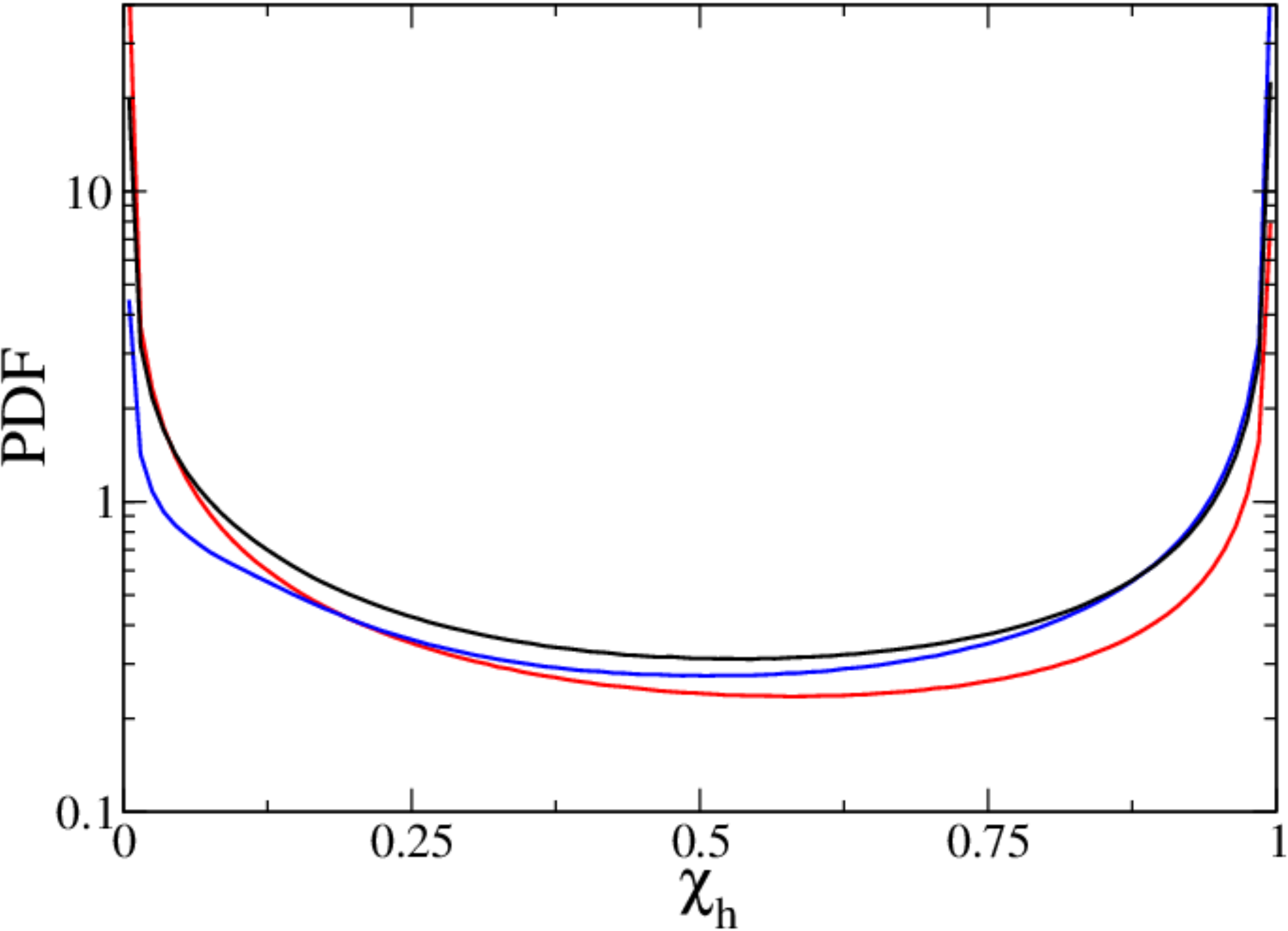}\\
    \rotatebox{90}{\hspace{1.5cm}$t/t_r=2.4$}\includegraphics[width=6cm]{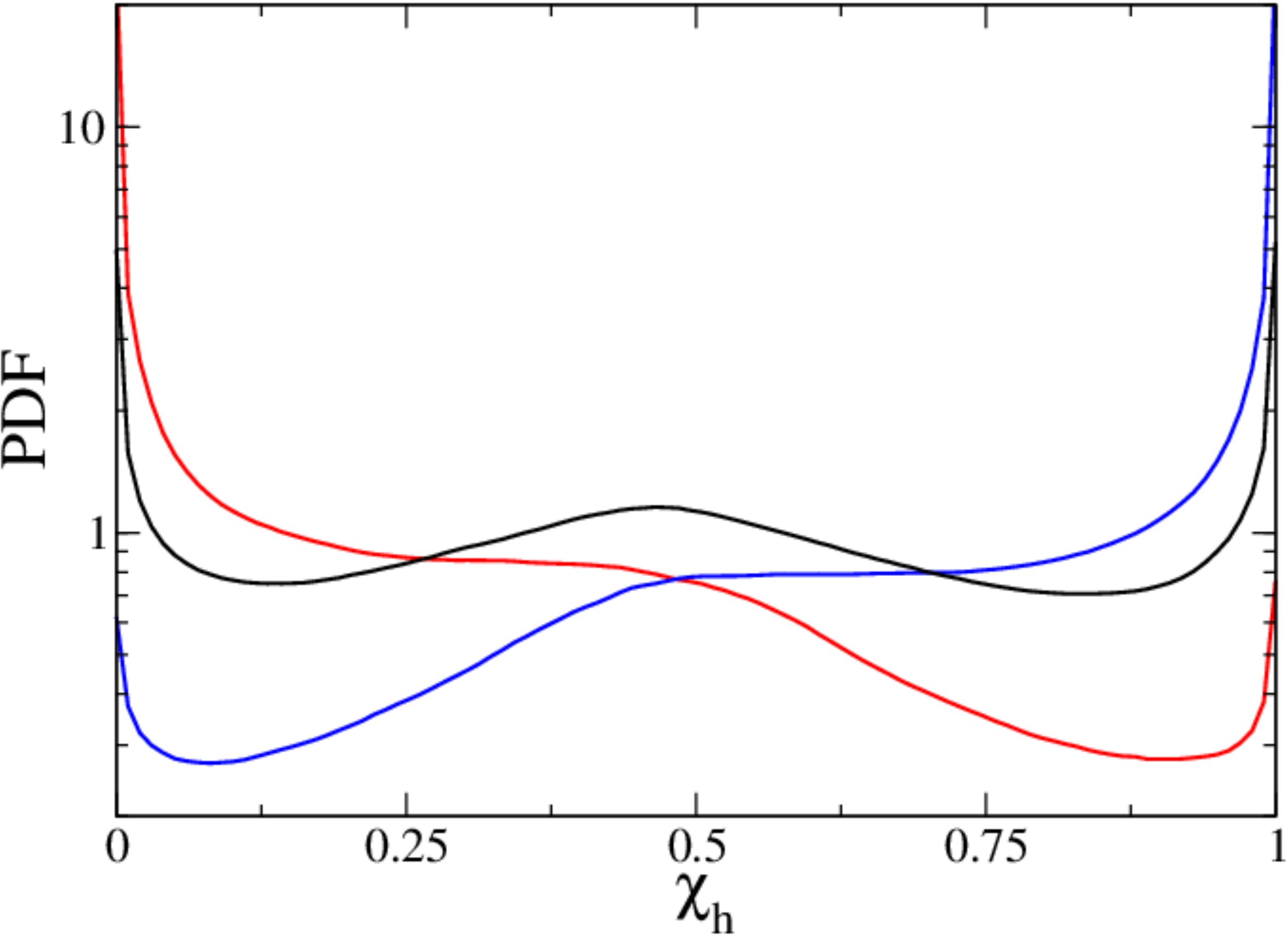}\includegraphics[width=6cm]{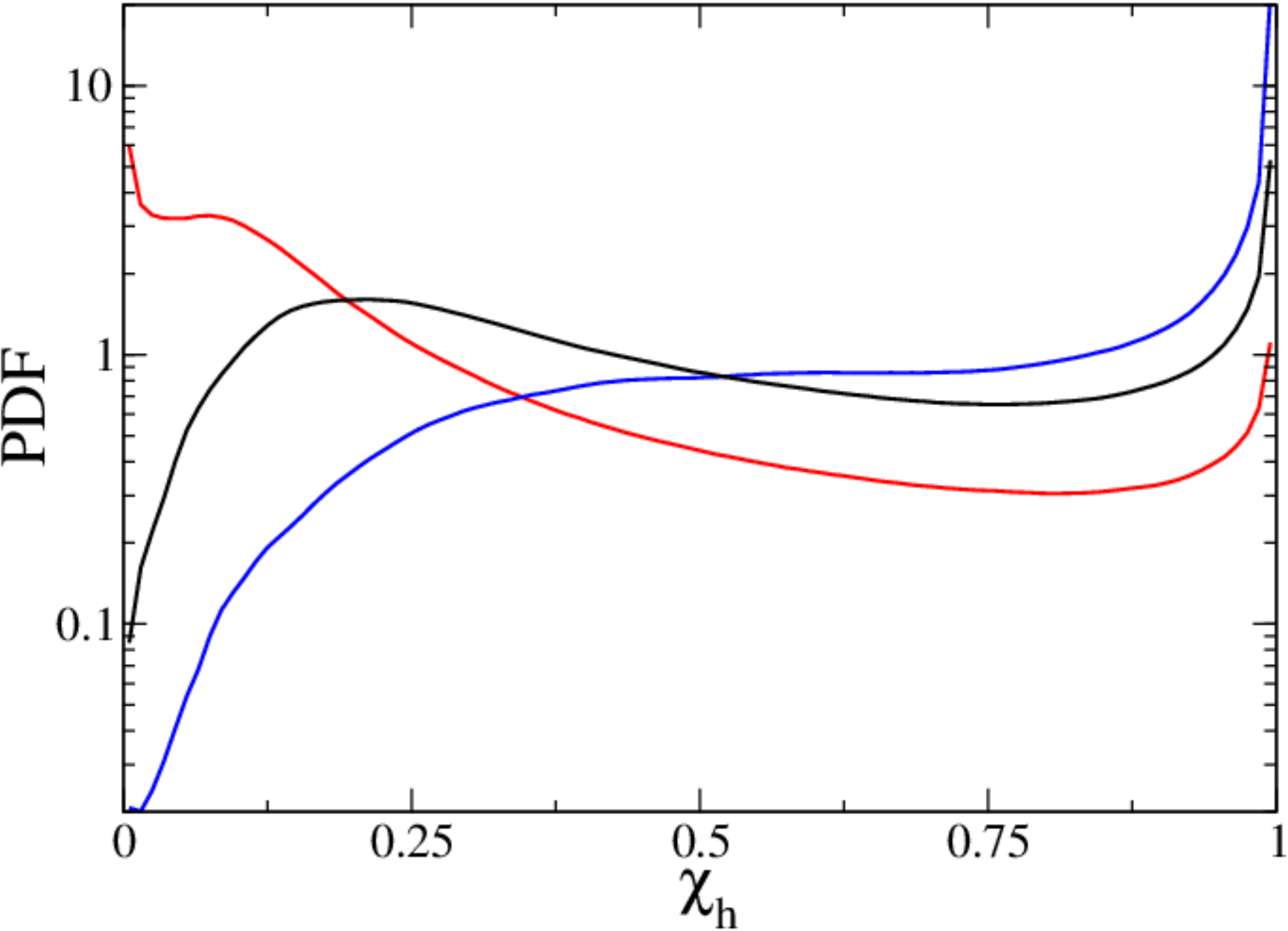}\\
    \rotatebox{90}{\hspace{1.5cm}$t/t_r=3.6$}\includegraphics[width=6cm]{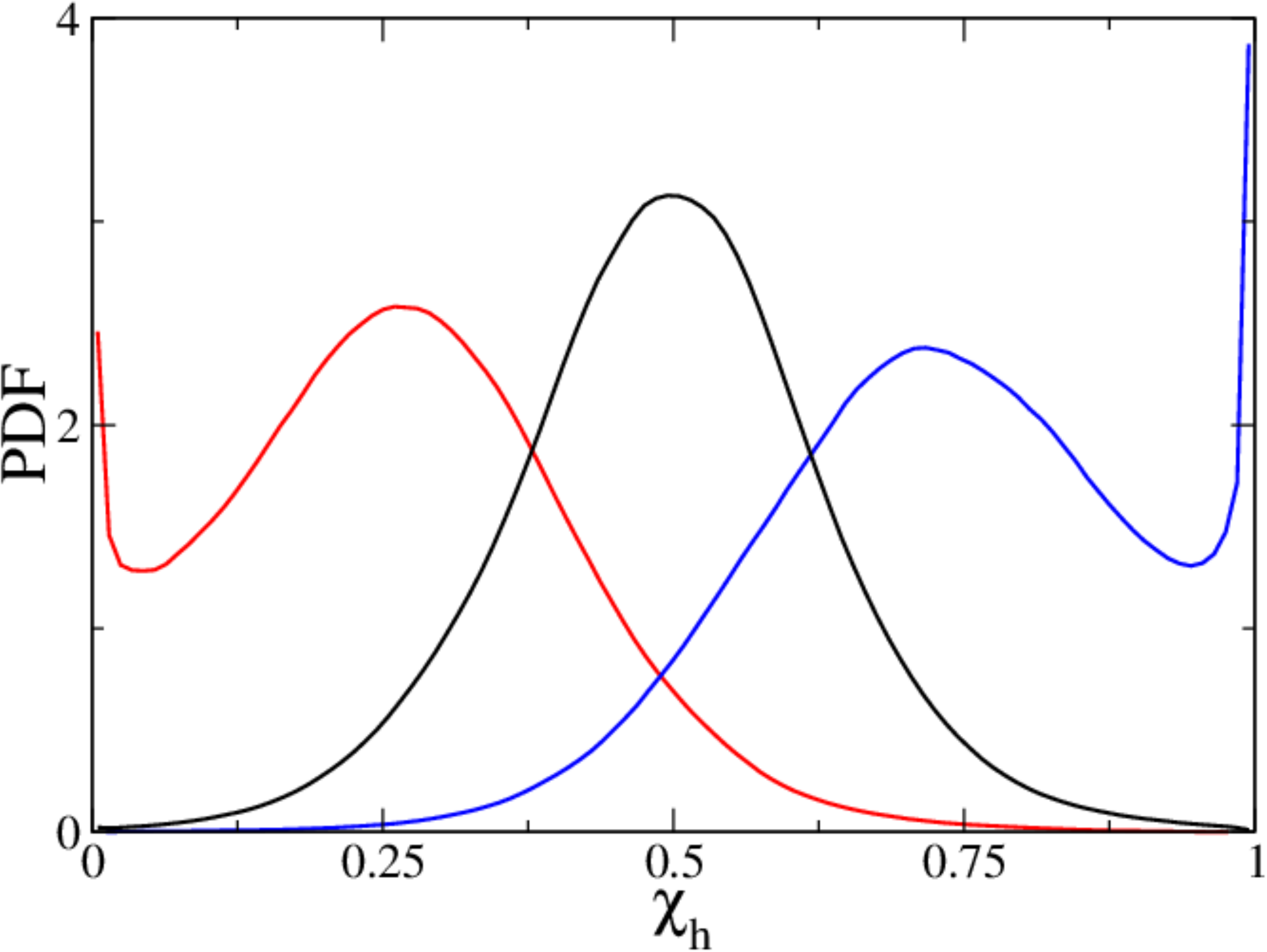}\includegraphics[width=6cm]{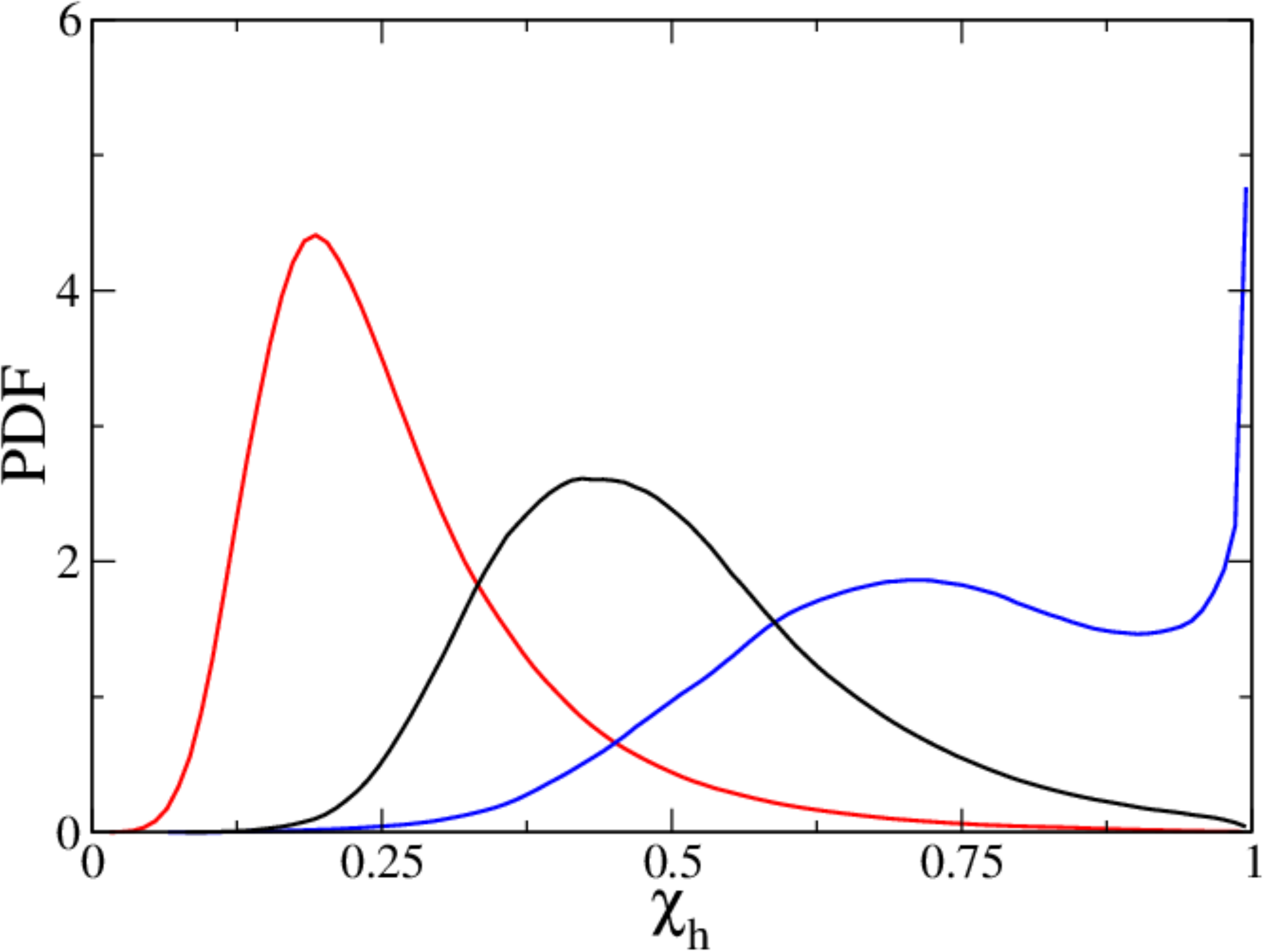}\\
    \rotatebox{90}{\hspace{1.5cm}$t/t_r=9.2$}\includegraphics[width=6cm]{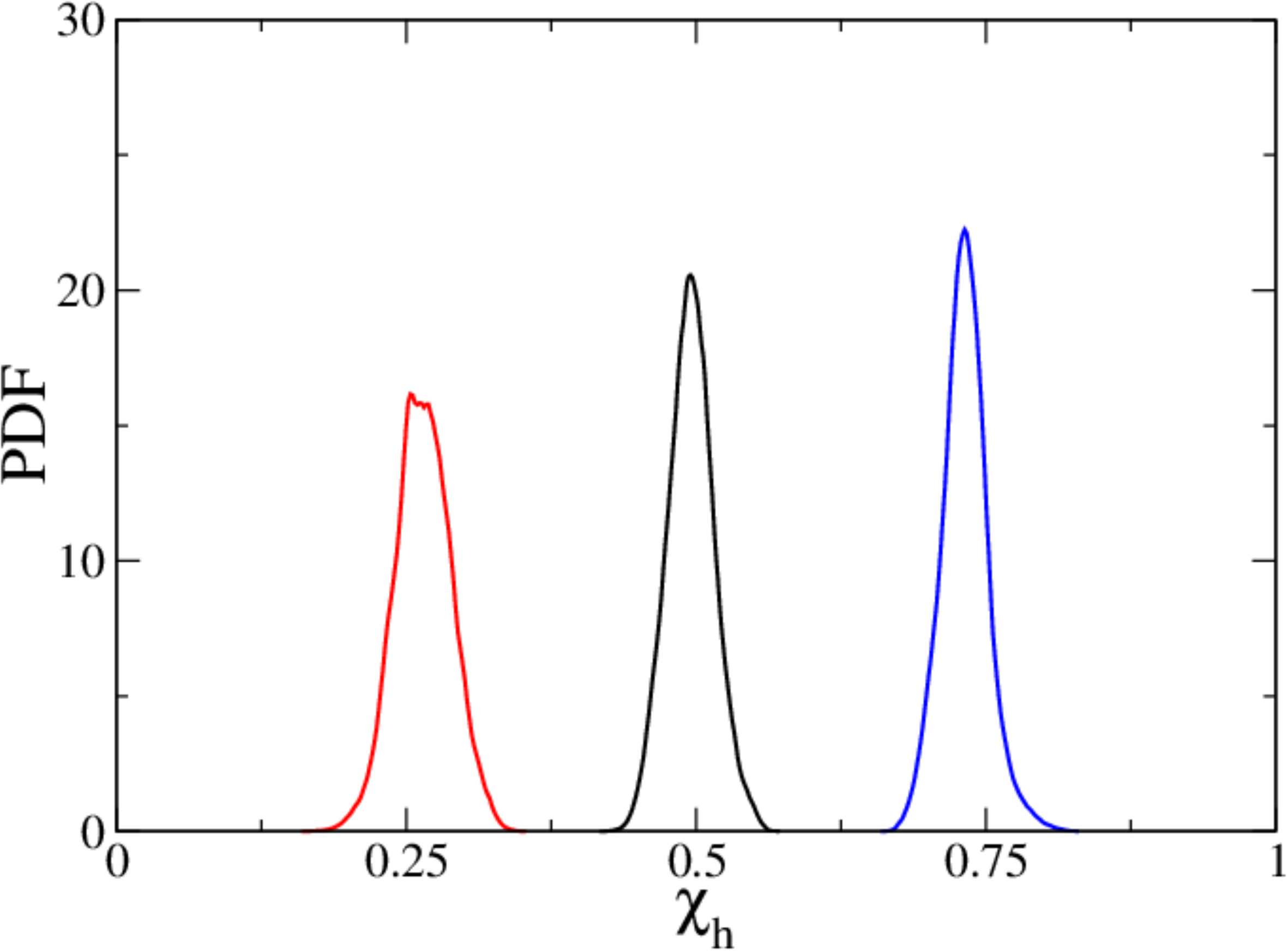}\includegraphics[width=6cm]{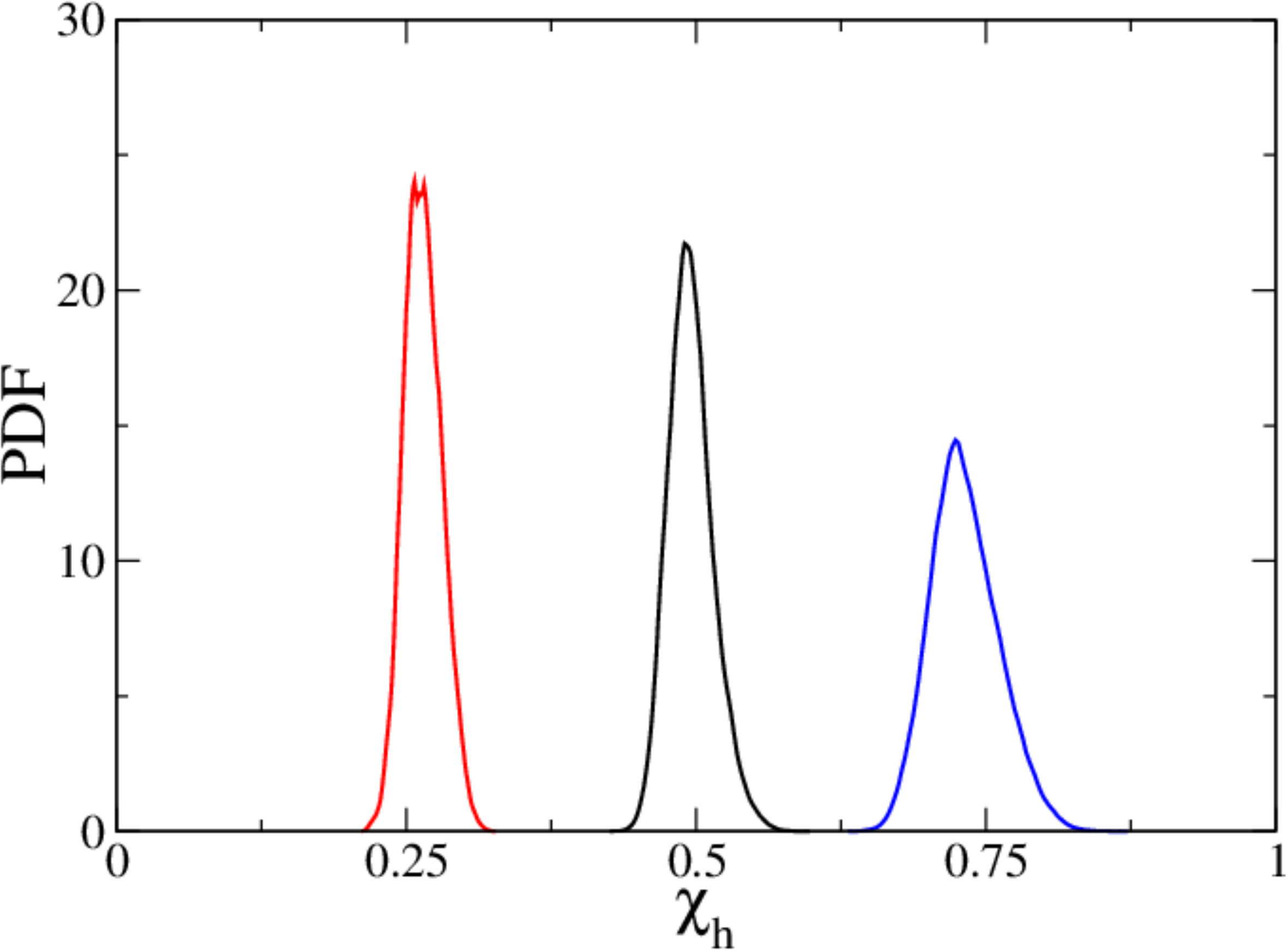}}
    \caption{Evolution of density PDFs at different time instants, where the red color represents the LF, blue color represents the HF, and black color represents the SF.}
    \label{fig:pdfs}
\end{figure*}

\begin{figure*}
\hspace{2cm}(\emph{A075HF}) \hspace{3.2cm}  (\emph{A075SF}) \hspace{3cm}  (\emph{A075LF})\\
    \centering{
    \rotatebox{90}{\hspace{1.7cm}$t/t_r=1.2$}\includegraphics[width=4.6cm]{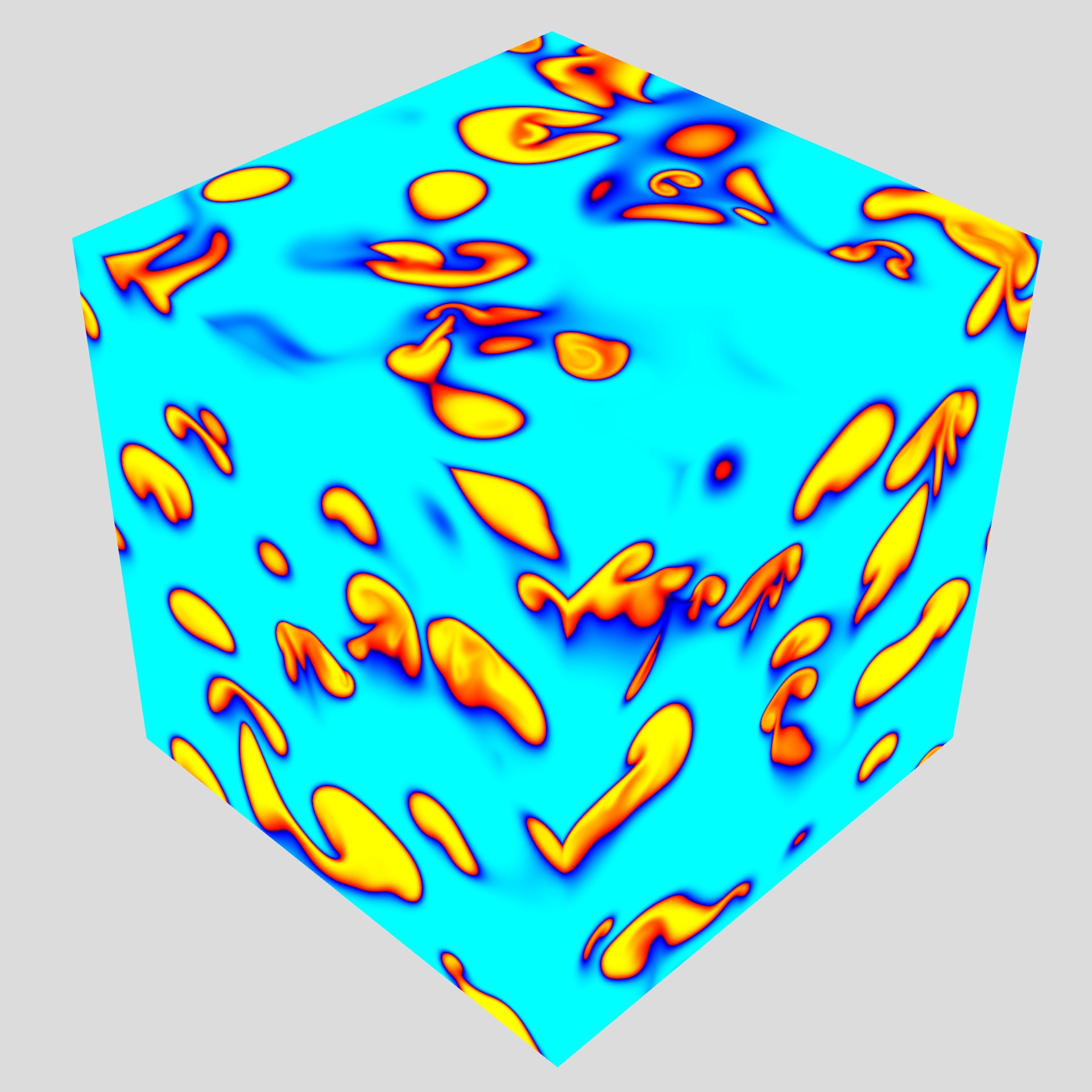}\includegraphics[width=4.6cm]{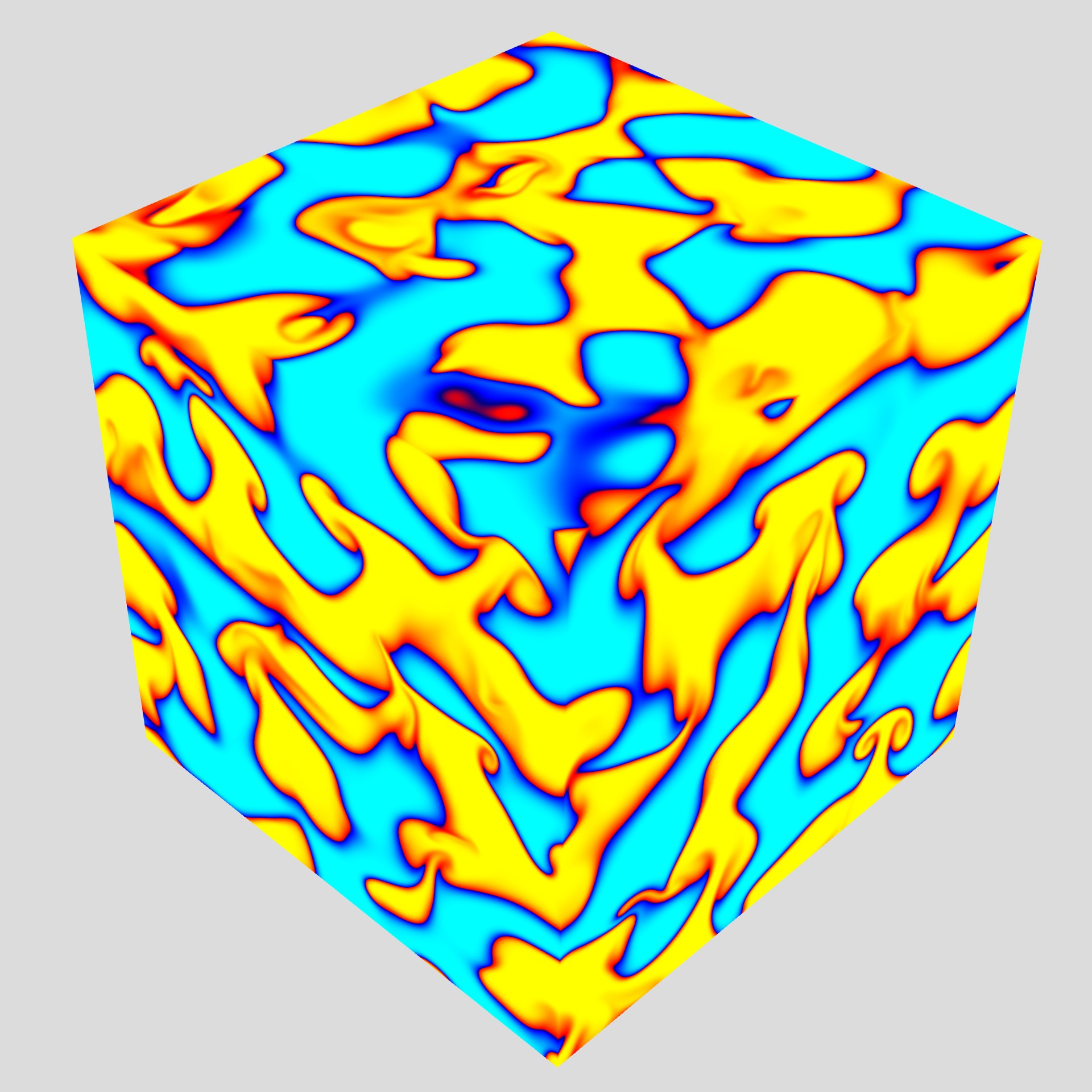}\includegraphics[width=4.6cm]{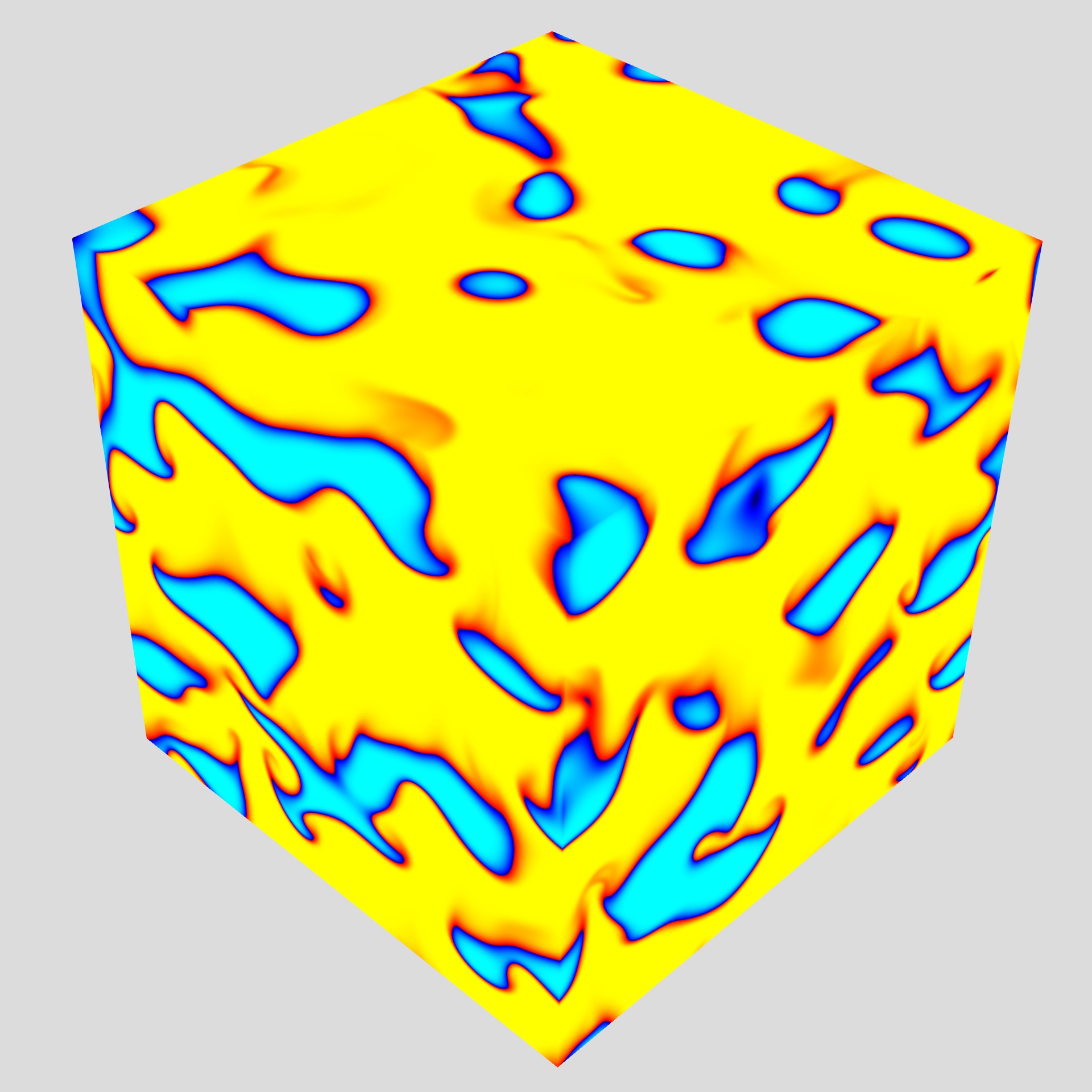}\\
    \rotatebox{90}{\hspace{1.7cm}$t/t_r=2.4$}\includegraphics[width=4.6cm]{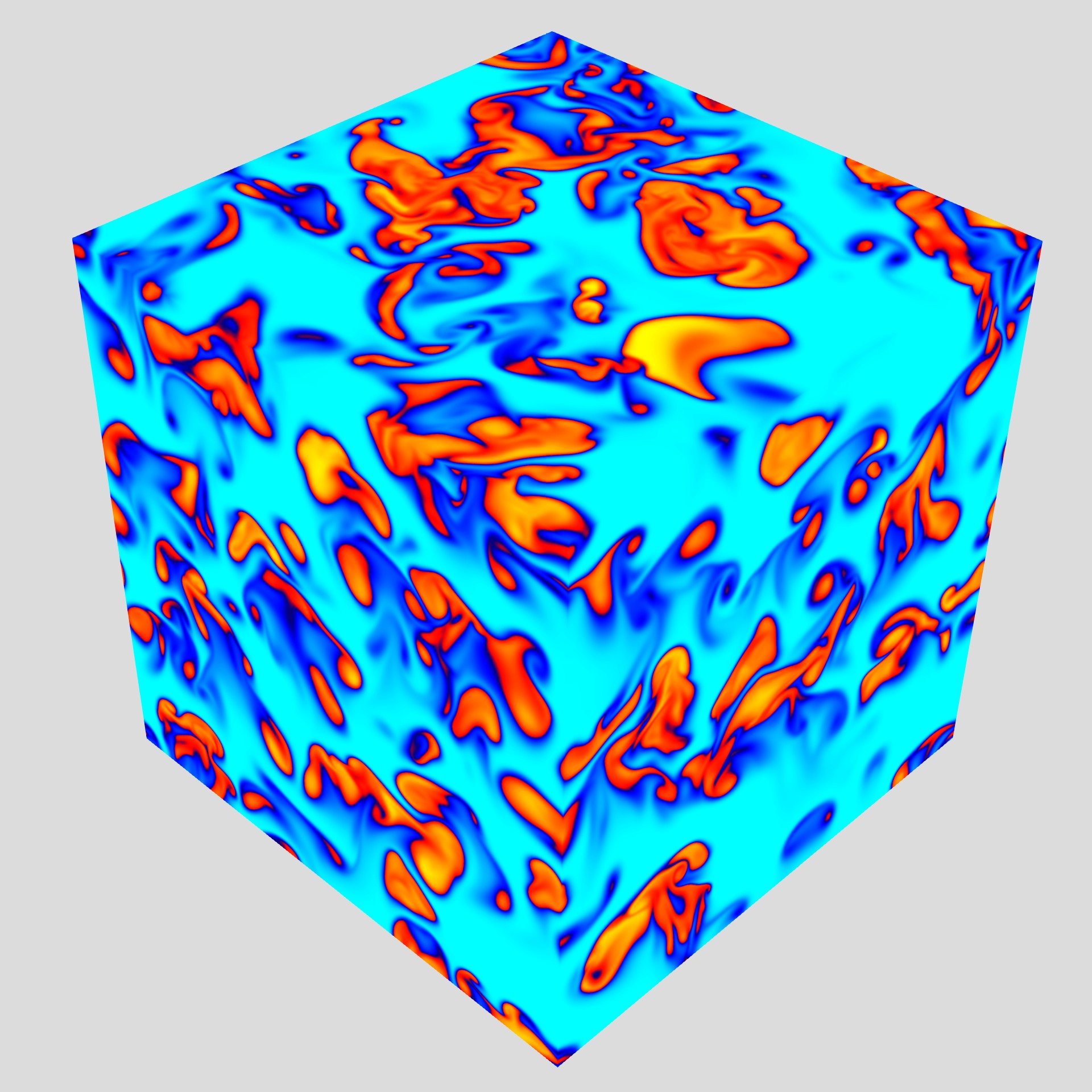}\includegraphics[width=4.6cm]{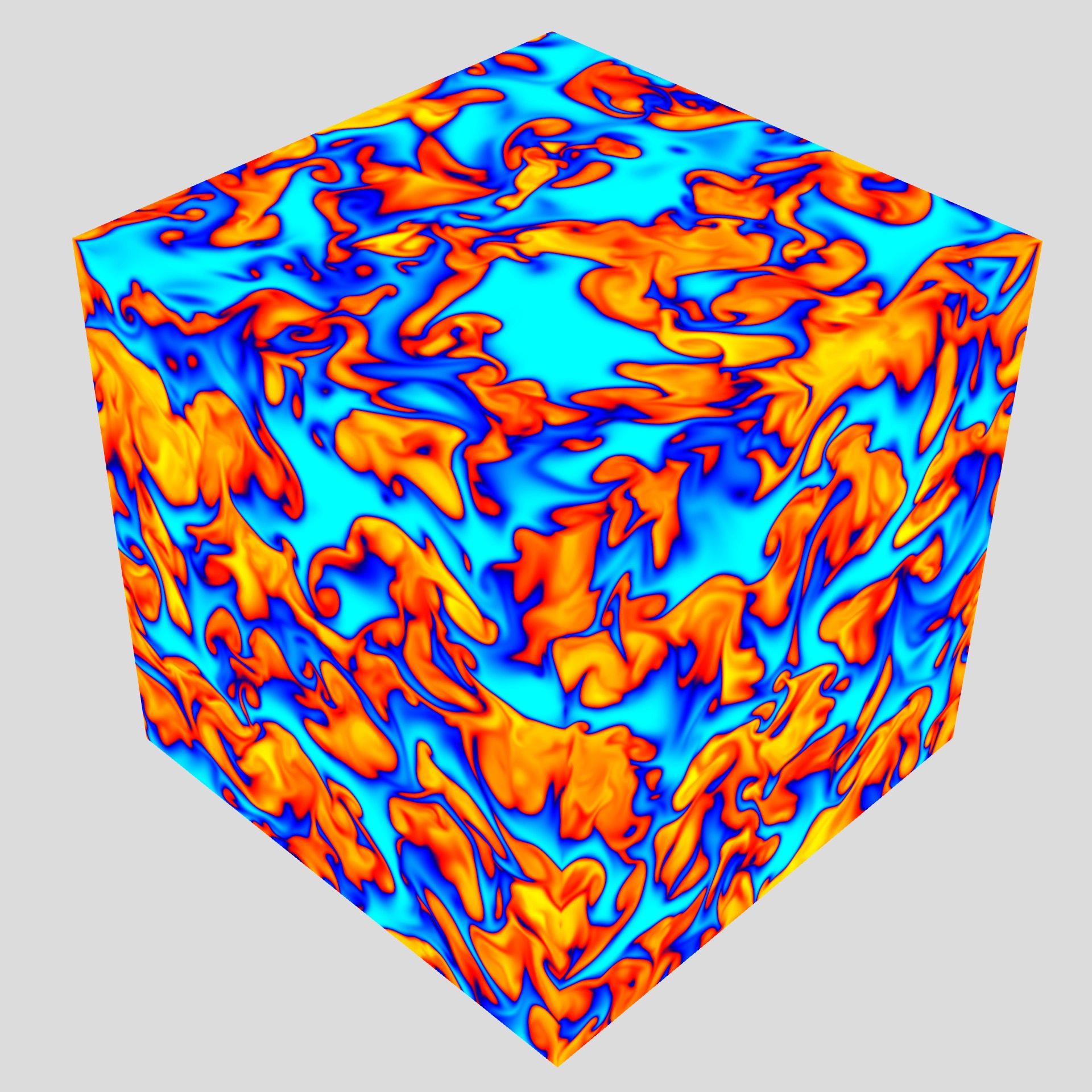}\includegraphics[width=4.6cm]{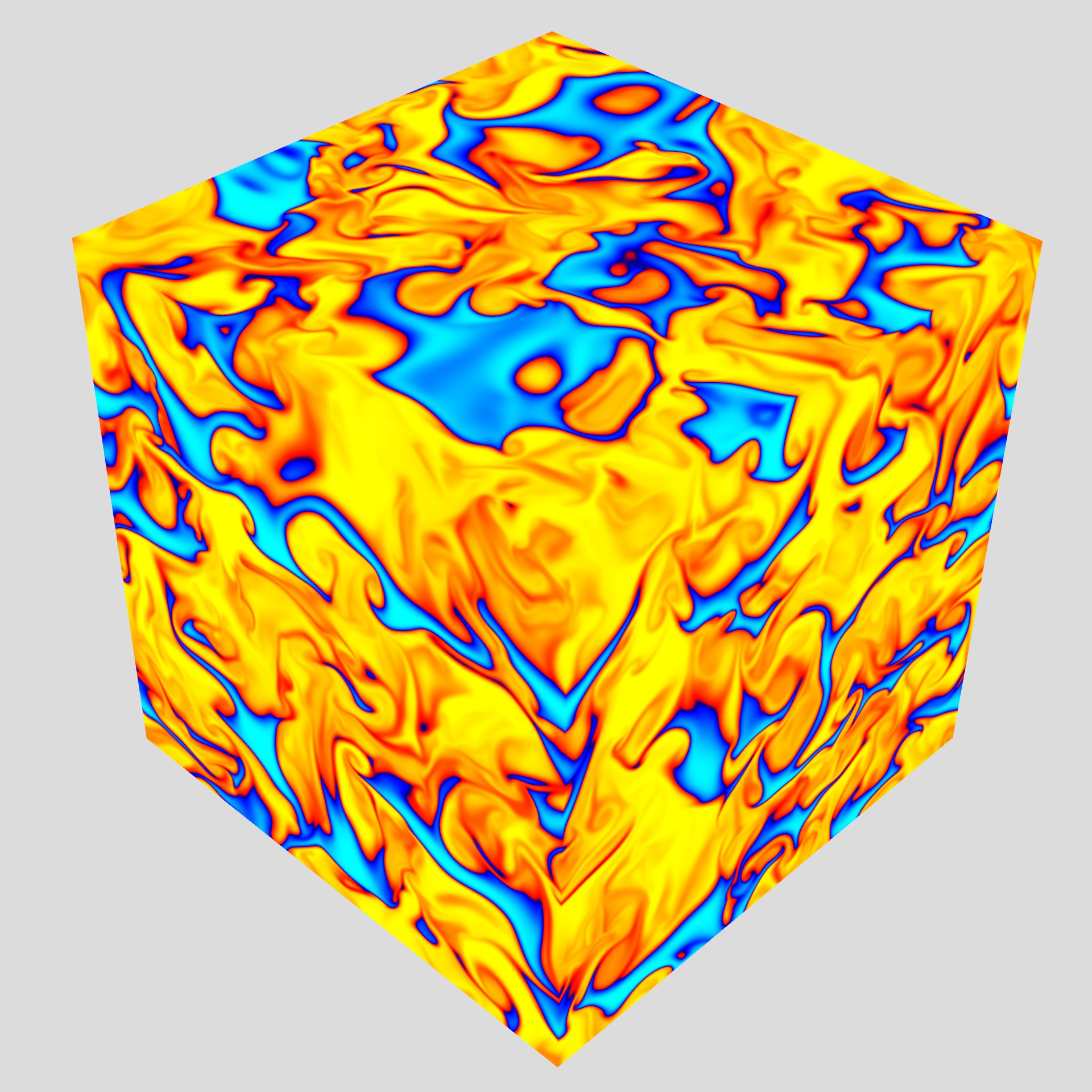}\\
    \rotatebox{90}{\hspace{1.7cm}$t/t_r=3.6$}\includegraphics[width=4.6cm]{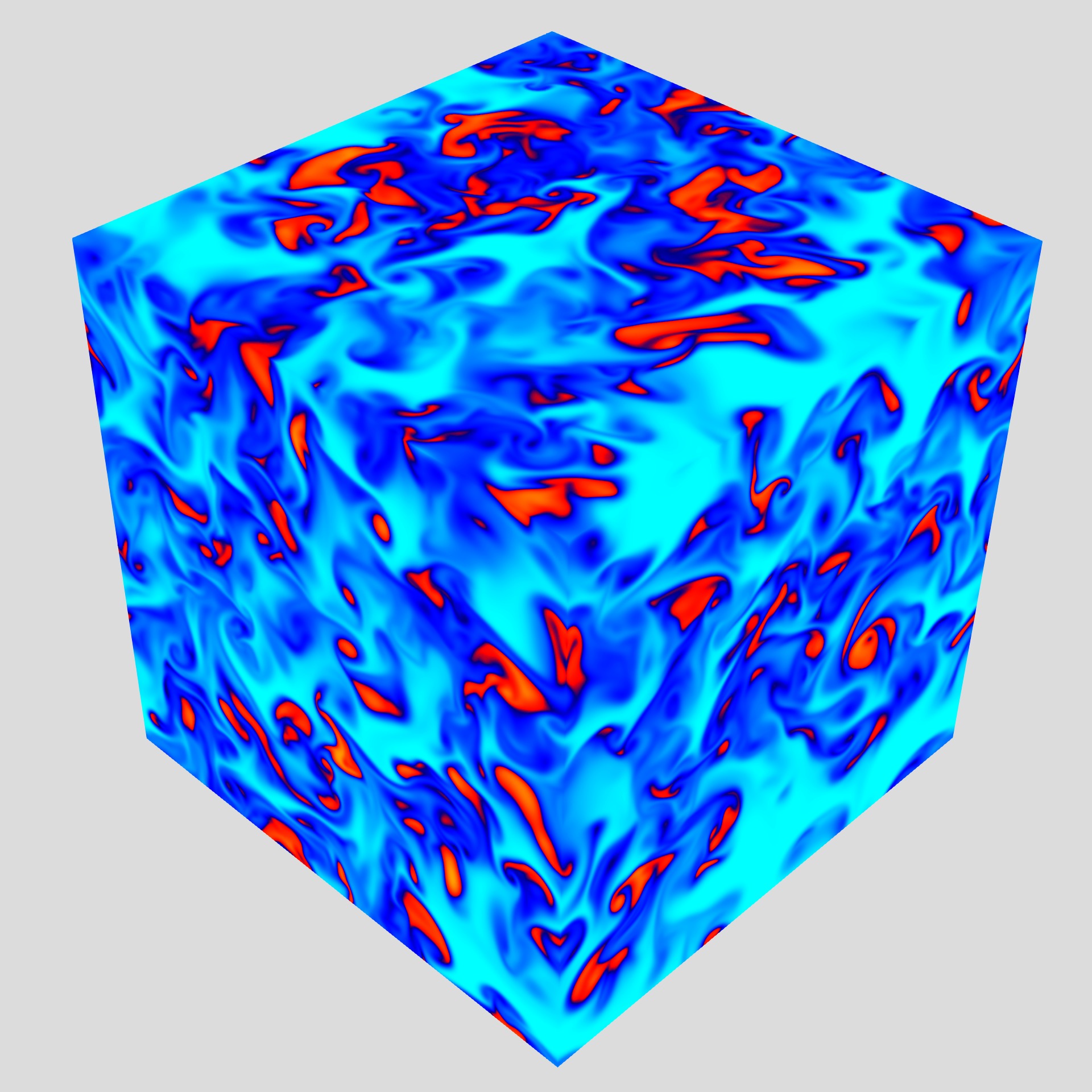}\includegraphics[width=4.6cm]{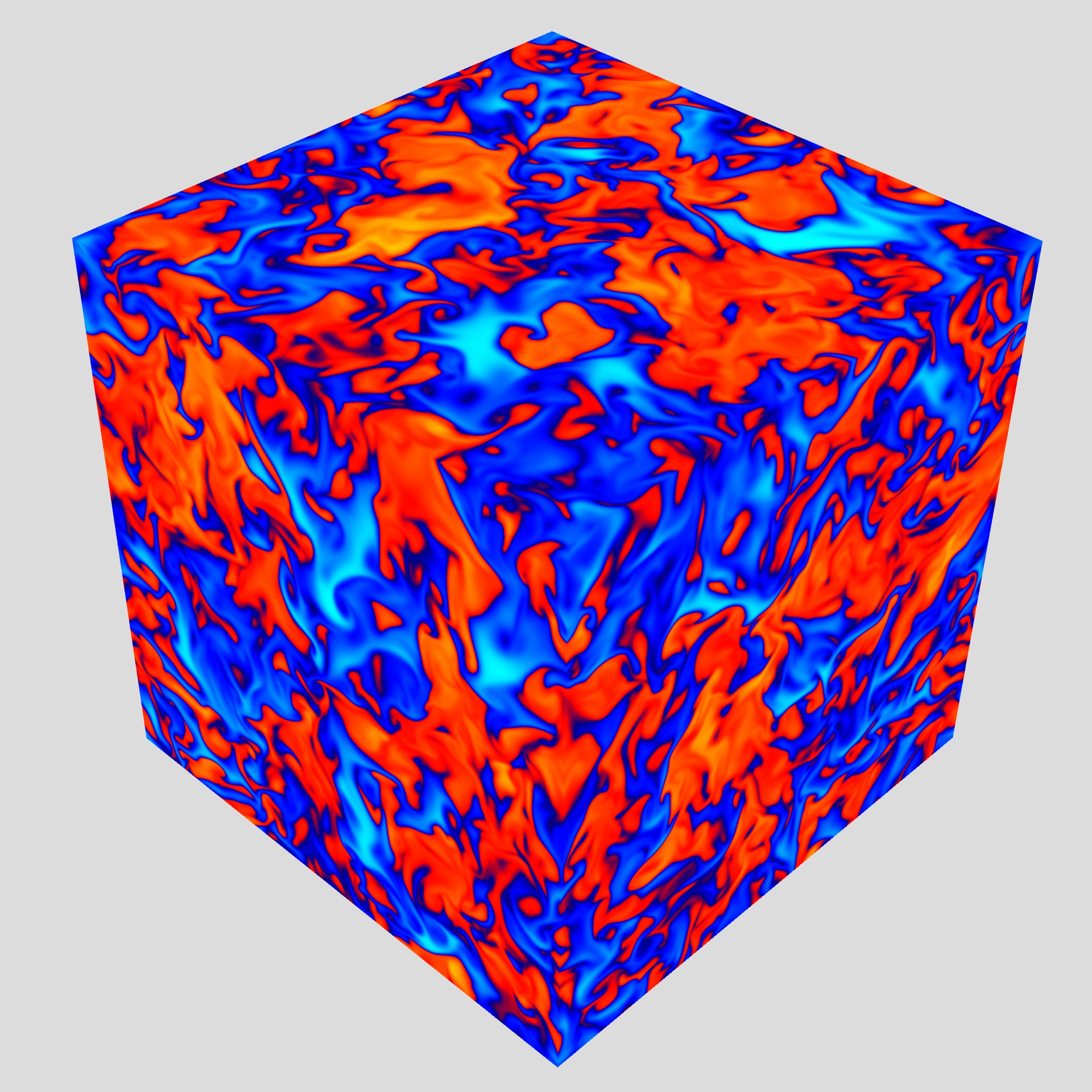}\includegraphics[width=4.6cm]{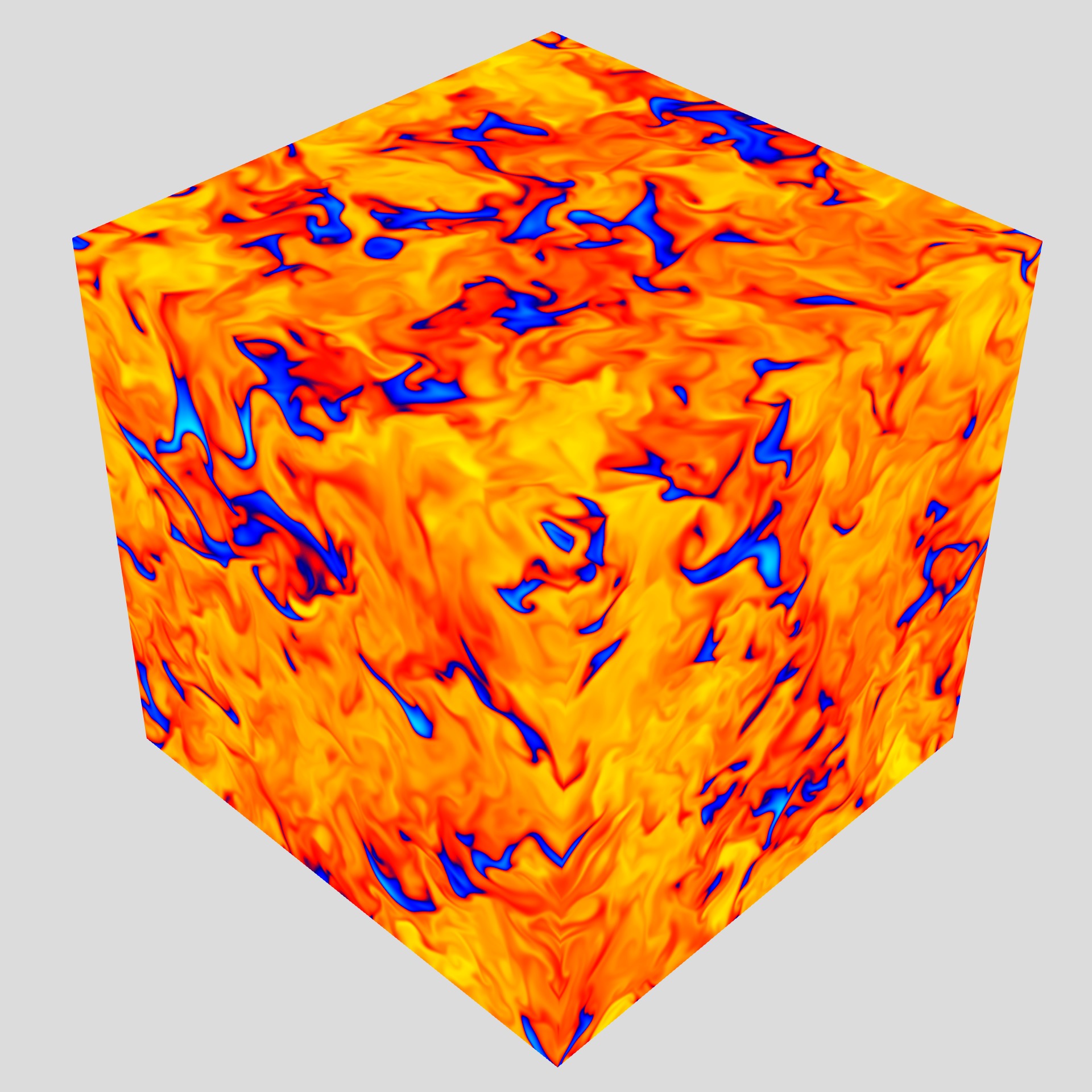}}
    \caption{3D visualization of the density field for the cases with $A=0.75$ at three different time instants $t/t_r=1.2$ (the first row), $t/t_r=2.4$ (the second row), and $t/t_r=3.6$ (the third row), where yellow color represents the pure light fluid ($\chi_h=0, \chi_l=1$) and light-blue color represent the pure heavy fluid ($\chi_h=1, \chi_l=0$).}
    \label{fig:3D_dens}
\end{figure*}

The corresponding density PDFs are shown in Figure \ref{fig:pdfs} at different normalized times running from $t/t_r=1.2$ (top row) to $t/t_r=9.2$ (bottom row); the low \At cases are plotted on the left and high \At cases are on the right. In the figure, the density field is represented by the mole fraction of the heavy fluid ($\chi_h$) defined as $\chi_h=(\rho^*-\rho_1)/(\rho_2-\rho_1)$; for the pure heavy fluid and pure light fluid $\chi_h$ is equal to $1$ and $0$, respectively. In addition, 3D visualization of the density field for the cases with $A=0.75$ at three different time instants: $t/t_r=1.2$ (the first row), $t/t_r=2.4$ (the second row), and $t/t_r=3.6$ (the third row) can be seen in Figure \ref{fig:3D_dens}. Density PDFs mostly conserve their initial shapes during the explosive growth (at $t/t_r=1/2$) as mixing is mostly localized \cite{aslangil2019}. For the low \At number cases, the PDF of the A005SF case remains almost symmetric throughout the flow evolution, and the PDFs of both A005LF and A005HF cases are symmetric to each other at the point where $\chi=0.5$, as both pure light and pure heavy fluids have similar mixing rates during the flow evolution for low \At cases (see Fig. \ref{fig:PureFluids} (a)). 

The density PDFs of the high \At cases are very different. During the flow evolution, not only is the evolution of the PDF of the A075SF case highly asymmetric (even though it starts with symmetric initial PDF), but also the PDF evolution of the density field shows no similarity between the HF and LF cases (even though they start with asymmetric PDFs, but similar to each other at $\chi=0.5$). During the $E_{TKE}$ peak, there is no pure light fluid remaining for the A075HF case; however, there is still a sufficient amount of pure heavy fluid left for the A075LF case, which indicates slower mixing rates for the heavy fluid. The pure light fluid is also more mixed for the A075LF case compared to the pure heavy fluid for the A075HF case. At the end of saturated growth, there are almost no pure fluids left for the A075LF case, but most of heavy fluid remains until very late time for the A075HF case (see also Fig. \ref{fig:PureFluids}). During gradual decay, all cases tend to reach a symmetric shape with different variance values. At late time (at $t/t_r=9.2$) the small variations in the density PDFs between the cases are attributed to the different amounts of the fully-mixed flow for that time (see also fig. \ref{fig:PureFluids}).

\subsubsection{Conditional expectation of turbulent kinetic energy}

\begin{figure}
    \centering
    (\emph{a})\\
    \includegraphics[height=3.6cm]{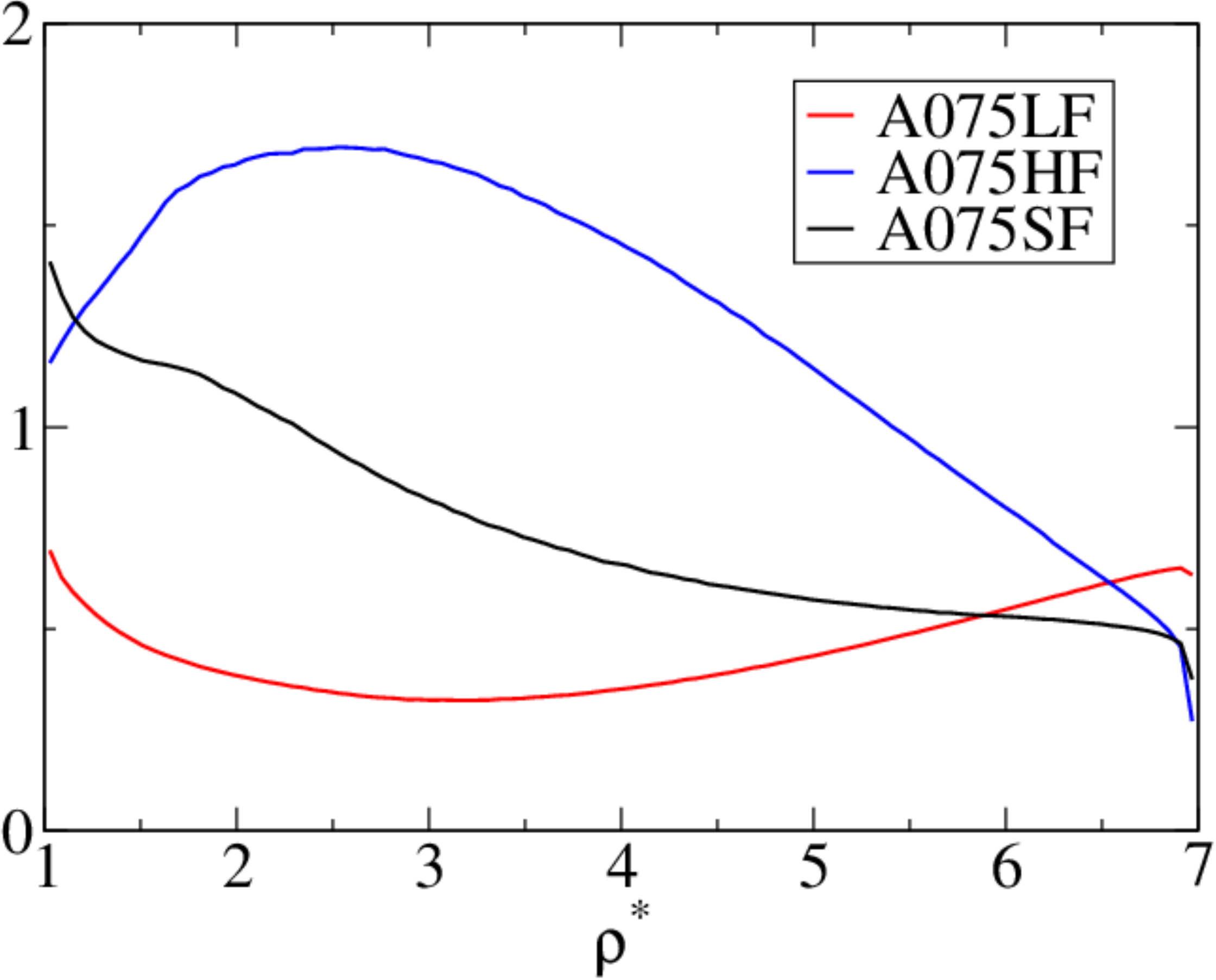}\\
    (\emph{b})\\
    \includegraphics[height=3.6cm]{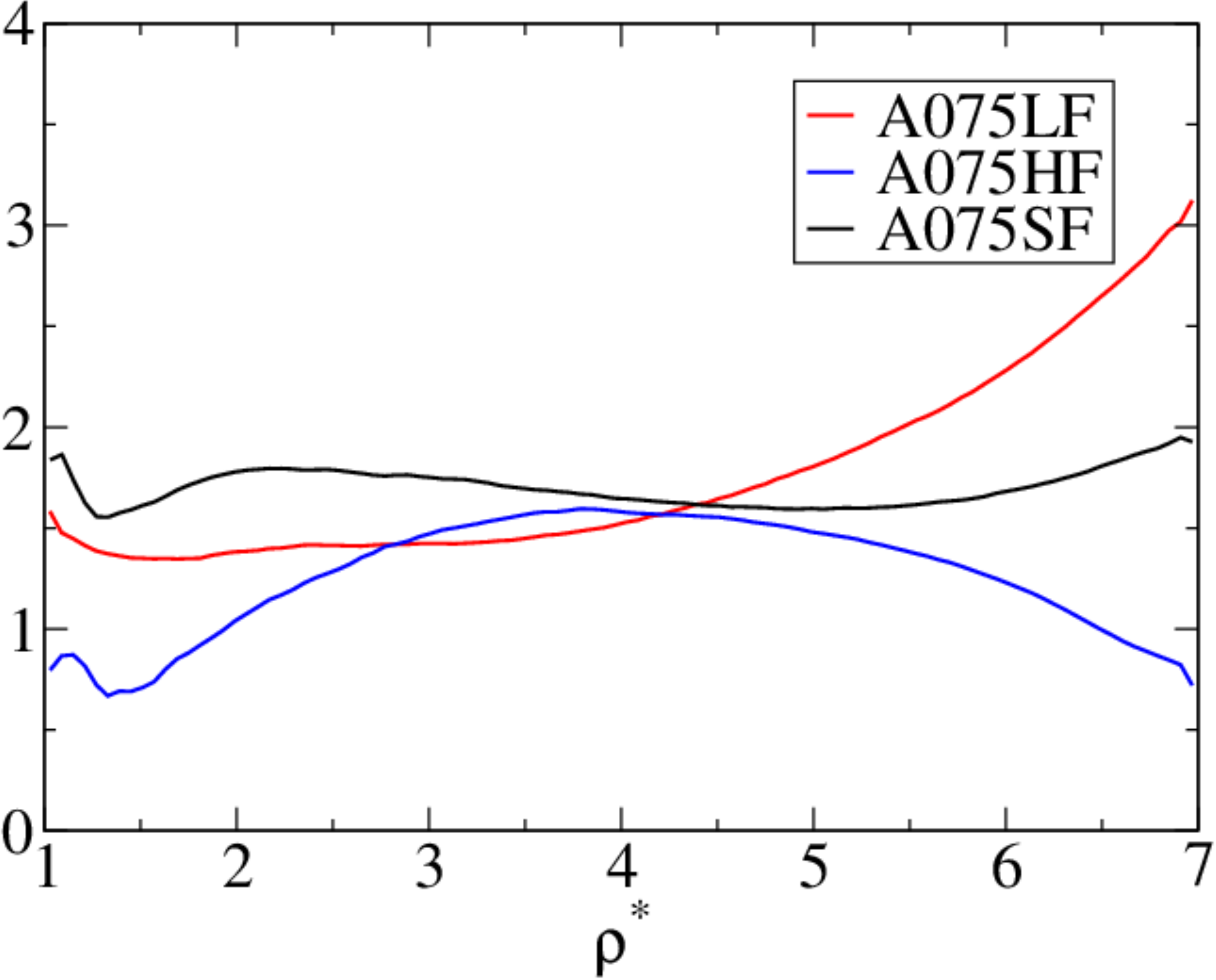}\\ 
    (\emph{c})\\
    \includegraphics[height=3.6cm]{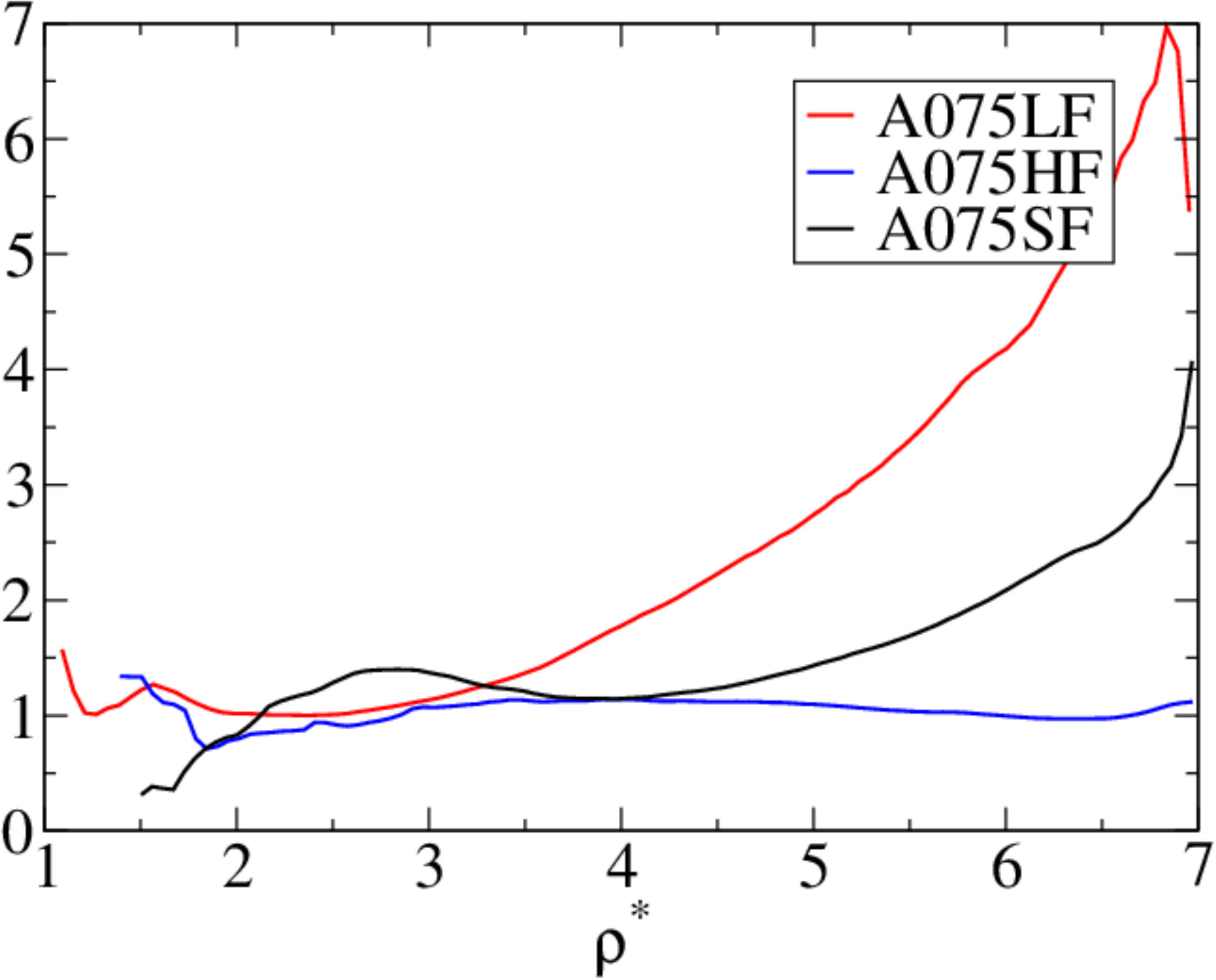}
    \caption{Conditional $E_{TKE}$ for the $A=0.75$ cases at (a) $t/t_r=1.2$, (b) $t/t_r=2.4$, (c) $t/t_r=3.6$.}
    \label{fig:Conditional}
\end{figure}

Figure \ref{fig:Conditional} presents the conditional expectation of the $E_{TKE}$ ($\langle E_{TKE}\rangle\vert_R$) where R is the given/chosen value of density. During explosive growth, $E_{TKE}$ values are larger within lighter fluid regions for the A075SF case, as it is easier to stir these regions due to their smaller inertia \cite{aslangil2019} (see figure \ref{fig:Conditional}a). As it is seen in Figure \ref{fig:Conditional}b) and c), around $E_{TKE}$ peak time and at the end of fast decay, the $E_{TKE}$ levels of the heavy fluid regions increase  upon decrease in initial amount of the pure heavy fluid. Since the initial volume of heavy fluid regions is the smallest, it is easier to disturb and comprehensively stir the heavy fluid regions for the A075LF case. Also, compared to the A075HF case, the initial volume of heavy fluid regions is smaller for the A075SF case. It is thus easier to disturb and stir the heavy fluid regions for the A075SF case compared to the A075HF case.

\subsection{Small scale features}\label{Sec:Small_scales}

\subsubsection{Strain-enstrophy angle}\label{Sec:strain}

\begin{figure}
\hspace{1.5cm}(\emph{a}) \hspace{3.5cm}  (\emph{b}) \\
    \centering
    {\includegraphics[height=4cm]{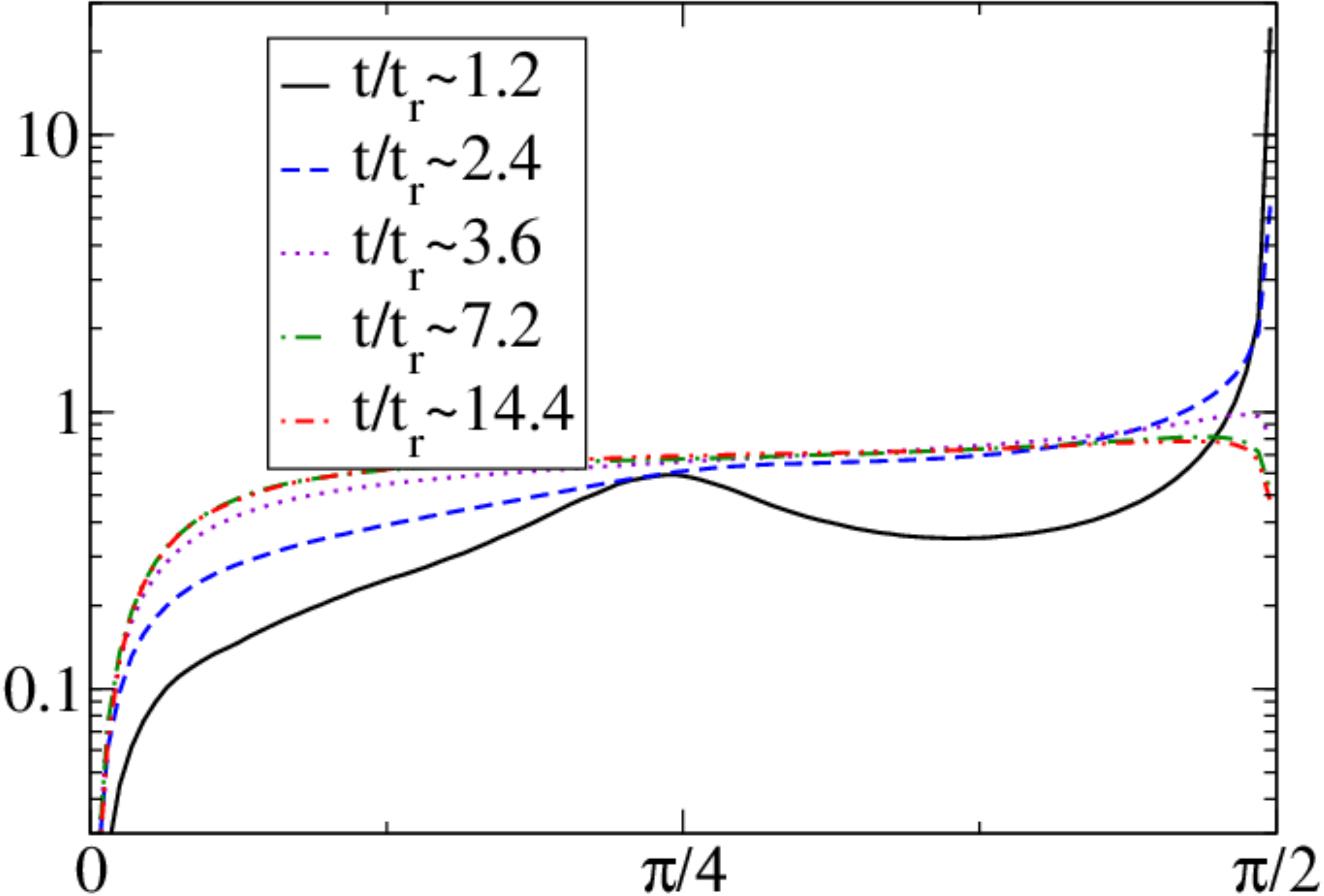}\\ \includegraphics[height=4cm]{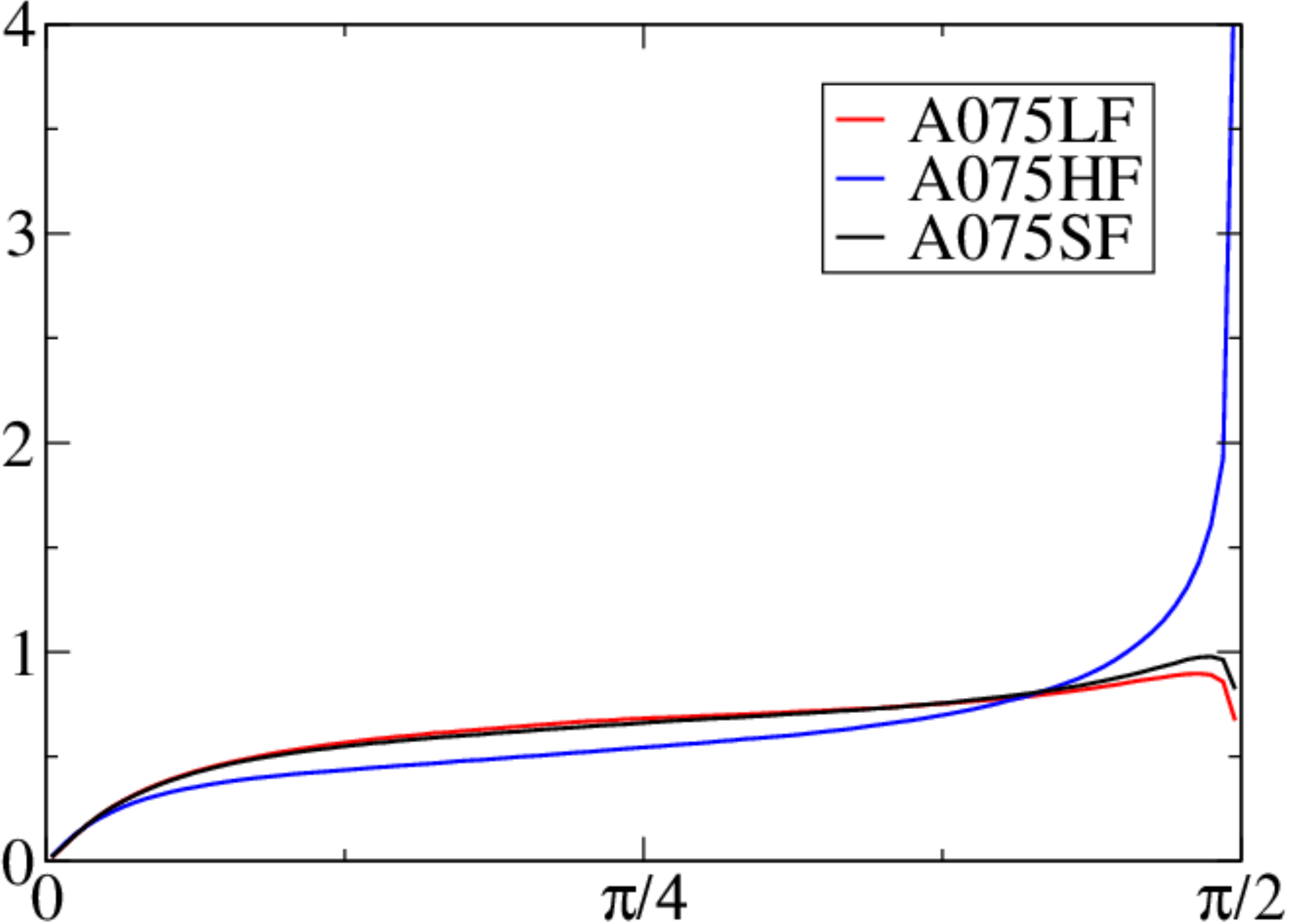}}
    \caption{PDF of the strain-enstrophy angle for (a) A075SF at different time instants, and (b) all cases with $A=0.75$ at $t/t_r=3.6$.}
    \label{fig:Psi}
\end{figure}

\begin{figure}
    \centering
    {\includegraphics[height=4.8cm]{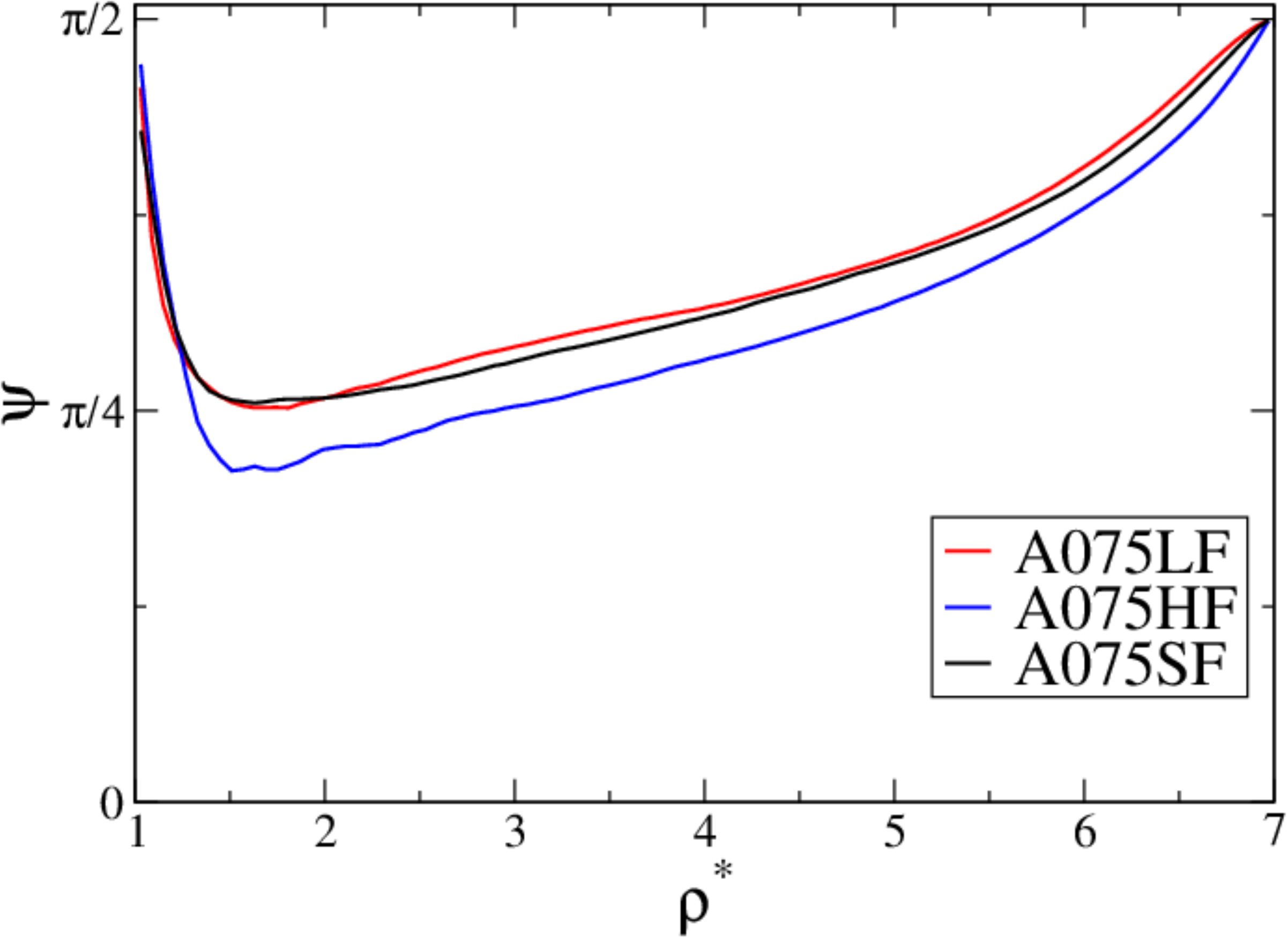}}
    \caption{Conditional expectations of the strain-enstrophy angle ($\langle \Psi\Vert R \rangle$) for the cases with $A=0.75$ at $t/t_r=2.4$.}
    \label{fig:Cond_Psi}
\end{figure}

The strain-enstrophy angle is defined as \cite{Boratav_onthealign_doi:10.1063/1.869747}:
\begin{equation}
    \Psi=tan^{-1}\frac{S_{ij}S_{ij}}{W_{ij}W_{ij}},
    \label{Eq:strain}
\end{equation}
where $S_{ij}=\frac{1}{2}(A_{ij}+A_{ji})$ is the rate of strain tensor, and $W_{ij}=\frac{1}{2}(A_{ij}-A_{ji})$ is the rate of rotation tensor, and the velocity gradient tensor is $A_{ij} = u^*_{i,j}$. Thus, $\Psi>\pi/4$ values represent the flow regions where the strain effects are more dominant than the rotation effects. Figure \ref{fig:Psi}a presents the PDF of $\Psi$ for the A075SF case at different time instants (note that y axis is in the log scale). As seen in the figure, during both explosive and saturated growth regimes, the PDF mostly accumulates at a value of $\pi/2$, which indicates that the flow is largely strain dominated. However, while the growth and mixing saturate, the peak of the PDF at the value of $\pi/2$ starts to decrease indicating that more regions are being affected by vorticity. 
Figure \ref{fig:Psi}(b), compares for the PDF of $\Psi$ for the high \At cases during fast decay (at $t/t_r=3.6$). For the A075HF case, the PDF reaches its asymptotic shape later compared to the A075LF and A075SF cases. The delay in reaching the long time behavior for the A075HF case is attributed to the absence of turbulence within the heaviest fluid regions. This can be also observed in Figure \ref{fig:3D_dens}, at $t/t_r=3.6$, the A075HF case still contains pure heavy fluid regions that are not comprehensively stirred yet. This is consistent with our earlier findings \cite{aslangil2019} as stirring heavy fluid regions is more difficult due to their larger inertia. During flow evolution, the behavior of $\psi$ is similar for the low \At cases and are not shown here for brevity. Figure \ref{fig:Cond_Psi} shows the conditional expectation of $\Psi$ ($\langle \Psi\Vert R \rangle$) at around $E_{TKE}$ peak (at $t/t_r=2.4$). As observed, $\Psi$ has significantly lower values within slightly mixed light fluid regions, where strain and rotation are almost balanced. On the contrary, within heavier than fluid regions, larger values were observed.

\subsubsection{Flow topology} \label{Sec:PQR}

\begin{figure*}
\centering{
\hspace{1cm}(\emph{a}) \hspace{3.5cm}  (\emph{b}) \\
\includegraphics[height=4.6cm]{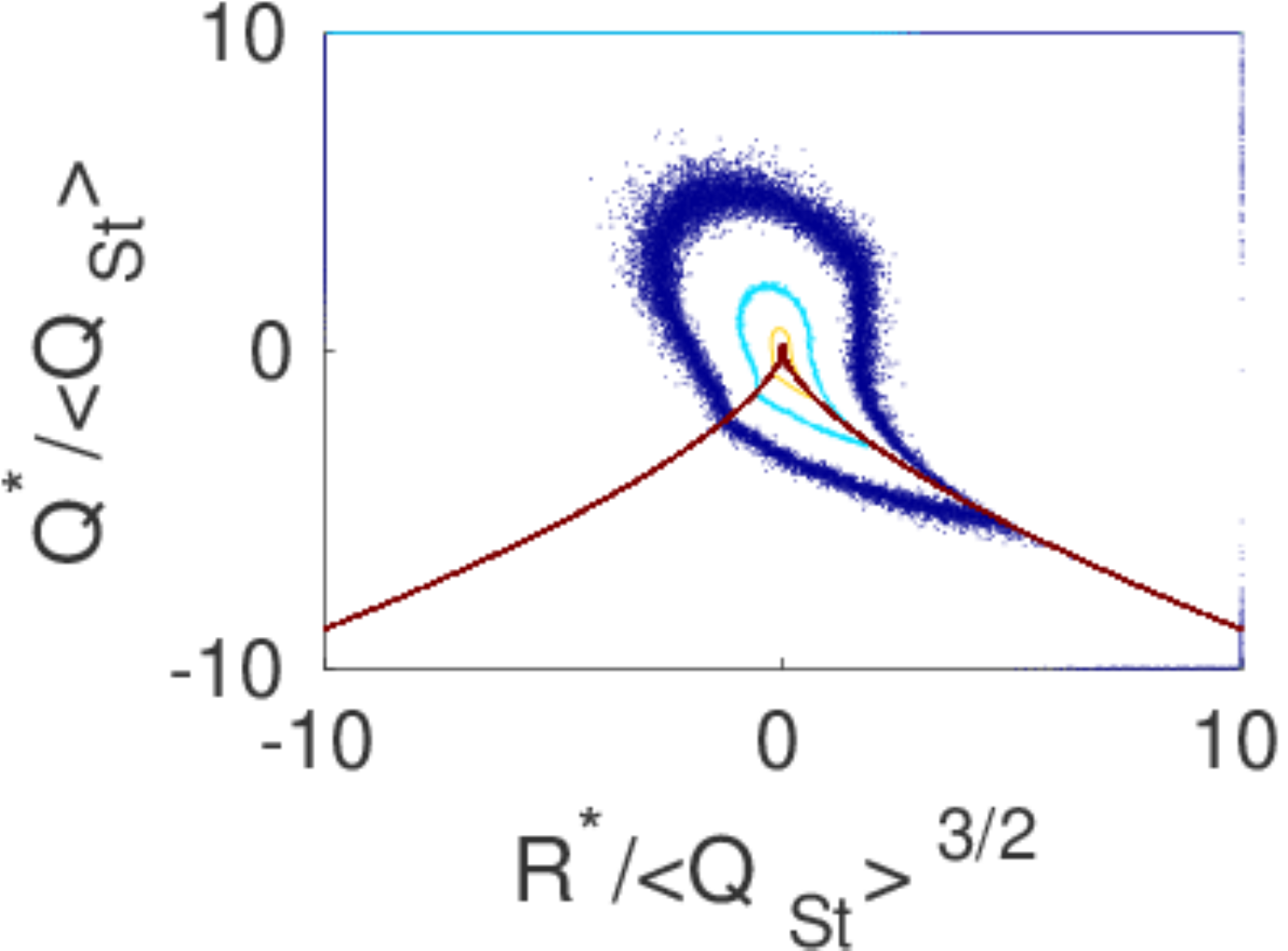}\includegraphics[height=4.6cm]{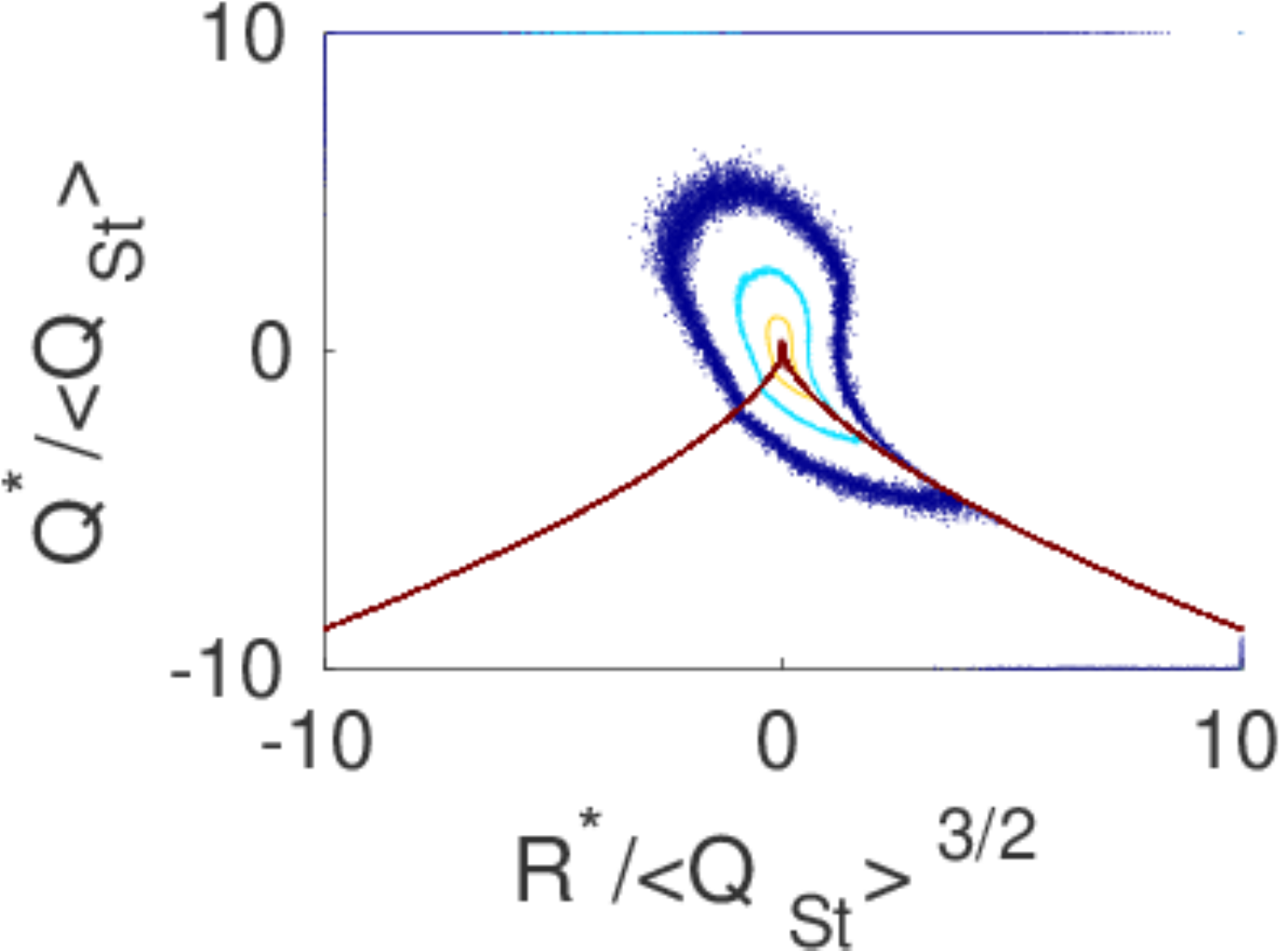}\\
\hspace{1cm}(\emph{c}) \hspace{3.5cm}  (\emph{d}) \\
\includegraphics[height=4.6cm]{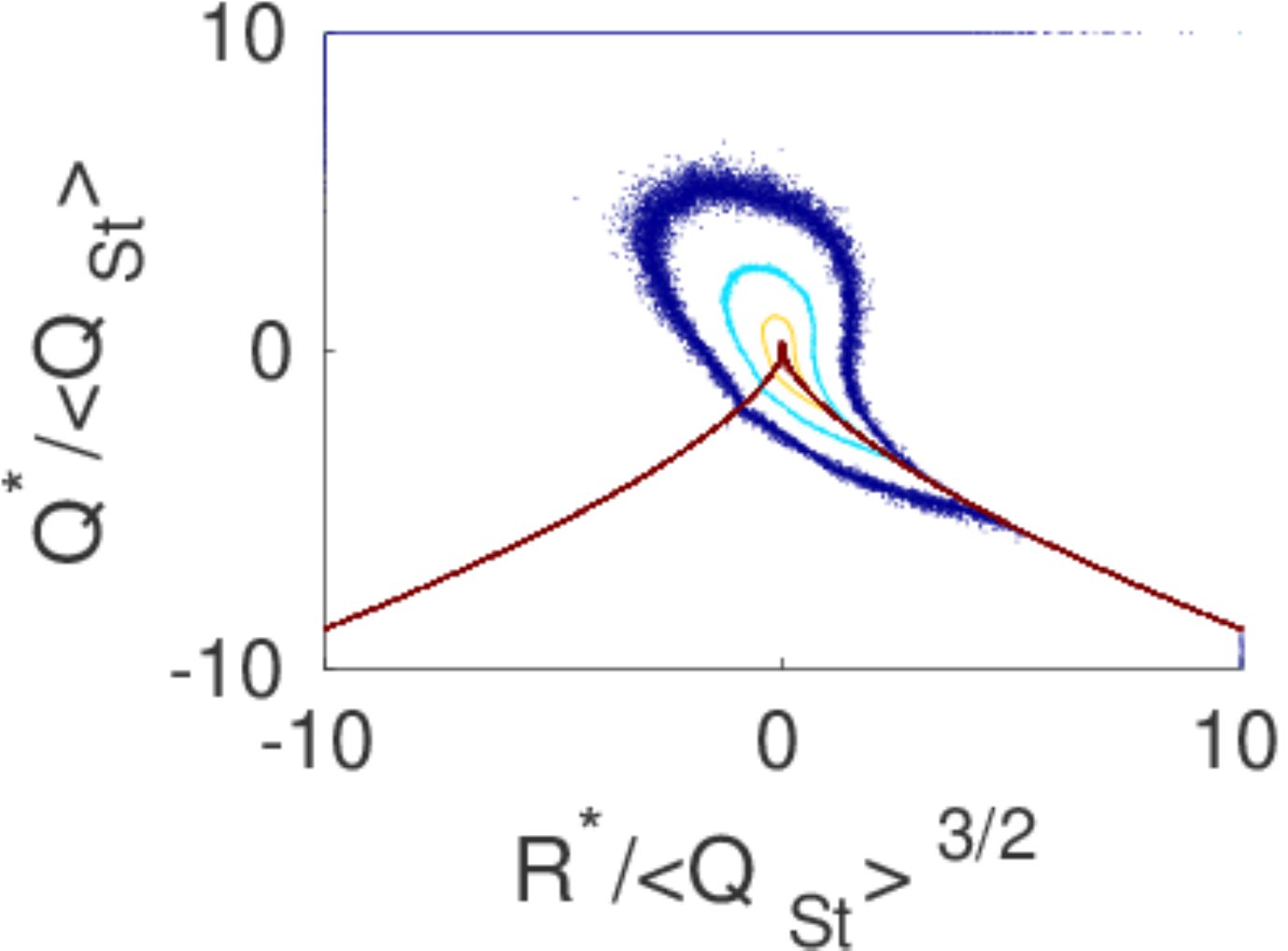}\includegraphics[height=4.6cm]{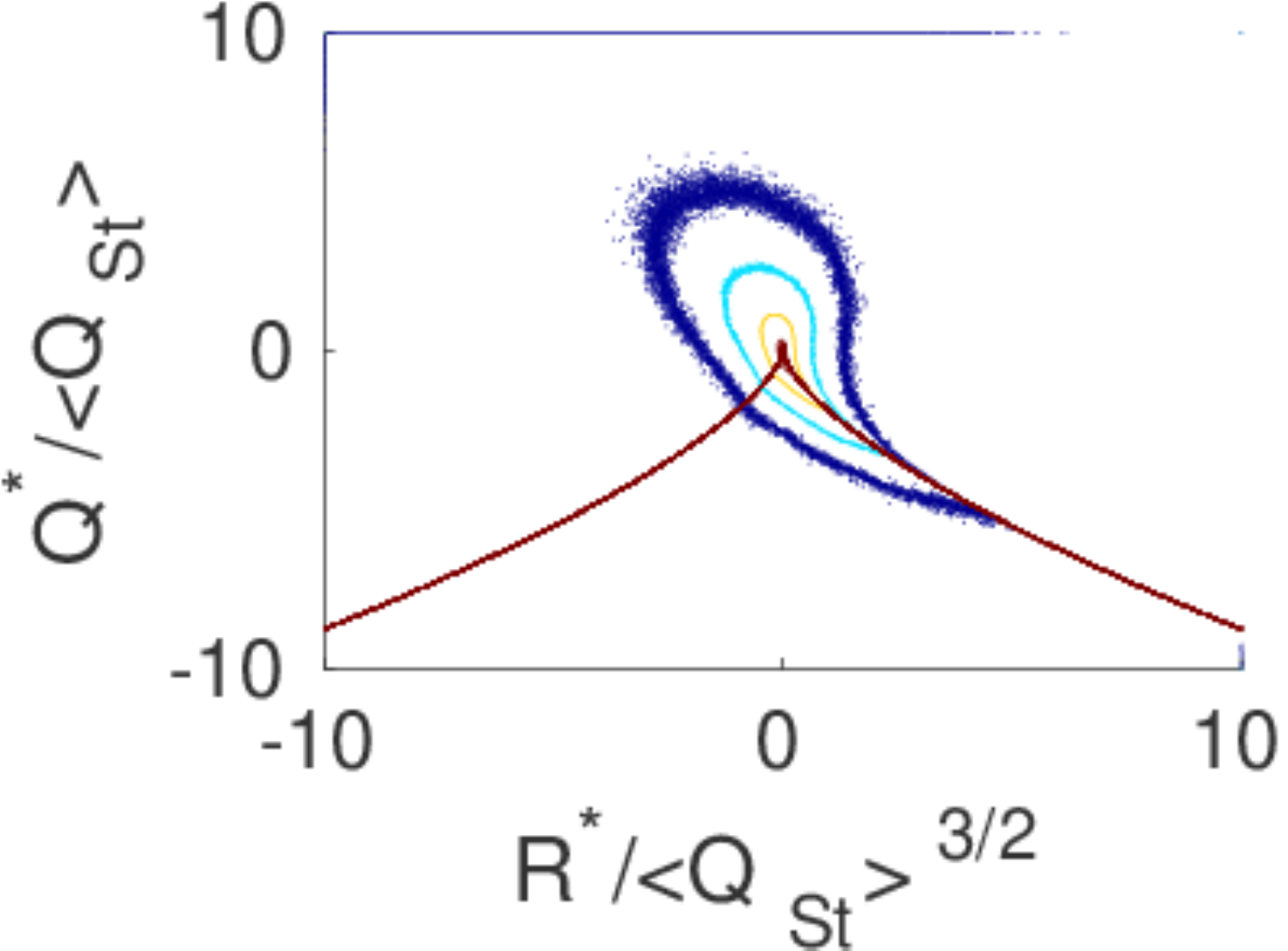}}
    \caption{Iso-contour lines of $\log_{10}$(jPDF)s of the normalized second and third invariants of the anisotropic part of the velocity gradient tensor, ($Q^*/\langle St \rangle$, $Q^*/\langle St  \rangle^3/2$) for the (a) A075HF at $t/t_r=2.4$, (b) A075LF at $t/t_r=2.4$, (c) A075HF at $t/t_r=4.8$, and (d) A075HF at $t/t_r=4.8$.}
    \label{fig:PQR}
\end{figure*}

\begin{figure}
    \centering
    {
    (\emph{a})\\
    \includegraphics[height=4.4cm]{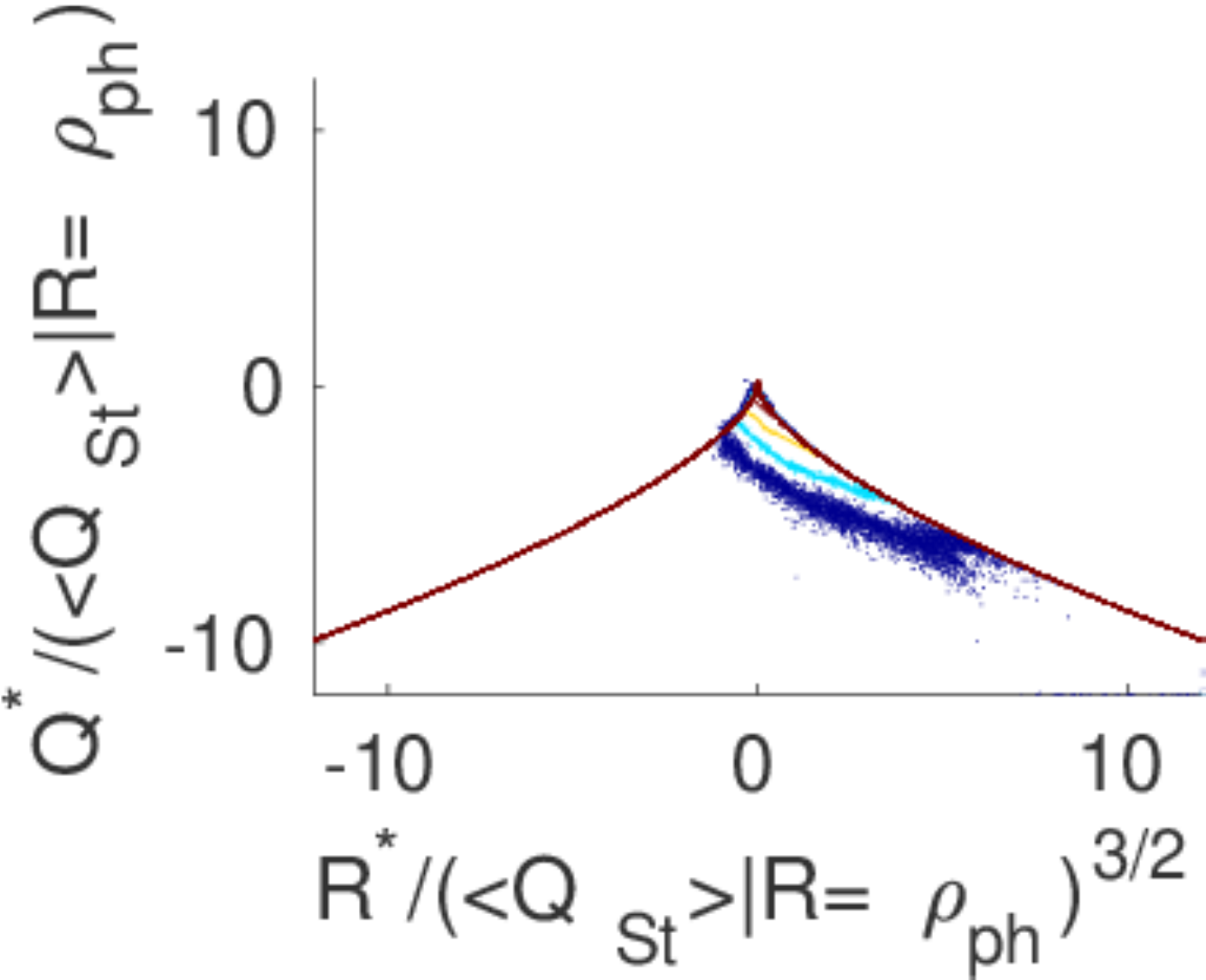}\\
    (\emph{b})\\
    \includegraphics[height=4.4cm]{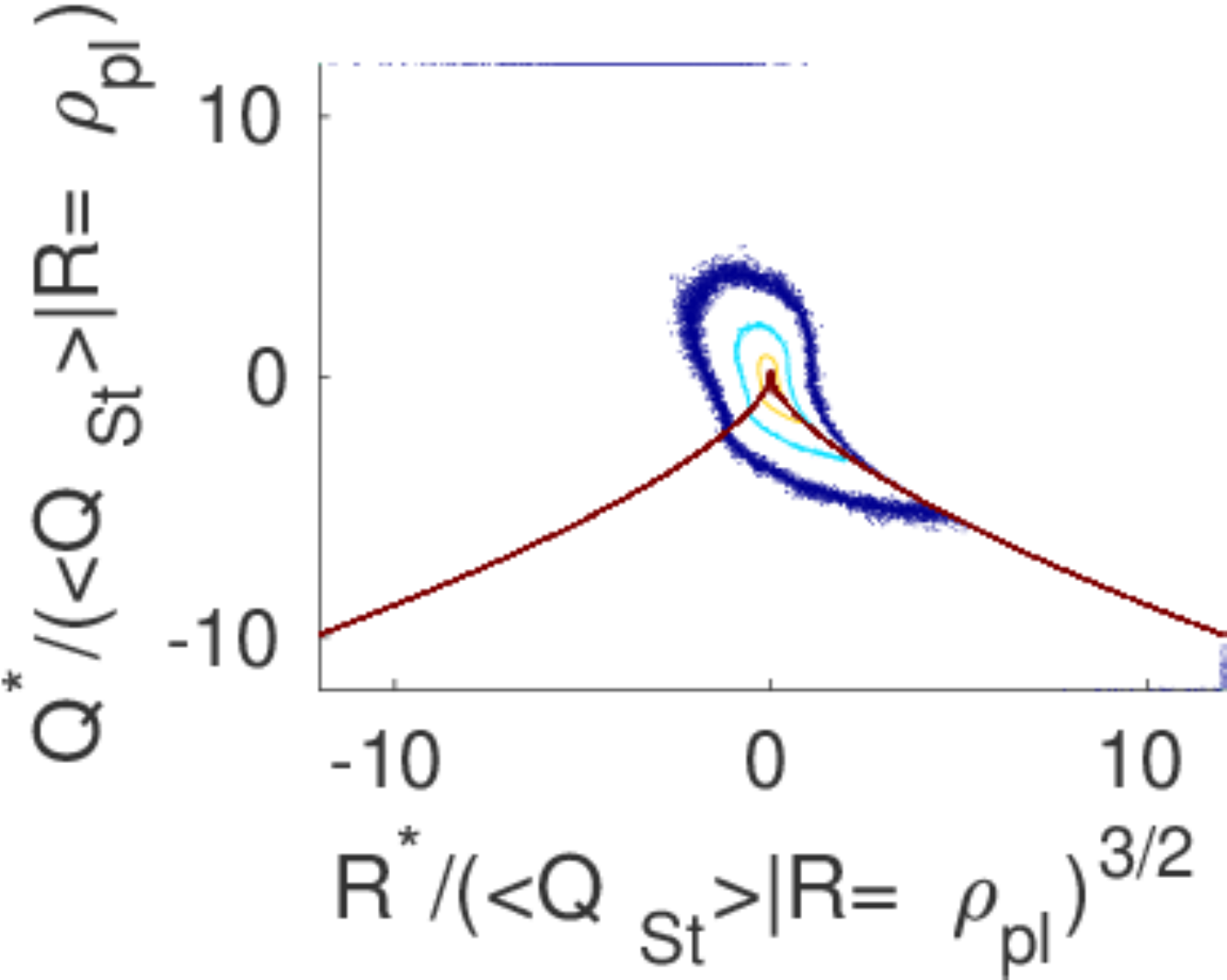}}
    \caption{Iso-contour lines of $\log_{10}$(jPDF)s of the normalized second and third invariants of the anisotropic part of the velocity gradient tensor, ($Q^*/\langle St \vert R \rangle$, $Q^*/\langle St \vert R  \rangle^3/2$) for the (a) pure heavy and (b) pure light fluid regions of the A075LF at $t/t_r=2.4$.}
    \label{fig:PQR_PF}
\end{figure}

Here, the topology of HVDT flow is investigated by using the second ($Q^*$) and third ($R^*$) invariants of the velocity gradient tensor, which are defined as:
\begin{equation}\label{Eq:QR}
    Q^*=-\frac{1}{2}A_{ij}A_{ji};~~~~~~~R^*=-\frac{1}{3}A_{ij}A_{jk}A_{ki}.
\end{equation}
The flow regions where $Q^*$ is large indicates rotation dominant regions, large values of $R^*$ indicate strain dominant regions. Figure \ref{fig:PQR} presents the iso-contour lines of the joint-PDF of the normalized second and third invariants of the anisotropic part of the velocity gradient tensor \cite{Book_turb_eng}. Note that, in previous studies, $Q^*$ and $R^*$ are normalized by $\langle Q_\omega \rangle= \langle W_{ij}W_{ij}/2\rangle$ and $\langle Q_\omega \rangle^{3/2}$, respectively. For the current work, $Q^*$ and $R^*$ are normalized by using $\langle Q_{St} \rangle= \langle S_{ij}S_{ij}/2\rangle$ and $\langle Q_{St} \rangle^{3/2}$ as the flow is mostly dominated by strain effects (see Figures \ref{fig:Psi} and \ref{fig:Cond_Psi}). As observed, jPDF of $Q^*$ and $R^*$ has a teardrop shape, similar to ther canonical turbulent flows, with some subtle variations among cases. However, jPDF of $Q^*$ and $R^*$ for the pure light and heavy fluid regions for the A075LF case at $E_{TKE}$ peak time ($t/t_r=2.4$) show significant differences. This time is chosen as it contains sufficient amounts of both pure light and heavy fluids at the $E_{TKE}$. The conditional expectation of the $\langle Q_{St} \rangle$ at the pure fluid regions are used for normalization; for example, in figure \ref{fig:PQR_PF}a, $Q^*$ is normalized by $\langle Q_{St} \vert R=\rho_{ph} \rangle$ where $\rho_{ph} \geq 0.95(\rho_2-\rho_1)$. While the pure light fluid regions maintain the tear-drop shape, within the heavy fluid regions, there are no points above the characteristic lines, indicating that the velocity gradient tensor has real eigenvalues. The absence of points in the focal regions of the flows, which are associated with vorticity production and attenuation is
consistent with observations in the conditional expectations of $\psi$ (see figure \ref{fig:Cond_Psi}) and also with our recent findings \cite{aslangil2019}, where it is seen that, for high \At number cases, turbulence is mostly generated within light fluid regions.

\begin{figure}
    \centering
    (\emph{a})\\
    \includegraphics[height=4.4cm]{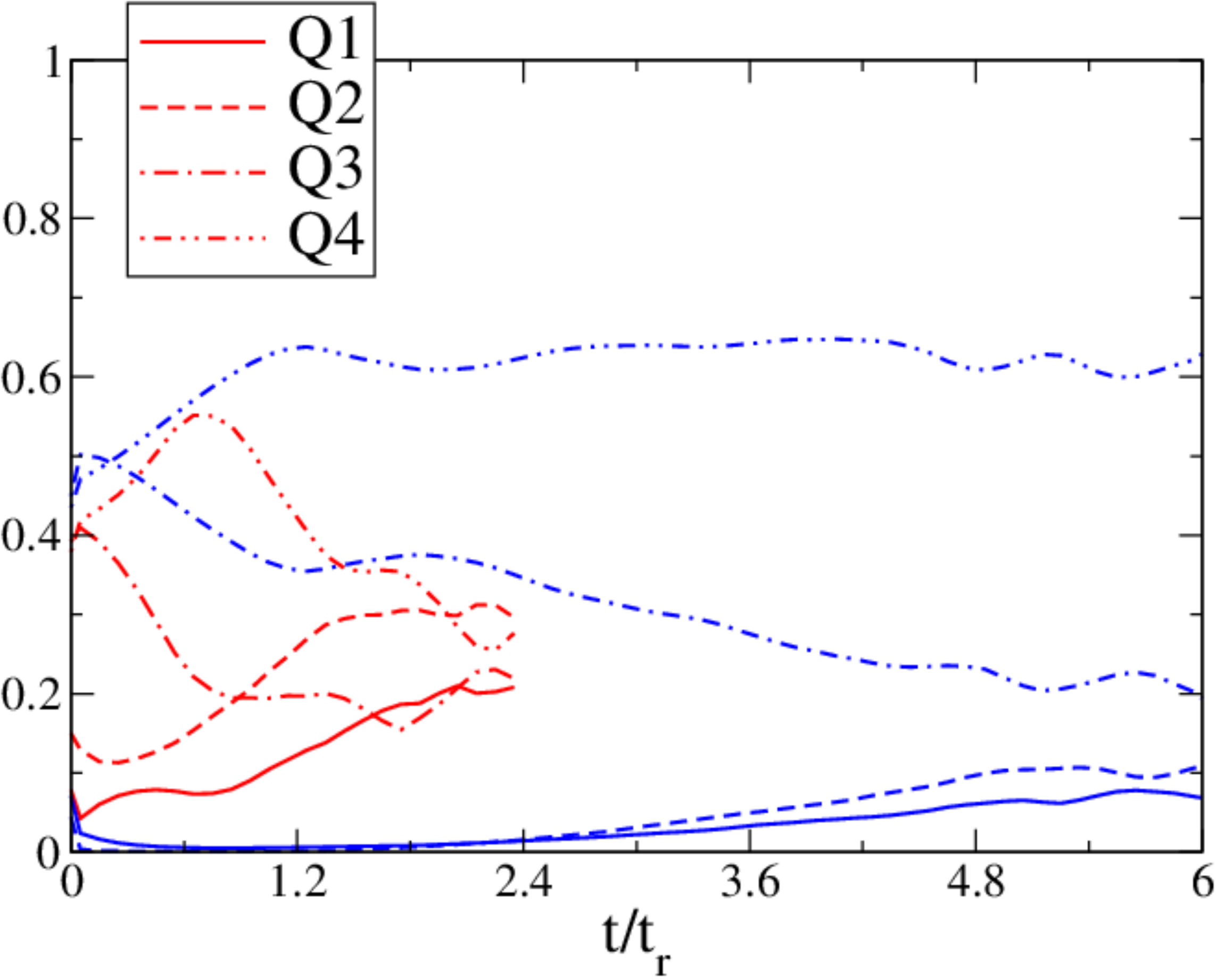}\\
    (\emph{b})\\
    \includegraphics[height=4.4cm]{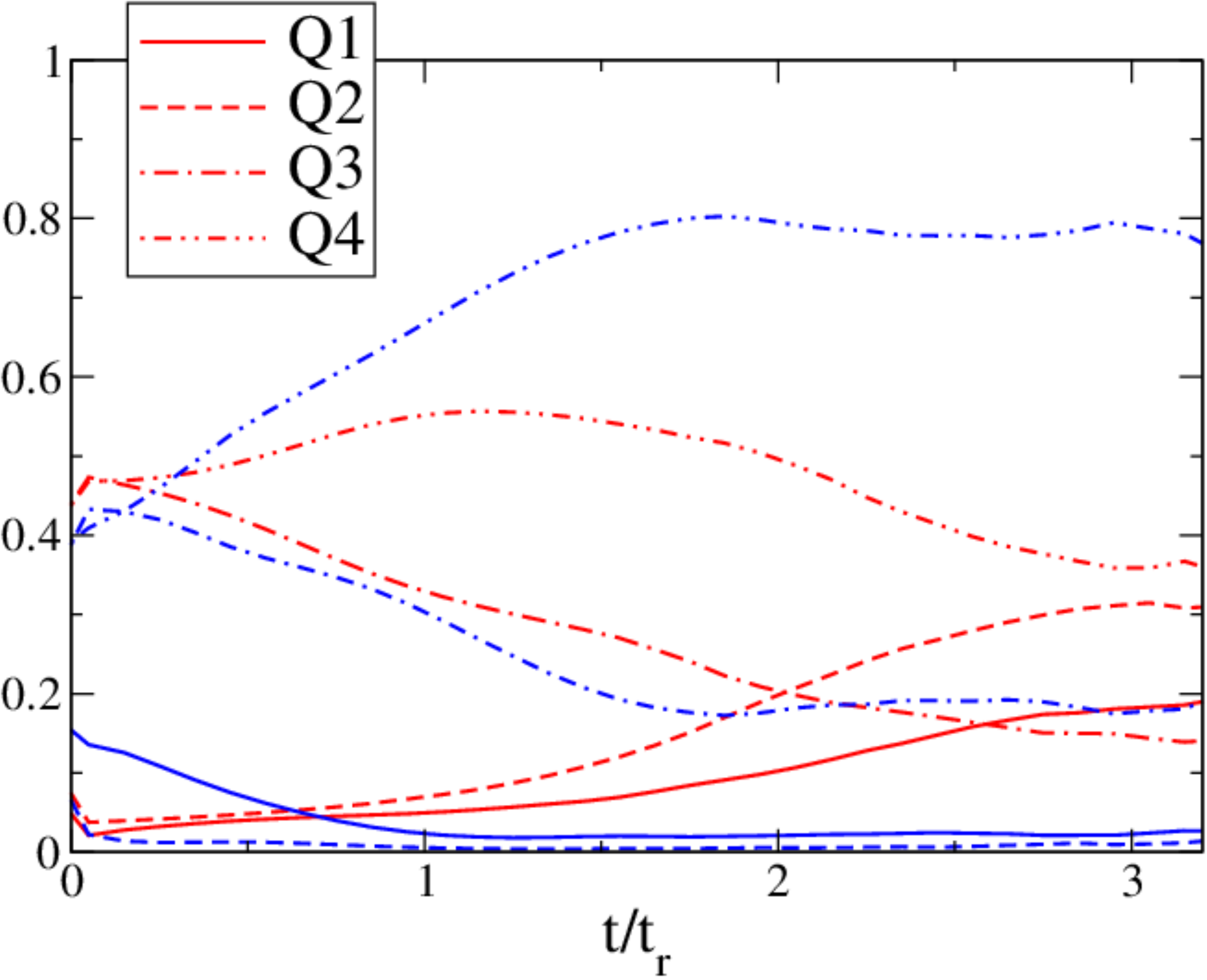}
  \caption{Evolution of the Q,R quadrant percentages with in pure light (red lines) and pure heavy (blue lines) regions for (a) A075HF and (b) A075LF cases.}
\label{fig:QR_amounts}
\end{figure}

Finally, the time evolutions of the percentage of the points that appear in region Q1 (above the zero discriminant line and $R^*>0$), Q2 (above the zero discriminant line and $R^*<0$),  Q3 (below the zero discriminant line and $R^*<0$) and Q4 (below the zero discriminant line and $R^*>0$) within the pure heavy and light fluid regions for A075LF and A075HF cases are plotted in Figure \ref{fig:QR_amounts}. The curves are shown up to the time instants when the amounts of the pure fluids become too small for reliable statistics. Regions Q1 and Q2 represent the focal regions (stable focus / contraction and stable focus / stretching, respectively) where vorticity is produced and attenuated, whereas  Q3 and Q4 represent the regions of stable node / saddle / saddle and unstable node / saddle / saddle. For both cases, the pure fluids regions are dominated by strain effects, which is consistent with figure \ref{fig:Cond_Psi}. In addition, there are almost no points that appear in Q1 and Q2 quadrants within the pure heavy fluid regions. This indicates a lack of vortical activity. This observation is also consistent with the behavior of $\Psi$ in Figure \ref{fig:Psi}, where it also reaches its long time asymptotic  shape later for the A075HF case, as it contains more regions with heavy fluid.

\section{Discussions and conclusions} \label{Sec:Discussions}

Effects of initial composition ratio on the evolution of the HVDT have been explored for low \At $=0.05$ and high \At $0.75$ cases. Three different cases with different initial compositions have been investigated for each Atwood number; these are the initially symmetric case SF, and the initially non-symmetric cases LF and HF. SF may be considered as the classical HVDT flow, where the initial amounts of both the pure fluids are identical and the initial density distribution is symmetric. LF is the light fluid dominated flow and HF is the heavy fluid dominated flow; for LF, $\approx 3/4$ of the flow is composed of pure light fluid and $\approx 1/4$ of the flow is composed of pure heavy fluid. The scenario is reversed for HF case. The DNS results presented here represent new test cases for turbulence models of variable density turbulence, which complement the existent studies. The main findings based on our simulations can be listed as:

\begin{itemize}
\item For low \At number cases, $E_{TKE}$ reaches slightly larger values for SF compared to non-SF cases, due to its larger initial density variance. Meanwhile, the evolution of the global parameters such as turbulent kinetic production to dissipation ratio, $P/\epsilon$, density-specific volume correlation, $b$, time scale ratio of mechanical to density variance dissipations, $\Upsilon$, and the mixing rates of the pure fluids are similar for all the cases. In addition, the duration of each of the HVDT flow regimes is not affected by different initial compositions.
\item For high \At number cases, due to large inertial differences between the light and heavy fluids, the flow evolution differs between the SF case and the non-SF cases. The main differences are:
\begin{itemize}
    \item  Upon increasing the initial amount of the pure light fluid, the turbulence kinetic energy generation is enhanced, whereas upon increasing the initial amount of the pure heavy fluid, the turbulence generation is suppressed.
    \item  Differential initial composition also changes the duration of the flow regimes. It is found that explosive growth is longer for the flows that are initially composed of more pure light fluid compared to those of more pure heavy fluid. For the A075LF case, having longer explosive growth, not only drives the $E_{TKE}$ to higher levels, but also postpones the time instant of transitional behavior of parameters such as $b$ and the kinetic energy production to dissipation rate. 
    \item It takes longer for turbulence to disperse into the regions of heavy fluid compared to regions of light fluid. When the amount of the pure heavy fluid is increased within the flow (as in A075HF), turbulence is not observed at the center of the pure heavy fluid regions as no local stirring occurs in those regions. Hence, the conditional expectation of the $E_{TKE}$ stays at lower levels during the flow evolution, and it takes longer to reach the asymptotic behavior of the strain-enstrophy angle.
\end{itemize}
 \item During gradual decay, the initial composition effects on the flow are minimal for both \At numbers. The flow becomes well-mixed, the decay behavior converges and the small scale features reach their long time self-similar stage for all the cases investigated in this work.
\end{itemize}

\section{Acknowledgement}\label{Acknowledgement}
 
Arindam Banerjee acknowledges financial support from DOE/NNSA SSAA Program (Grant No. DE-NA0003195) and the U.S. National Science Foundation Early CAREER Program (Grant No. 1453056 from CBET-Fluid Dynamics). This work is co-authored by an employee of Triad National Security, LLC which operates Los Alamos National Laboratory under Contract No. 89233218CNA000001 with the U.S. Department of Energy/National Nuclear Security Administration. Computational resources were provided by the Institutional Computing  Program at Los Alamos National Laboratory and the Argonne Leadership Computing Facility at Argonne National Laboratory through a 2017 ALCC Award.

\bibliography{HVDT}
\end{document}